%% file: main.tex
\documentclass[manuscript,screen,nonacm]{acmart}

\usepackage{nicefrac}

\usepackage{amsfonts}
\usepackage{amsmath}
\usepackage[inkscapelatex=false]{svg}
\usepackage{graphicx}
\graphicspath{ {images/} }
\usepackage[boxruled,vlined,linesnumbered]{algorithm2e}
\usepackage{float} 
\usepackage{adjustbox}
\usepackage{multirow}
\usepackage{multicol}
\usepackage{arydshln}
\usepackage{subfig}
\usepackage[n ,advantage ,operators ,sets, adversary,landau,probability,notions,logic,ff,mm,primitives,events,complexity,asymptotics,keys]{cryptocode}
\usepackage{tikz}
\usepackage{float}
\usetikzlibrary{matrix,shapes,arrows,positioning,chains,calc,fit}
\usepackage[n ,advantage ,operators ,sets, adversary,landau,probability,notions,logic,ff,mm,primitives,events,complexity,asymptotics,keys]{cryptocode}
\usepackage{caption}
\usepackage{subcaption}
\usepackage{bm}
\usetikzlibrary{fit}
\usepackage{enumitem}

\newcommand{\KEM}{\ensuremath{\mathtt{KEM}} }

\AtBeginDocument{%
  \providecommand\BibTeX{{%
    \normalfont B\kern-0.5em{\scshape i\kern-0.25em b}\kern-0.8em\TeX}}}

\setcopyright{acmlicensed}
\copyrightyear{2024}
\acmYear{2024}
\acmDOI{XXXXXXX.XXXXXXX}

\acmISBN{978-1-4503-XXXX-X/18/06}

\begin{document}

\title[Scabbard: An Exploratory Study on Hardware Aware Design Choices of LWR-based KEMs]{Scabbard: An Exploratory Study on Hardware Aware Design Choices of Learning with Rounding-based Key Encapsulation Mechanisms}

\author{Suparna Kundu}
\email{Suparna.Kundu@esat.kuleuven.be}
\orcid{0000-0003-4354-852X}
\author{Quinten Norga}
\email{Quinten.Norga@esat.kuleuven.be}
\orcid{0000-0003-0983-5664}
\affiliation{%
  \institution{COSIC, KU Leuven}
  \streetaddress{Kasteelpark Arenberg 10, Bus 2452}
  \city{B-3001 Leuven-Heverlee}
  \country{Belgium}
}

\author{Angshuman Karmakar}
\email{angshuman@cse.iitk.ac.in}
\orcid{0000-0003-2594-588X}
\affiliation{%
  \institution{IIT Kanpur}
  \country{India}
}

\author{Shreya Gangopadhyay}
\email{gangopadhyay.shreya09@gmail.com}
\orcid{0009-0003-0718-8384}
\affiliation{%
  \institution{IIT Kharagpur}
  \country{India}
}

\author{Jose Maria Bermudo Mera}
\email{josebmera@gmail.com}
\orcid{0000-0003-0457-5728}
\affiliation{%
  \institution{PQShield, Oxford}
  \country{UK}
}

\author{Ingrid Verbauwhede}
\email{Ingrid.Verbauwhede@esat.kuleuven.be}
\orcid{0000-0002-0879-076X}
\affiliation{%
  \institution{COSIC, KU Leuven}
  \streetaddress{Kasteelpark Arenberg 10, Bus 2452}
  \city{B-3001 Leuven-Heverlee}
  \country{Belgium}
}

\renewcommand{\shortauthors}{Kundu et al.}

\begin{abstract}
  Recently, the construction of cryptographic schemes based on hard lattice problems has gained immense popularity. Apart from being quantum resistant, lattice-based cryptography allows a wide range of variations in the underlying hard problem. As cryptographic schemes can work in different environments under different operational constraints such as memory footprint, silicon area, efficiency, power requirement, etc., such variations in the underlying hard problem are very useful for designers to construct different cryptographic schemes.
  In this work, we explore various design choices of lattice-based cryptography and their impact on performance in the real world. In particular, we propose a suite of key-encapsulation mechanisms based on the learning with rounding problem with a focus on improving different performance aspects of lattice-based cryptography. Our suite consists of three schemes. Our first scheme is Florete, which is designed for efficiency. The second scheme is Espada, which is aimed at improving parallelization, flexibility, and memory footprint. The last scheme is Sable, which can be considered an improved version in terms of key sizes and parameters of the Saber key-encapsulation mechanism, one of the finalists in the National Institute of Standards and Technology's post-quantum standardization procedure. In this work, we have described our design rationale behind each scheme.
  
  Further, to demonstrate the justification of our design decisions, we have provided software and hardware implementations. Our results show Florete is faster than most state-of-the-art KEMs on software platforms. For example, the key-generation algorithm of high-security version Florete outperforms the National Institute of Standards and Technology's standard Kyber by $47\%$, the Federal Office for Information Security's standard Frodo by $99\%$, and Saber by $57\%$ on the ARM Cortex-M4 platform. Similarly, in hardware, Florete outperforms Frodo and NTRU Prime for all KEM operations. The scheme Espada requires less memory and area than the implementation of most state-of-the-art schemes. For example, the encapsulation algorithm of high-security version Espada uses $30\%$ less stack memory than Kyber, $57\%$ less stack memory than Frodo, and $67\%$ less stack memory than Saber on the ARM Cortex-M4 platform. The implementations of Sable maintain a trade-off between Florete and Espada regarding software performance and memory requirements. Sable outperforms Saber at least by $6\%$ and Frodo by $99\%$. Through an efficient polynomial multiplier design, which exploits the small secret size, Sable outperforms most state-of-the-art KEMs, including Saber, Frodo, and NTRU Prime. The implementations of Sable that use number theoretic transform-based polynomial multiplication (SableNTT) surpass all the state-of-the-art schemes in performance, which are optimized for speed on the Cortext M4 platform. The performance benefit of SableNTT against Kyber lies in between $7-29\%$, $2-13\%$ for Saber, and around $99\%$ for Frodo. 
  
\end{abstract}

\ccsdesc[500]{Security and privacy~Public key encryption}
\ccsdesc[100]{Security and privacy~Hardware-based security protocols}

\keywords{Post-quantum cryptography, Lattice-based cryptography, Learning with rounding, Key-encapsulation mechanism, Software implementations, AVX2, Cortex-M4, Hardware implementations, FPGA}


\maketitle

\input{sections/introduction}

\input{sections/preliminaries}

\input{sections/scabbard_suite}

\input{sections/parameters}

\input{sections/software_implementation}
\input{sections/hardware_implementation}

\input{sections/sca}

\input{sections/conclusion}

\begin{acks}
This work was partially supported by Horizon 2020 ERC Advanced Grant (101020005 Belfort), CyberSecurity Research Flanders with reference number VR20192203, BE QCI: Belgian-QCI (3E230370) (see beqci.eu), Intel Corporation, Secure Implementation of Post-Quantum Cryptosystems (SECPQC) DST-India and BELSPO.  
\end{acks}

\bibliographystyle{ACM-Reference-Format}
\bibliography{main}

\appendix
\input{sections/appendix}

\end{document}

%% file: sections/introduction.tex
\section{Introduction}\label{sec:intro}

Lattice-based cryptography has been one of the most discussed topics in public-key cryptography (PKC) for the past several years. Apart from being resistant to quantum attacks and hence a possible alternative for integer-factorization (IF) and discrete-log problem (DLP)-based cryptographic constructions, lattice-based cryptographic constructions are relatively simpler. Moreover, compared to IF and DLP, lattices offer lots of variations of underlying hard problems. This provides cryptographic designers with lots of maneuvering space to explore different designs to optimize and curate their cryptographic constructions for different applications. For example, from Ajtai's short-integer solution (SIS)~\cite{Ajtai96generatinghard} and Hoffstein et al.'s NTRU~\cite{NTRU-Kem} in 1996 to Regev's learning with errors (LWE)~\cite{Regev:2004} in 2005 and its subsequent variations such as Ring-LWE~\cite{ringlwe_2010}, Module-LWE~\cite{Langlois2015}, learning with rounding~\cite{Banerjee_lwr, DBLP:journals/iacr/Alperin-Sheriff16}, and recently discovered PLWE~\cite{plwe}, CLWE~\cite{clwe}, etc., the choice of computationally hard problems to design cryptographic schemes is in galore. Nevertheless, lattice-based cryptography has not always been the most preferred choice for cryptographers. IF and DLP-based cryptography, which were invented a couple of decades earlier, had already well-established themselves in the existing public-key infrastructure. Due to lots of research on their implementation, side-channel security, cryptanalysis, etc., their theoretical and implementation aspects were well understood. Therefore, there was little incentive for theorists and practitioners alike to replace these classical cryptosystems with lattice-based cryptography even though Shor's~\cite{Shor_1994, Proos_Zalka_2003} algorithm and its detrimental effect on IF and DLP-based cryptography was known since 1994. This happened partly because quantum computing research was mostly restricted to the realm of theory, and there was skepticism about its physical existence in the future.

However, as the research on developing large-scale quantum computers gained momentum, the future of IF and DLP-based cryptosystems as mainstream PKC algorithms started looking bleak proportionately. Due to the recent advancements in the field of quantum computing, the adverse effects of quantum computers on our existing public-key infrastructure have become too hard to ignore further. Although the research in quantum-resistant PKC or post-quantum cryptography started a couple of decades ago, the watershed moment in the process of transitioning from classical PKC to PQC is the National Institute of Standards and Technology's (NIST) conclusion of a long and multi-staged standardization procedure~\cite{nist_final_report} in 2022. NIST standardized PQC primitives such as public-key encryption (PKE) or key-encapsulation mechanism (KEM) Kyber~\cite{Kyber-Kem}, and digital signature schemes CRYSTALS-Dilithium~\cite{dilithium}, FALCON~\cite{web:falcon}, and SPHINCS+~\cite{web:sphincs}. 

During NIST's standardization process, the cryptographic community witnessed many innovations in the design and implementation of PQC. Such as the introduction of module lattices~\cite{Langlois2015} instead of more traditional standard~\cite{Regev:2004} or ideal~\cite{ringlwe_2010} lattices as a trade-off between speed and security, usage of central binomial distributions~\cite{newhope} instead of discrete Gaussian distribution for protection against potential side-channel attacks, just-in-time matrix generation in module lattices~\cite{saber_on_arm} and improvements in polynomial multiplications~\cite{NttFriendly,Saber_Time-memory,high_speed_NTT_HW,plantard_arithmetic_NTT} algorithms to improve efficiency, the introduction of error-correcting codes~\cite{round5} to reduce the decryption failure rates of lattice-based cryptography, etc. These different designs went through a thorough and rigorous evaluation. For example, the non constant-time behavior of error-correcting codes was found to be highly vulnerable to side-channel attacks; similarly, the sparse distributions used in schemes such as LAC~\cite{LAC} were found to be unsuitable for generating secret polynomials. On the other hand, improvements in the NTT polynomial multiplications, such as using K-reduction algorithm~\cite{longa_ntt}, or just-in-time generation of module lattices, almost became the standard choice. Therefore, the NIST standardization process does not mark a zenith in the research and development of lattice-based or PQC; rather, it has established a framework and set a course for future advancement in PQC.

In this work, we have decided to evaluate various design choices for constructing PQ KEM. We have particularly chosen the hard lattice problem learning with rounding (LWR). LWR is a relatively less used hard problem when designing lattice-based cryptographic schemes. 
\textcolor{black}{The LWR problem is a de-randomized variant of the LWE problem where a deterministic rounding to a smaller modulus replaces the error sampling. This problem was introduced by Banerjee et al.~\cite{Banerjee_lwr} in 2012. Several works have been done on the hardness of the LWR problem and deduced that the LWR problem is as hard as the LWE problem~\cite{DBLP:conf/crypto/AlwenKPW13,DBLP:conf/tcc/BogdanovGMRR16,DBLP:journals/iacr/Alperin-Sheriff16}.}
Nevertheless, in the context of PQ KEM Saber~\cite{saber_round_3}, one of the finalists in the NIST procedure, we have seen quite some intriguing results on the LWR-based schemes. In particular, the major reasons that motivated us to explore the LWR-based PQ KEMs further are described below.

LWR-based schemes require fewer pseudo-random numbers than LWE-based schemes, as errors are not required to be sampled explicitly here. The error is generated inherently from rounding operations, which helps to gain better performance. 
\textcolor{black}{The rounding modulus is smaller than the modulus of the LWE problem. Therefore for similar security levels, it results in smaller public-key sizes and ciphertext sizes. This also implies lesser bandwidth compared to the schemes based on the LWE problem. Although Kyber is based on the module-LWE (MLWE) problem, it also uses rounding on the encapsulation procedure (Compress function) to reduce the ciphertext size.}
In terms of performance, module-LWR (MLWR) based scheme Saber outperforms MLWE-based scheme Kyber in the cortex-M4  platform when an NTT-based multiplier is used for Saber (shown in Table~\ref{tab:M4_performance_stack}). As NTT-based polynomial multiplication takes a similar amount of cycles for Saber and Kyber, Saber doesn't require expensive error sampling. Also, Saber's auxiliary functions, such as compress, decompress, encode, and decode operations, are simple and cheaper than Kyber's, thanks to its power-of-two moduli.  

The choice of power-of-two moduli helps Saber achieve efficient hardware implementation as well. LWR-based schemes, in general, use Toom-Cook based polynomial multiplication instead of NTT-based polynomial multiplication. It helps to reduce the area requirements to implement LWR-based schemes in hardware compared to the LWE-based schemes. To perform NTT multiplication efficiently, the twiddle factors need to be stored in the memory. Also, the secret polynomial is smaller in the LWR-based scheme than LWE-based scheme, because in LWE-based scheme NTT needs to be performed on the secret polynomial, and it increases the memory requirement to store the secret key after the NTT. LWE-based schemes need to use a prime reduction algorithm (for Kyber, it is Montgomery and Barrett reduction), which also costs memory. However, no fancy reduction algorithm is required for LWR-based schemes due to its power-of-two moduli. All the factor mentioned above helps LWR-based schemes to a resource-scarce cryptographic application. There is already an implementation of MLWR-based KEM Saber in the application-specific integrated circuit (ASIC) available that uses the lowest area, lowest power, and low energy~\cite{DBLP:journals/jssc/GhoshMKDGVS23,DBLP:conf/cicc/GhoshMKDGVS22}.  

\textcolor{black}{There exist several physical attack \textit{i.e.} side-channel attacks (SCA) and fault injection attacks(FIA)~\cite{HCY2019,RBS+2020,ACL+2020,GJN2020,catinca_sca,RBS+2020,puja-rowhammer-fault,carry-your-fault} on both lattice-based signatures and KEMs. Since in this work, we are mostly concerned with KEMs, we will keep our discussion regarding physical attacks on PQC limited to KEMs only}. Masking~\cite{ChaJutRaoRoh1999} is a well-known and provably secure countermeasure against SCA.
The integration of the masking technique into a KEM scheme incorporates huge performance overhead. However, the performance overhead of the LWR-based KEM Saber after masking is comparatively less than the LWE-based scheme Kyber. State-of-the-art first-order masked Saber performs slightly better ($4\%$) than Kyber~\cite{FO-masked-kyber-HeinzKLPSS22,FO-masked-saber}, but the performance difference between Saber and Kyber is considerably much for higher-order masking. State-of-the-art second-order and third-order masked Saber perform $53\%$ and $48\%$ better than Kyber, respectively~\cite{HO_mask_Saber,BronchainC22}. There are several components, such as arithmetic-to-Boolean conversion, Boolean-to-arithmetic conversion, compress, encode, and decode, that are cheaper when power-of-two modulus (used in LWR-based schemes) is used instead of prime modulus (used in LWE-based schemes).
The secret and error of the LWE instances in the LWE-based schemes are generated from the same seed. There is a fault attack on LWE-based KEM where the successfully injected fault in the seed results in an LWE instance where the error and secret are the same~\cite{DBLP:conf/cosade/RaviRBCM19}. It breaks the hardness assumption of the LWE problem. However, as the LWR problem has no explicit error sampling, LWR-based schemes are naturally protected against this kind of fault attack.

Although NIST has selected Kyber as their first post-quantum secure KEM standard, several other standardization efforts are still in process, such as the Korean post-quantum competition (KPQC)~\cite{kpqc}. It is currently in its second round. \textcolor{black}{Smaug} is based on a combination of MLWE and MLWR problems, and its design is inspired by the initial Scabbard paper~\cite{Bermudo_Mera_Karmakar_Kundu_Verbauwhede_2021}. Smaug~\cite{smaug_kem} is one of the second-round KEM candidates. 
\textcolor{black}{
The currently selected PQC algorithms have been designed to address a variety of problems for many different applications. In particular, they have not been designed specifically for resource-constrained or Internet of Things (IoT) devices. Due to the rapid proliferation of these devices in almost every part of our digital ecosystem, they have become ubiquitous. However, due to their small sizes, it is often difficult to equip them with strong security measures. Due to these reasons often they become the weakest part of any security protocol. Therefore, there is an urgent need to design PQC schemes specifically for these devices. }

\textcolor{black}{There are two ways to design lightweight schemes for resource-constrained devices: design new schemes (different design components, different parameters, etc.) from scratch for resource-constrained devices or implement the existing schemes in a lightweight manner by probably trading off efficiency or reducing the security. The work on designing lightweight PQC has just begun to gain attention~\cite{DBLP:conf/ccs/BuchmannGGOP16,9027941}. Recently, a lightweight MLWE-based KEM, Rudraksh, has been proposed for resource-constrained devices~\cite{cryptoeprint:2024/1170}. Therefore, we believe that our current work on studying the exploration of various design and parameter choices will have a valuable positive impact on the seamless transition from classical to PQC. 
}
We also think that this study will help construct efficient schemes and improve state-of-the-art practices. In fact, NIST historically includes and updates their already standardized cryptographic primitives with efficient ones. For example, NIST first standardized the elliptic curves digital signature algorithm in 1999~\cite{nist_ecc_1999} and recommended 15 elliptic curves. After that, throughout the 2.5 decades, NIST has been modifying its recommended list of elliptic curves~\cite{fips_ecc_2009,fips_ecc_2013}, and the last modification was performed in 2023~\cite{fips_ecc_2023}. Therefore, the progress achieved in this work will benefit to improve the state-of-the-art of post-quantum cryptography.

\noindent\textbf{Contribution:} We propose Scabbard, a suite with three new LWR-based key-encapsulation mechanisms (KEMs): Florete, Espada, and Sable. To implement these three schemes efficiently, we utilized one of the NIST's finalist KEMs, Saber's optimized software and hardware implementations, and modified it according to our scheme requirements. In this paper, we extend our earlier work, Scabbard's initial suite~\cite{Bermudo_Mera_Karmakar_Kundu_Verbauwhede_2021} by proposing parameters for several new security versions of previously proposed schemes and also implementing them in software. We also present unified hardware implementations of a (medium) security version of all these three schemes. Below, we briefly elaborate on all of our contributions.

\begin{itemize}
 \item Florete is the first candidate of the Scabbard suite, and it is based on the ring-LWR (RLWR) hard problem. This KEM is designed to provide performance efficiency over the other LWE/LWR-based schemes. We proposed only the medium (NIST-3) security version in the initial paper. As an extension, we propose a low (NIST-1) and a high (NIST-5) security version of the Florete scheme in this paper. We show that all three security versions of Florete maintain the initial design rationale, and Florete performs better than most of the other lattice-based KEMs on the software platforms. 

 \item Espada is the second scheme of this suite, and its hardness depends on the MLWR problem. However, the size of the polynomial used in this scheme is as small as $64$, which is the first of its kind. The small polynomial size makes this scheme suitable for resource-constraint devices and also highly parallelizable in hardware. In this paper, we propose a low and high-security version of the Espada. Together with the previous medium security version of Espada~\cite{Bermudo_Mera_Karmakar_Kundu_Verbauwhede_2021}, this scheme now has three security versions, which broadens its applicability. We also show that all security versions of this scheme use less stack memory than most of the other lattice-based KEMs on the software platforms.  

 \item We explored parameter sets similar to Saber's and obtained slightly reduced parameter sets that provide similar security. This new variant of Saber is the third scheme of our suite, Sable. It is based on the MLWR problem, and the polynomial used in this scheme is of the same size as Saber ($256$). This scheme can also be considered an efficient variant of Saber. We implemented all three security versions of Sable and show that it performs better than Saber and requires less stack memory on software platforms. We also show that Sable performs better than Saber and has less memory footprint when implemented on hardware platforms.  

 \item We provide efficient implementations of all the schemes of Scabbard (also for all the security variants) on Intel's general-purpose processor and further optimize them with advanced vector instructions (AVX2). We also provide efficient implementations of Scabbard's schemes on the ARM Cortex-M4 platform. Low and high-security versions of Florete and Espada are implemented efficiently on all these software platforms for this paper. We compare our schemes with state-of-the-art schemes, such as the NIST standard Kyber, the Federal Office for Information Security's (BSI) standard Frodo, and several KPQC schemes on software platforms. We show that Florete performs better and Espada uses less amount of stack memory than most of the state-of-the-art KEMs on the Cortex-M4 platforms for all the security versions.
 
 \item MLWR-based scheme Saber is implemented using Toom-Cook (TC) based polynomial multiplication, but Chung et al.~\cite{chung_kannwischer_NTT_friendly} improved its performance with a number-theoretic transformation (NTT) based polynomial multiplication and then  Abdulrahman et al.~\cite{Saber_ntt_opt_implementation} improved it even more. To show that this result can be extended for the schemes of our suite, we propose an implementation of Sable with number theoretic transformation (NTT) based polynomial multiplication on the Cortex-M4 platform and call it SableNTT. Our SableNTT not only performs better than SaberNTT (Saber with NTT-based polynomial multiplication) but also performs better than Kyber-Speed (Kyber's implementation optimized for speed).

 \item We implement Florete, Espada, and Sable as full instruction-set coprocessor architectures on hardware (medium security version, NIST-3). By integrating and optimizing all building blocks, our design can compute all KEM operations in hardware: key generation, encapsulation, and decapsulation. As most individual components use non-multiples of 8-bit operands, hardware implementations become increasingly complex. We discuss our approach and optimized design, leading to reduced cycle counts and area counts. We utilize the polynomial multiplier architectures proposed in the initial paper. We show that our Sable implementation outperforms Saber, and all of the Scabbard schemes have comparable performance with state-of-the-art KEMs on hardware platforms.

\end{itemize}

%% file: sections/preliminaries.tex
\section{Preliminaries}\label{sec:preliminaries}

\subsection{Notation}

We represent the set of integers modulo $q$ by $\mathbb{Z}_q$ for a positive integer $q$. We use $\mathcal{R}_q^n$ to denote the quotient ring $\mathbb{Z}_q[x]/(x^n+1)$ or $\mathbb{Z}_q[x]/(x^n-x^{n/2}+1)$. The ring with $l$ length vectors over $\mathcal{R}_q^n$ is denoted by $(\mathcal{R}_q^n)^l$, and the ring of $m\times l$ matrices over $\mathcal{R}_q^n$ is refereed by $(\mathcal{R}_q^n)^{m\times l}$. We denote single polynomials by lower case letters and matrices by upper case letters. We denote by $\{x_i\}_{0 \leq i \leq t}$ to the set of $t+1$ elements $\{ x_0, x_1, \ldots, x_t \}$ from the same ring $\mathcal{R}$. If $x$ is sampled from the set $S$ according to the distribution $\chi$, then we use $x \leftarrow \mathcal{\chi} (S)$. When $x$ is generated from a seed $\mathtt{seed}_x$ using some pseudo-random number generator according to the distribution $\chi$ over the set $S$, then we denote it by $x \leftarrow \mathcal{\chi} (S;\ \mathtt{seed_x})$.
We denote the uniform distribution by $\mathcal{U}$. The centered binomial distribution (CBD) with standard deviation $\sqrt{\mu/2}$ is refereed as $\beta_\mu$. We use $\cdot$ to indicate matrix-vector and vector-vector multiplications. Here, we use scaling down function $\lfloor \cdot \rceil_p: \mathbb{Z}_q \longrightarrow \mathbb{Z}_p$ defined by $\lfloor x \rceil_p = \lfloor (q/p)x \rceil$, where $q>p$ and the rounding function $\lfloor y \rceil$ outputs the closest integer to the real number $y$, and during ties rounded upwards e.g. $\lfloor 1/2 \rceil = 1$. 
The operations $\lfloor \cdot \rceil_p$ can be extended for matrices and vectors by applying them coefficient-wise. Throughout this paper, the multiplication of two $n$ degree polynomials over the ring $\mathcal{R}_q^n$ is mentioned as $n\times n$ polynomial multiplication. We use $\cdot$ to represent multiplication between two polynomials, two vectors of polynomials, or one matrix and one vector of polynomials, depending on the context.    

\subsection{Learning with Rounding Problem}

The decision version of LWE problem \cite{Regev:2004} states that given $\mathbf{A}\leftarrow \mathcal{U}(\mathbb{Z}_q^{m\times l})$ and $\mathbf{s}$ and $\mathbf{e}$ are sampled according to following respective small distributions $\mathcal{\beta_{\mu}}(\mathbb{Z}_q^{l})$ and $\mathcal{\beta_{\mu}}(\mathbb{Z}_q^{m})$, distinguishing between the LWE sample $(\mathbf{A},\mathbf{b} = \mathbf{A}\cdot \mathbf{s}+\mathbf{e})\in \mathbb{Z}_q^{m\times l}\times \mathbb{Z}_q^{m}$ and $(\mathbf{A},\mathbf{b'})\in \mathbb{Z}_q^{m\times l}\times \mathbb{Z}_q^{m}$ is hard, when $\mathbf{b'}$ is sampled uniformly from $\mathbb{Z}_q^{m}$. The LWR problem~\cite{Banerjee_lwr,DBLP:journals/iacr/Alperin-Sheriff16} is a variation of the LWE problem, and the LWR sample is constructed as $(\mathbf{A},\mathbf{b} = \lfloor\mathbf{A}\cdot \mathbf{s}\rceil_p = \lfloor (q/p)\mathbf{A}\cdot \mathbf{s} \rceil)\in \mathbb{Z}_q^{m\times l}\times \mathbb{Z}_p^{m}$, where $\mathbf{s}\leftarrow\mathcal{\beta_\mu}(\mathbb{Z}_q^{l})$. Here, we do not need an explicit sampling of $\mathbf{e}$ rather, it generates implicitly from rounding. The decision version of LWR problem states that given $\mathbf{A}\in \mathbb{Z}_q^{m\times l}$, it is hard to differentiate between the LWR sample $(\mathbf{A},\mathbf{b} = \lfloor\mathbf{A}\cdot \mathbf{s}\rceil_p)\in \mathbb{Z}_q^{m\times l}\times \mathbb{Z}_p^{m}$ and $(\mathbf{A},\mathbf{b'})\in \mathbb{Z}_q^{m\times l}\times \mathbb{Z}_p^{m}$, where $\mathbf{s}\leftarrow\mathcal{\beta_\mu}(\mathbb{Z}_q^{l})$ and $\mathbf{b'}\leftarrow\mathcal{U}(\mathbb{Z}_p^{m})$.

Ring-LWE (RLWE) problem is a variant of the LWE problem based on structure lattice and is proposed in~\cite{ringlwe_2010} to improve the practicality and efficiency of cryptographic schemes. In the RLWE, $\mathbf{A},\ \mathbf{s},\ \mathbf{e},$ and $\mathbf{b}$ of the LWE are all replaced by polynomials of the ring $({\mathcal{R}_q^n})$. Similar to the RLWE, we can define the decision version of the RLWR problem, which states that the RLWR sample $(\mathbf{a},\mathbf{b} = \lfloor\mathbf{a}\cdot \mathbf{s}\rceil_p)\in {\mathcal{R}_q^n}\times {\mathcal{R}_p^n}$ and $(\mathbf{a},\mathbf{b'})\in {\mathcal{R}_q^n}\times {\mathcal{R}_p^n}$ are hard to distinguish, where $\mathbf{s}\leftarrow\mathcal{\beta_\mu}({\mathcal{R}_q^n})$ and $\mathbf{b'}\leftarrow\mathcal{U}({\mathcal{R}_p^n})$. The ring version offers better efficiency and practicality compared to the cryptosystem based on standard lattices of similar security. However, due to the presence of additional structures, many researchers are skeptical about their hardness. Therefore, as a trade-off between security and efficiency, the MLWR problem was introduced~\cite{Langlois2015}, which states that it is hard to differentiate between the MLWR sample $(\mathbf{A},\mathbf{b} = \lfloor\mathbf{A}\cdot \mathbf{s}\rceil_p)\in ({\mathcal{R}_q^n})^{l\times l}\times ({\mathcal{R}_p^n})^{l}$ and $(\mathbf{A},\mathbf{b'})\in ({\mathcal{R}_q^n})^{l\times l}\times ({\mathcal{R}_p^n})^{l}$, where $\mathbf{s}\leftarrow\mathcal{\beta_\mu}(({\mathcal{R}_q^n})^{l})$ and $\mathbf{b'}\leftarrow\mathcal{U}(({\mathcal{R}_p^n})^{l})$. 
The rank of the underlying lattice of this MLWR problem is $n\times l=n'$. MLWR has less structure than RLWR, and the expensive matrix-vector multiplication of the standard LWR problem is replaced by efficient polynomial multiplication in the MLWR problem.

The module lattice-based problem can be used as generic construction, as all the MLWR problems with $n=n'$ and $l=1$ are classified as RLWR problems, and all the MLWR problems with $n=1$ and $l=n'$ are categorized as standard LWR problems.  At this moment, there is no attack that provides any advantage to the adversary for MLWR or RLWR problems over the standard LWR problem. Therefore, if the rank of the underlying lattice problem is the same, then the security provided by the problem is the same. Henceforth, we will use the MLWR problem to denote different variations of the LWR problem.


\subsection{Construction of Generic LWR-based KEM}\label{sec:genericLWR}

The LWR-based public-key encryption (PKE) scheme is used to construct an LWR-based key-encapsulation mechanism (KEM), and we illustrate the PKE scheme in Fig.~\ref{fig:lwr_pke}. It consists of three algorithms (i) key-generation (\texttt{LWR.PKE.KeyGen}), (ii) encryption (\texttt{LWR.PKE.Enc}), and (iii) decryption (\texttt{LWR.PKE.Dec}). Firstly, the \texttt{LWR.PKE.KeyGen} algorithm generates the public key and secret key pair. Secondly, the \texttt{LWR.PKE.Enc} algorithm uses the public key to encrypt the message $m$ and to produce ciphertext. Lastly, the \texttt{LWR.PKE.Dec} algorithm decrypts the received ciphertext to the message $m'$.

Three quotient rings ${\mathcal{R}_q^n}, {\mathcal{R}_p^n}, {\mathcal{R}_t^n}$ has been used here, and $t<p<q$. The constants $\epsilon_q=\log_2(q), \epsilon_p=\log_2(p), \epsilon_t=\log_2(t)$ are used to construct the constant polynomials $h_2$, $h_3$, and by the vector of constant polynomials $\mathbf{h_1}$. Each coefficient of the constant polynomials $h_2$ and $h_3$ are $2^{(\epsilon_q-\epsilon_p-1)}$ and  $(2^{(\epsilon_p-B-1)}-2^{(\epsilon_p-\epsilon_t-1)})$, respectively. The value of each coefficient of the vector with constant polynomials $\mathbf{h_1}$ is $2^{(\epsilon_q-\epsilon_p-1)}$. The extendable-output function $\mathtt{XOF}: \{0,\ 1\}^{256} \longrightarrow \{0,\ 1\}^*$ is a pseudorandom number generator realized with \texttt{SHAKE-128}. 

As the LWR-based schemes are not perfect, there is always a possibility of a decryption failure, i.e. the encrypted message $m$ and decrypted message $m'$ are not equal even when the scheme is executed properly. The decryption failure probability depends on the decryption noise $d = v''-v'$. The decryption failure will not occur if the decryption noise $d$ satisfies the following relation $| d | \leq \frac{p}{2^{B+1}}(1-\frac{1}{t})$~\cite{DBLP:conf/ccs/BosCDMNNRS16}. Therefore, if $t= 2^\epsilon_t$ is large enough, then the decryption failure probability, $\delta$, becomes negligible. Eventually, it makes the corresponding LWR-based PKE scheme $(1-\delta)$ correct. 

\input{image/LWREnc}


The aforementioned LWR-based PKE scheme is indistinguishable against chosen plaintext attacks (IND-CPA) and can be converted to an indistinguishable under adaptive chosen ciphertext attacks (IND-CCA) KEM with a modified version of Fujisaki-Okamoto transformation~\cite{DBLP:journals/joc/FujisakiO13} proposed by Hofheinz \textit{et al\,.}~\cite{DBLP:conf/tcc/HofheinzHK17}. Authors show that if the underlying PKE scheme is $(1-\delta)$ correct, then the KEM is also $(1-\delta)$ correct. The KEM is $S$ bit post-quantum secure only when the failure probability $\delta<2^{-S}$~\cite{Jiang_2017}.       

\input{image/LWRKEM}

The CCA-secure LWR-based KEM based on the CPA-secure LWR-based PKE is presented in Fig.~\ref{fig:lwr_kem}. This KEM consists of three algorithms: (i) key-generation (\texttt{LWR.KEM.KeyGen}), (ii) encapsulation (\texttt{LWR.KEM.Encaps}), and (iii) decapsulation (\texttt{LWR.KEM.Decaps}). Here, the key-generation algorithm generates the public key and secret key pair ($\overline{pk},\ \overline{sk}$). Secondly, the encapsulation algorithm uses the public key $\overline{pk}$ to encrypt the message and to produce ciphertext $c$ and the session key $K$. Lastly, in the decapsulation algorithm, we decrypt the received ciphertext to the message then we re-encrypt the decrypted message using the public key. If the re-encrypted ciphertext is equal to the received ciphertext, then the algorithm outputs the session key $K$; else outputs a random key. In these algorithms, we use two hash functions $H: \{0,\ 1\}^* \longrightarrow \{0,\ 1\}^{256}$ realized by \texttt{SHA3-256} and $\mathcal{G}: \{0,\ 1\}^* \longrightarrow \{0,\ 1\}^{512}$ implemented with \texttt{SHA3-512}. In \texttt{LWR.KEM.Encaps} and \texttt{LWR.KEM.Decaps}, the $\mathtt{arrange\_msg} : \{0,\ 1\}^{256} \longrightarrow {\mathcal{R}_q^n}$ function is used, which converts the $256$ bits message to the message polynomial in ${\mathcal{R}_q^n}$. The inverse of $\mathtt{arrange\_msg}$ function $\mathtt{original\_msg} : {\mathcal{R}_q^n} \longrightarrow \{0,\ 1\}^{256}$ is required in \texttt{LWR.KEM.Decaps}. It converts the message polynomial in ${\mathcal{R}_q^n}$ to the $256$ bits message.

%% file: image/LWREnc.tex
\begin{figure}[ht]
\fbox{\begin{varwidth}{\textwidth}
    \raggedright 
    $\underline{\mathtt{LWR}{.}\mathtt{PKE}{.}\mathtt{KeyGen} ()}$
    \begin{enumerate}[wide=0em, itemsep=0pt, parsep=0pt, font=\scriptsize\tt\color{gray}]
            \item $\mathtt{seed_{\pmb{A}}} \leftarrow  \mathcal{U}(\{0,\ 1\}^{256})$ \hfill $\rhd$ seed of the public matrix $\pmb{A}$\\
            \item $\pmb{A} \leftarrow \mathcal{U}({\text{(}\mathcal{R}_q^n\text{)}}^{l\times l};\ \mathtt{seed_{\pmb{A}}}) $  \hfill $\rhd$ $\pmb{A}$ is generated using \texttt{XOF} function over $\mathtt{seed_{\pmb{A}}}$\\ 
            \item $\mathtt{seed_{\pmb{s}}} \leftarrow \mathcal{U}(\{0,\ 1\}^{256})$ \hfill $\rhd$ seed of the secret vector $\pmb{s}$\\
            \item $\pmb{s} \leftarrow \beta_\mu ({\text{(}\mathcal{R}_q^n\text{)}}^l;\ \mathtt{seed_{\pmb{s}}})$  \hfill $\rhd$ $\pmb{s}$ is generated using CBD $\beta_\mu$ over the \texttt{XOF}($\mathtt{seed_{\pmb{s}}}$)\\
            \item $\pmb{b} \leftarrow  ((\pmb{A}^T \cdot \pmb{s} + \pmb{h_1}) \bmod{q}) \gg (\epsilon_q - \epsilon_p) \in {\text{(}\mathcal{R}_p^n\text{)}}^l$   \hfill $\rhd$ Performing rounding operation on $\pmb{A}^T \cdot \pmb{s}$ to create $\pmb{b}$\\
            \item \textbf{return} $(pk = (\mathtt{seed_{\pmb{A}}},\ \pmb{b}),\ sk =  (\pmb{s}))$   \hfill $\rhd$ $pk$ public key and $sk$ secret key\\
    \end{enumerate} 
    \vspace{2mm}
    \raggedright
    $\underline{\mathtt{LWR}{.}\mathtt{PKE}{.}\mathtt{Enc} (pk = (\mathtt{seed_{\pmb{A}}}, \pmb{b}), m \in R_2; r )}$ \hfill $\rhd$ $pk$ is sent via insecure channel\\
    \begin{enumerate}[wide=0em, itemsep=0pt, parsep=0pt, font=\scriptsize\tt\color{gray}]
        \item $\pmb{A} \leftarrow \mathcal{U}({\text{(}\mathcal{R}_q^n\text{)}}^{l\times l};\ \mathtt{seed_{\pmb{A}}}) $  \hfill $\rhd$ $\pmb{A}$ is re-generated using \texttt{XOF} function over public $\mathtt{seed_{\pmb{A}}}$\\
        \item \textbf{if: } $r$  is not specified:
        \item \quad $ r \leftarrow \mathcal{U}(\{0,\ 1\}^{256})$ \hfill $\rhd$ seed of the $\pmb{s'}$ \\
        \item $\pmb{s'} \leftarrow \beta_\mu ( {\text{(}\mathcal{R}_q^n\text{)}}^l ;\ r)$ \hfill $\rhd$ $\pmb{s'}$ is produced using CBD $\beta_\mu$ over the \texttt{XOF}($\mathtt{seed_{\pmb{s'}}}$)\\
        \item $\pmb{u} \leftarrow ( ( \pmb{A} \cdot \pmb{s}' + \pmb{h_1}) \bmod{q})  \gg (\epsilon_q - \epsilon_p)  \in {\text{(}\mathcal{R}_p^n\text{)}}^l$ \hfill $\rhd$ Creates $\pmb{b'}$ key contained part of the ciphertext\\ 
        \item $ v' \leftarrow \pmb{b}^T \cdot (\pmb{s}' \bmod{p}) + h_2  \in \mathcal{R}_p^n$\\
        \item $v \leftarrow (v' - 2^{\epsilon_p-B} m \bmod{p}) \gg (\epsilon_p - \epsilon_t - B) \in \mathcal{R}_{2^{B}t}^n $  \hfill $\rhd$ $\pmb{v}$ message contained part of the ciphertext\\
        \item \textbf{return} $c =(\pmb{u},\ v)$  \hfill $\rhd$ Ciphertext c\\
    \end{enumerate}
    \vspace{2mm}
    \raggedright
    \underline{$\mathtt{LWR}{.}\mathtt{PKE}{.}\mathtt{Dec}(sk=\pmb{s},c=(\pmb{u}, v))$}\hfill $\rhd$ $sk$ is stored and $c$ is delivered via insecure channel\\
    \begin{enumerate}[wide=0em, itemsep=0pt, parsep=0pt, font=\scriptsize\tt\color{gray}]
        \item $v'' \leftarrow \pmb{u}^{T} \cdot (\pmb{s} \bmod{p})+h_2\in \mathcal{R}_p^n$ \\
        \item $m' \leftarrow (v''-2^{\epsilon_p-\epsilon_t-B}v + h_3) \bmod{p}\gg (\epsilon_p-B)\in \mathcal{R}_{2^{B}t}^n$ \hfill $\rhd$ Recover message by decoding the ciphertext $c$ \\
        \item \textbf{return} $m'$\\
    \end{enumerate}
\end{varwidth}}
\caption{Generic $\mathtt{LWR}{.}\mathtt{PKE}$}
\label{fig:lwr_pke}
\end{figure}

%% file: image/LWRKEM.tex
\begin{figure}[ht]
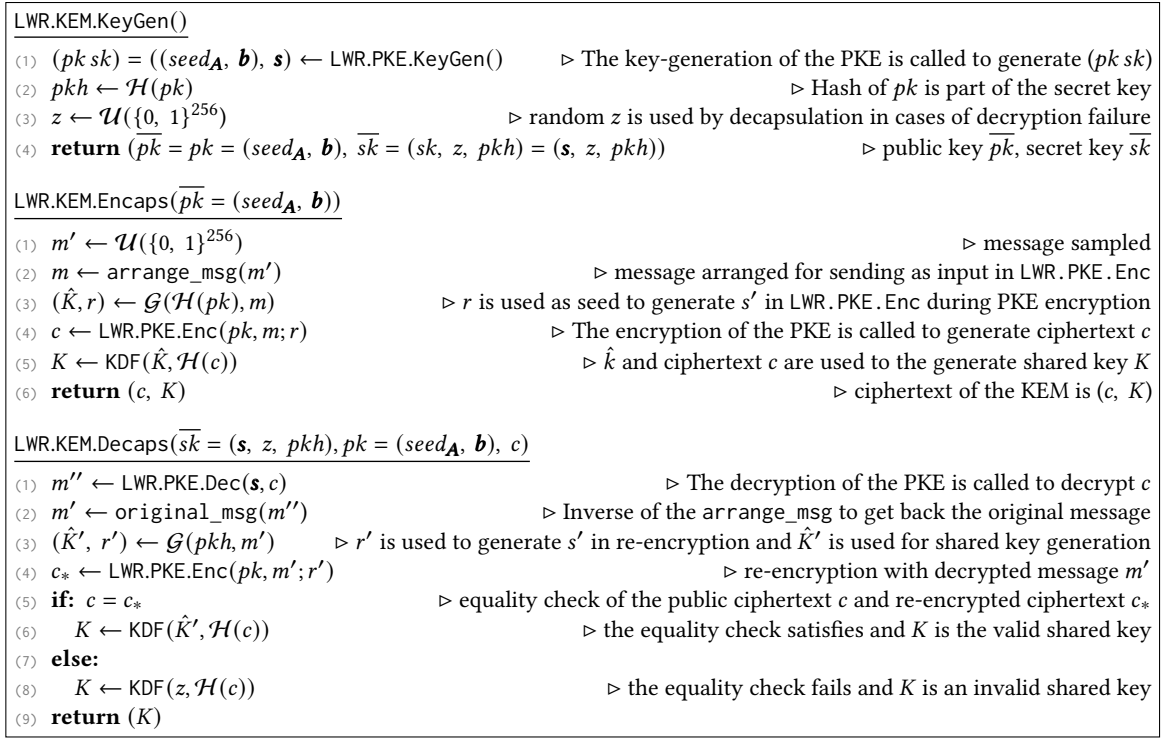

\fbox{\begin{varwidth}{\textwidth}
    \raggedright 
    $\underline{\mathtt{LWR}{.}\mathtt{KEM}{.}\mathtt{KeyGen} ()}$
    \begin{enumerate}[wide=0em, itemsep=0pt, parsep=0pt, font=\scriptsize\tt\color{gray}]
        \item $(pk\, sk) = ((seed_{\pmb{A}},\ \pmb{b}),\ \pmb{s})  \leftarrow \mathtt{LWR}{.}\mathtt{PKE}{.}\mathtt{KeyGen} ()$ \hfill $\rhd$ The key-generation of the PKE is called to generate ($pk\, sk$)\\
        \item $pkh \leftarrow \mathcal{H}(pk) $ \hfill $\rhd$ Hash of $pk$ is part of the secret key\\
        \item $z \leftarrow \mathcal{U}(\{0,\ 1\}^{256}$)  \hfill $\rhd$ random $z$ is used by decapsulation in cases of decryption failure \\
        \item \textbf{return} $(\overline{pk} = pk = (seed_{\pmb{A}},\ \pmb{b}),\ \overline{sk} =  (sk,\ z,\ pkh ) =(\pmb{s},\ z,\ pkh))$  \hfill $\rhd$ public key $\overline{pk}$, secret key $\overline{sk}$\\
    \end{enumerate} 
    \vspace{2mm}
    \raggedright
    $\underline{\mathtt{LWR}{.}\mathtt{KEM}{.}\mathtt{Encaps} (\overline{pk} = (seed_{\pmb{A}},\ \pmb{b}))}$
    \begin{enumerate}[wide=0em, itemsep=0pt, parsep=0pt, font=\scriptsize\tt\color{gray}]
        \item $m^\prime  \leftarrow  \mathcal{U}(\{0,\ 1\}^{256}) $  \hfill $\rhd$ message sampled\\
        \item $m \leftarrow \mathtt{arrange\_msg}(m^\prime)$ \hfill $\rhd$ message arranged for sending as input in \texttt{LWR.PKE.Enc}\\
        \item $(\hat{K}, r) \leftarrow \mathcal{G}(\mathcal{H}(pk), m)$ \hfill $\rhd$ $r$ is used as seed to generate $s'$ in \texttt{LWR.PKE.Enc} during PKE encryption\\
        \item $c \leftarrow \mathtt{LWR}{.}\mathtt{PKE}{.}\mathtt{Enc} (pk, m; r )$ \hfill $\rhd$ The encryption of the PKE is called to generate ciphertext $c$\\
        \item $K \leftarrow \mathtt{KDF}(\hat{K}, \mathcal{H}(c))$ \hfill $\rhd$ $\hat{k}$ and ciphertext $c$ are used to the generate shared key $K$\\
        \item \textbf{return} $(c,\ K)$ \hfill $\rhd$ ciphertext of the KEM is ($c,\ K$)
    \end{enumerate} 
    \vspace{2mm}
    \raggedright
    $\underline{\mathtt{LWR}{.}\mathtt{KEM}{.}\mathtt{Decaps} (\overline{sk} = (\pmb{s},\ z,\ pkh),pk = (seed_{\pmb{A}},\ \pmb{b}),\ c)}$
    \begin{enumerate}[wide=0em, itemsep=0pt, parsep=0pt, font=\scriptsize\tt\color{gray}]
        \item $m''  \leftarrow  \mathtt{LWR}{.}\mathtt{PKE}{.}\mathtt{Dec} (\pmb{s}, c )$ \hfill $\rhd$ The decryption of the PKE is called to decrypt $c$\\
        \item $m' \leftarrow \mathtt{original\_msg}(m''$) \hfill $\rhd$ Inverse of the $\mathtt{arrange\_msg}$ to get back the original message\\
        \item $(\hat{K}',\ r') \leftarrow \mathcal{G}(pkh, m')$ \hfill $\rhd$ $r'$ is used to generate $s'$ in re-encryption and $\hat{K}'$ is used for shared key generation\\
        \item $c_* \leftarrow \mathtt{LWR}{.}\mathtt{PKE}{.}\mathtt{Enc} (pk, m'; r')$ \hfill $\rhd$ re-encryption with decrypted message $m'$\\
        \item \textbf{if: } $c=c_*$ \hfill $\rhd$ equality check of the public ciphertext $c$ and re-encrypted ciphertext $c_*$ \\
        \item \quad $ K \leftarrow \mathtt{KDF}(\hat{K}', \mathcal{H}(c))$ \hfill $\rhd$ the equality check satisfies and $K$ is the valid shared key\\
        \item \textbf{else: }
        \item \quad $ K \leftarrow \mathtt{KDF}(z, \mathcal{H}(c))$ \hfill $\rhd$ the equality check fails and $K$ is an invalid shared key\\
        \item \textbf{return} $(K)$
    \end{enumerate} 
\end{varwidth}}
\caption{Generic $\mathtt{LWR}{.}\KEM$}
\label{fig:lwr_kem}
\end{figure}

%% file: sections/scabbard_suite.tex
\section{Scabbard suite of LWR-based KEMs}\label{section:scabbard}

We present the schemes of the suite Scabbard in this section. These schemes have been designed to improve the state-of-the-art of the efficient lattice-based KEM. This suite consists of three different designs of LWR-based KEMs (i) Florete, (ii) Espada, and (iii) Sable. All three schemes follow the generic LWR-based KEM construction. Florete is designed to achieve better performance. Espada is designed to use less memory footprint when implemented for resource-constrained devices and can also be implemented efficiently in hardware by using its high parallelism. Sable is designed to provide a trade-off between performance and memory usage. 
As we can see from Fig.~\ref{fig:lwr_pke}, for a LWR-based KEM, everything is the same except the choice of the ring/ module parameters $n$ and $l$, the CBD parameter $\mu$, moduli $q,\ p,\ t$. Therefore, the polynomial multiplication, message encoding and decoding, and the secret sampler used in these three schemes are different. We will discuss these different aspects of the KEMs and describe their design rationale in the following sections. In Scabbard's KEMs design, one of the important aspects is we tried to maximally utilize already developed optimized software and hardware modules of LWR-based schemes.

\subsection{Florete: RLWR based KEM}\label{sec:florete}

This scheme is based on the RLWR hard problem, and therefore the ring/modulus parameter $l$ is equal to $1$. The public matrix $\mathbf{A}$, secret vector $\mathbf{s}$, and public key vector $\mathbf{b}$ are all polynomials and are elements of the ring $\mathcal{R}_q^n$. The parameter $n$ and the ring $\mathcal{R}_q^n$ vary for different security versions of Florete. We choose $n=512$ for the low-security version, $n=768$ for the medium-security version, and $n=1024$ for the high-security version. We are required to use an irreducible polynomial to construct the ring $R_q^n$ for the RLWR problem~\cite{ringlwe_2010} (otherwise hardness of the RLWR problem reduces). Generally, $x^n+1$ is chosen as an irreducible polynomial to construct the ring $R_q^n$, but $x^{768}+1$ is not an irreducible polynomial. Therefore, the irreducible polynomial $(x^{768}-x^{384}+1)$ is applied to construct $R_q^n$ for $n=768$. 
\newline

\noindent\textbf{Polynomial Multiplication}

\noindent Polynomial multiplication is one of the fundamental operations performed during all three algorithms of a KEM, and it is one of the most time-consuming operations. The procedure of this multiplication depends on the two parameters of $\mathcal{R}_q^n$, which are $n$ and modulus $q$. As mentioned earlier, we plan to utilize the optimized software and hardware modules developed for LWR-based schemes (e.g., Saber) during the NIST competition for easier adaptation.
Polynomial multiplication is one of the modules whose implementation has been optimized in several works~\cite{acns_saber, Saber_Time-memory, Sujoy_Saber_h/w, NeonNTT}. LWR-based schemes can not use fast number theoretic transformation for polynomial multiplication because the modulus $q$ and $n$ are not co-prime ($\mathtt{gcd}(q,n)>1$). The next best option for the $n\times n$ polynomial multiplication is utilizing Toom-Cook or Karatsuba multiplication, which has been used and optimized for the MLWR-based KEM Saber~\cite{Saber_Time-memory, Sujoy_Saber_h/w}. Therefore, we have decided to re-purpose Saber's efficient $256\times 256$ multiplier for Florete's $n\times n$ multiplication.   
%
We use Saber's efficient $256\times 256$ multiplier for implementing all three polynomial multiplications of Florete. The $256\times 256$ polynomial multiplication of Saber is implemented by using a layer of Toom-Cook4 multiplication followed by two layers of Karatsuba multiplication, and the last stage is $16\times 16$ schoolbook multiplication. 

There is a small problem in using Saber's multiplier in Florete. We target to fit a coefficient of the multiplier polynomial fit into $16$ bit space for efficient implementation in vector processors, small microcontrollers (e.g. Cortex-M4), etc. Even though the coefficients of the multiplier are less than or equal to $16$ bit, we need to save some extra space when performing division by some $g$, which is a divisor of $2$. The reason is $\mathtt{gcd}(g,\ q)\geq 2$, and the inverse of $g$ does not exist in $\mathbb{Z}_q$. In this case, let us assume $y/g$ needs to be computed. If $g=h*2^w$, where $\mathtt{gcd}(h,\ 2)= 1$, then $\mathtt{gcd}(h,\ q)= 1$ (as $q$ is a power-of-2 modulus). We first compute $h^{-1}$, and then we multiply it with $g$ and perform $w$ right shift afterward. Therefore we need to store $w$ extra bits while performing $g\cdot h^{-1}$. There are several such divisions by $g$, where $g=h*2^w$ are needed while using Toom-Cook multiplications. The maximum value of such $w$ is equal to $1$ for Toom-Cook $3$-way multiplication, whereas $w$ is equal to $3$ for Toom-Cook $4$-way multiplication. For the $512\times 512$ polynomial multiplication, we use one extra layer of Karatsuba multiplication on top of Saber's $256\times 256$ multiplication. Therefore, the $\log_2$ of the modulus $q$, $\epsilon_q$ need to be $\leq13$ for the low-security version of Florete. For the $768\times 768$ polynomial multiplication, we add an extra layer of Toom-Cook $3$-way multiplication on top of the $256\times 256$ multiplication. Here, the $\epsilon_q$ need to be $\leq12\ (=16-3-1)$ (It is not possible with Saber's modulus $q$, which is $2^{13}$). We apply Toom-Cook $4$-way multiplication on top of the $256\times 256$ multiplication for the $1024\times 1024$ polynomial multiplication. The modulus $\epsilon_q$ need to be $\leq10\ (=16-3-3)$ in this case.

Now, we will compare the number of $256\times 256$ polynomial multiplications used in Saber with Florete. As Saber is based on the MLWR problem, it has a module structure and the parameter $l>1$. Therefore several $256\times 256$ polynomial multiplications are needed for matrix-vector multiplications (e.g. $\pmb{A}\cdot\pmb{s}$) and vector-vector multiplications (e.g. $\pmb{b}^T\cdot \pmb{s'}$), which are used in all three (key-generation, encapsulation, and decapsulation) algorithms of all three security versions of Saber. These exact numbers are provided in Table~\ref{table:mult_comp_florete}. For example, the key-generation, encapsulation, and decapsulation algorithm of the medium-security version of Saber requires $9$, $12$, and $15$ polynomial multiplication ($256\times 256$), respectively. 

\begin{figure}[ht!]
\centering
  \includegraphics[scale=.5]{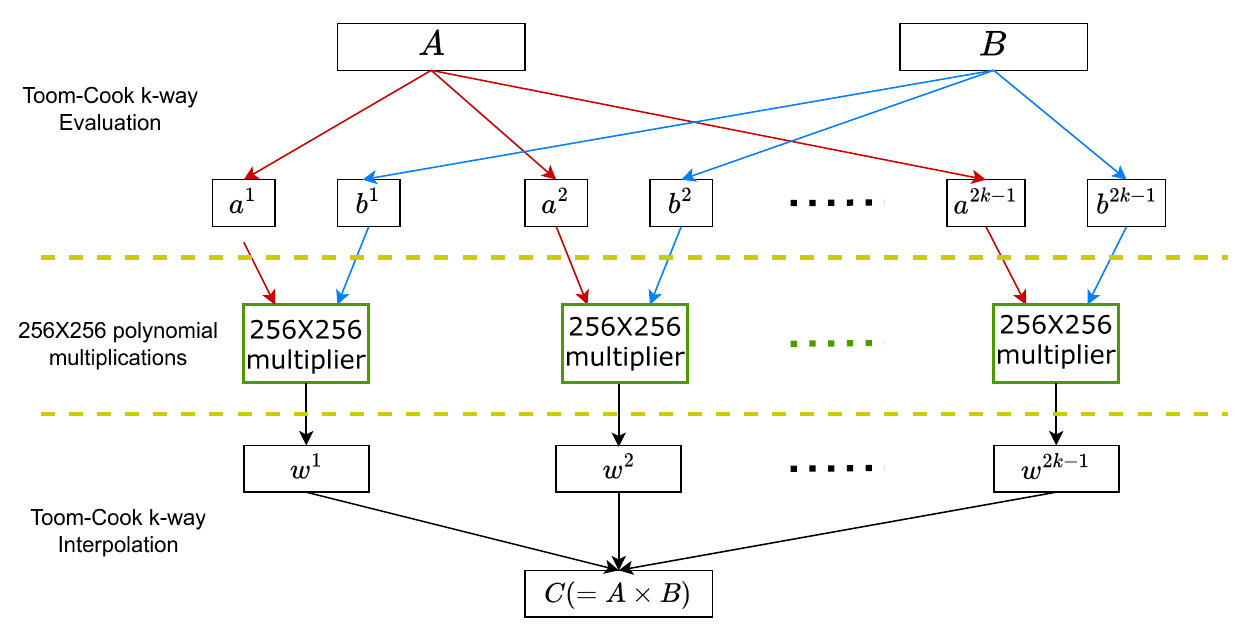}
  \caption{Polynomial multiplication used in Florete. The values of $k$ for low, medium, and security versions of Florete are $2$, $3$, and $4$, respectively.}
  \label{fig:florete_multiplication}
\end{figure}

Florete is an RLWR-based scheme. So, all matrix-vector multiplications (e.g. $\pmb{A}\cdot\pmb{s}$) and vector-vector multiplications (e.g. $\pmb{b}^T\cdot \pmb{s'}$) are just a single polynomial multiplication in Florete. The key-generation, encapsulation, and decapsulation algorithms of Florete require $1$, $2$, and $3$ $n\times n$ polynomial multiplications, respectively. As mentioned earlier, we apply an extra layer of Karatsuba multiplication on top of Saber's $256\times 256$ multiplication for $512\times 512$ polynomial multiplication of the low-security version of Florete. Here, we perform $3$ $256\times 256$ polynomial multiplications for a $512\times 512$ polynomial multiplication (as displayed in Fig.~\ref{fig:florete_multiplication}). There are some other steps which are interpolation and reduction, to complete the whole multiplication. However, the performance cost of these steps is negligible compared to the total number of polynomial multiplications.
Therefore, the low-security version of Florete needs $3$, $6$, and $9$ $256\times 256$ multiplications for the key-generation, encapsulation, and decapsulation algorithms, respectively. For a $768\times768$ polynomial multiplication, an extra layer of Toom-Cook $3$-way multiplication is added on top of $256\times 256$ multiplication. As we are applying Toom-Cook $3$-way multiplication for $768\times 768$ polynomial multiplication, we need to perform $5 =(2*3-1)$, $256\times 256$ polynomial multiplications (as portrayed in Fig.~\ref{fig:florete_multiplication}). We are applying an extra layer of Toom-Cook $4$-way multiplication for $1024\times 1024$ polynomial multiplication. So, we need to perform $7 =(2*4-1)$, $256\times 256$ polynomial multiplications for a single $1024\times 1024$ multiplication (as shown in Fig.~\ref{fig:florete_multiplication}). We provide the number of $256\times 256$ polynomial multiplications required by all the algorithms of all the security versions of Florete in Table~\ref{table:mult_comp_florete}. 
\textcolor{black}{We have included the performance of multiplications used in Florete and Saber on the Cortex-M4 platform. More detailed performance results are given in Sec.~\ref{sec:SW_impl}.}   
\begin{table}[!ht]
\centering
\caption{\textcolor{black}{Comparison of the usage of $256\times 256$ multiplications in the algorithms of Florete with Saber.}}
\label{table:mult_comp_florete}
\begin{tabular}{ccccc|rrr}
\hline
{\color[HTML]{3531FF} } & {\color[HTML]{3531FF} } & \multicolumn{3}{c|}{{\color[HTML]{3531FF} }} & \multicolumn{3}{c}{{\color[HTML]{3531FF} Multiplication on Cortex-M4}} \\
{\color[HTML]{3531FF} } & {\color[HTML]{3531FF} } & \multicolumn{3}{c|}{\multirow{-2}{*}{{\color[HTML]{3531FF} $\# 256\times 256$ multiplications}}} & \multicolumn{3}{c}{{\color[HTML]{3531FF} (x1000 clock cycles)}} \\ \cline{3-8} 
\multirow{-3}{*}{{\color[HTML]{3531FF} \begin{tabular}[c]{@{}c@{}}Scheme \\ Name\end{tabular}}} & \multirow{-3}{*}{{\color[HTML]{3531FF} \begin{tabular}[c]{@{}c@{}}Security \\ level\end{tabular}}} & {\color[HTML]{3531FF} KeyGen} & {\color[HTML]{3531FF} Encaps} & {\color[HTML]{3531FF} Decaps} & \multicolumn{1}{c}{{\color[HTML]{3531FF} KeyGen}} & \multicolumn{1}{c}{{\color[HTML]{3531FF} Encaps}} & \multicolumn{1}{c}{{\color[HTML]{3531FF} Decaps}} \\ \hline
\multicolumn{1}{l}{} & {\color[HTML]{FD6864} Low} & {\color[HTML]{000000} \textbf{3}} & {\color[HTML]{000000} 6} & {\color[HTML]{000000} 9} & {\color[HTML]{333333} \textbf{121}} & {\color[HTML]{333333} 243} & {\color[HTML]{333333} 364} \\
{\color[HTML]{036400} Florete} & {\color[HTML]{CB0000} Medium} & {\color[HTML]{000000} \textbf{5}} & {\color[HTML]{000000} \textbf{10}} & {\color[HTML]{000000} 15} & {\color[HTML]{333333} \textbf{202}} & {\color[HTML]{333333} \textbf{405}} & {\color[HTML]{333333} 607} \\
\multicolumn{1}{l}{} & {\color[HTML]{680100} High} & {\color[HTML]{000000} \textbf{7}} & {\color[HTML]{000000} \textbf{14}} & {\color[HTML]{000000} \textbf{21}} & {\color[HTML]{333333} \textbf{300}} & {\color[HTML]{333333} \textbf{600}} & {\color[HTML]{333333} 900} \\ \hline
\multicolumn{1}{l}{} & {\color[HTML]{FD6864} Low} & {\color[HTML]{000000} 4} & {\color[HTML]{000000} 6} & {\color[HTML]{000000} 8} & {\color[HTML]{333333} 149} & {\color[HTML]{333333} 223} & {\color[HTML]{333333} 298} \\
{\color[HTML]{036400} Saber} & {\color[HTML]{CB0000} Medium} & {\color[HTML]{000000} 9} & {\color[HTML]{000000} 12} & {\color[HTML]{000000} 15} & {\color[HTML]{333333} 334} & {\color[HTML]{333333} 446} & {\color[HTML]{333333} 557} \\
\multicolumn{1}{l}{} & {\color[HTML]{680100} High} & {\color[HTML]{000000} 16} & {\color[HTML]{000000} 20} & {\color[HTML]{000000} 24} & {\color[HTML]{333333} 594} & {\color[HTML]{333333} 743} & {\color[HTML]{333333} 891} \\ \hline
\end{tabular}
\end{table}
We can see from Table~\ref{table:mult_comp_florete} that the number of $256\times 256$ multiplications used in the key-generation algorithm of the low-security version of Florete is less than the low-security version of Saber. The number of $256\times 256$ multiplications used in the encapsulation algorithm of the low-security version of Florete is the same as the low-security version of Saber, whereas the decapsulation algorithm of the low-security version of Florete uses $1$ more $256\times 256$ polynomial multiplication than Saber. For the medium security, the number of $256\times 256$ multiplications used in the key-generation and encapsulation algorithms of Florete is less than Saber. The number of $256\times 256$ multiplications used in the decapsulation algorithm of the medium-security version of Florete is the same as Saber. Lastly, for the high security, the number of $256\times 256$ multiplications used in all the algorithms of Florete is less than Saber. 
\newline

\noindent\textbf{Message Encoding and Decoding}

\noindent Message encoding and decoding are done in the LWR-based KEM described in Fig.~\ref{fig:lwr_kem} by using the $\mathtt{arrange\_msg}$ and $\mathtt{original\_msg}$, respectively. 
The secret payload/message ($m'$) size is $256$ bits in all the security versions of Florete, and the size of the polynomial is at least twice. So, we repeat the secret payload multiple times with the help of the $\mathtt{arrange\_msg}\{0,\ 1\}^{256} \longrightarrow \{0,\ 1\}^{n}$ function and make its bit size the same as the size of any polynomial in the corresponding security version of Florete. 
\begin{equation*}
\mathtt{arrange\_msg}(m') =
\begin{cases}
m'||m' & \text{if } n = 512 \\
m'||m'||m' & \text{if } n = 768 \\
m'||m'||m'||m' & \text{if } n = 1024 \\
\end{cases}\,.
\end{equation*}
The $\mathtt{original\_msg}$ function is the counter function of $\mathtt{arrange\_msg}$ function and is used in the decryption algorithm. We define the $\mathtt{original\_msg}$ function for each security version of Florete below. 
For the low security version of Florete, the $\mathtt{original\_msg}: \{0,\ 1\}^{512} \longrightarrow \{0,\ 1\}^{256} $ is $\mathtt{original\_msg}(m'') = m'$ and $b\in \{ 0, 1, \ldots, 255 \}$ 
\begin{equation*}
m'[b] =
\begin{cases}
0 & \text{if } m''[b]+m''[b+256]\leq 0 \\
1 & \text{else } \\
\end{cases}\,.
\end{equation*}
For the medium security version of Florete, the $\mathtt{original\_msg}: \{0,\ 1\}^{768} \longrightarrow \{0,\ 1\}^{256} $ is $\mathtt{original\_msg}(m'') = m'$ and $b\in \{ 0, 1, \ldots, 255 \}$ 
\begin{equation*}
m'[b] =
\begin{cases}
0 & \text{if } m''[b]+m''[b+256]+m''[b+512]\leq 1 \\
1 & \text{else } \\
\end{cases}\,.
\end{equation*}
For the high security version of Florete the $\mathtt{original\_msg}: \{0,\ 1\}^{1024} \longrightarrow \{0,\ 1\}^{256} $ is $\mathtt{original\_msg}(m'') = m'$ and $b\in \{ 0, 1, \ldots, 255 \}$ 
\begin{equation*}
m'[b] =
\begin{cases}
0 & \text{if } m''[b]+m''[b+256] +m''[b+512]\\ & +m''[b+768] \leq 2 \\
1 & \text{else } \\
\end{cases}\,.
\end{equation*}
The repetition of message bits during encoding helps to reduce the failure probability and eventually helps to achieve more security. Therefore, Florete can achieve the same level of security with a smaller modulus.
Therefore, we can reduce the three modulus size $\epsilon_q<\epsilon_p<\epsilon_t$ even further than Saber (or Kyber). It reduces the requirement of pseudo-random bytes to create the public matrix $\mathbf{A}$ in Florete compared to Saber, which has been supplied in Table~\ref{table:pseudo-rand-florete-saber}. It eventually helps to reduce the public key size of Florete compared to Kyber (the exact public key sizes are shown in Table~\ref{table:parameters}). Kindly note that we have not applied any error-correction code to reduce failure probability RLWE-based scheme LAC~\cite{LAC}, as it leads to several attacks~\cite{LAC_attack_Aurelien,LAC_attack_Jan-Pieter,LAC_attack_Qian}.    
\newline

\noindent\textbf{Secret Distribution}

\noindent The coefficient of the secret $\mathbf{s}$ (or $\mathbf{s'}$) is sampled from centered binomial distribution, $\beta_1$. Therefore possible values of a coefficient of $\mathbf{s}$ ($\mathbf{s'}$) are $\{-1,\ 0,\ 1\}$. It enables the possibility of very fast multiplication in the processors. In this case, multiplication can be replaced by addition and subtraction only. This method is highly advantageous to the processor, where multiplication is way more costlier than addition or subtraction (e.g. MSP430 microcontrollers). 
Saber's secret coefficients are from $\beta_5$, $\beta_4$, and $\beta_3$ for low, medium, and high-security versions, respectively. In comparison, the secret coefficients of Florete are from $\beta_1$ for all the security versions.
It leads less pseudo-random number requirements for Florete than Saber, which has been provided in Table~\ref{table:pseudo-rand-florete-saber}. 
\textcolor{black}{We have shown the required clock cycles to generate the matrix $\mathbf{A}$ and the secret $\mathbf{s}$ for all the versions of Florete and Saber on the Cortex-M4 platform. More detailed performance analyses are shown in Sec.~\ref{sec:SW_impl}.} 
The coefficients of the secret can be represented by $2$ bits, which reduces the memory requirement to store the secret $\mathbf{s}$ ($\mathbf{s'}$) in hardware compared to Saber (Saber needs $4$ bits to store a coefficient of $\mathbf{s}$). It ultimately helps Florete to have smaller secret key sizes than Saber for all the security versions (the exact numbers are shown in Table~\ref{table:parameters}). 

\begin{table}[!ht]
\centering
\caption{\textcolor{black}{Comparison of pseudo-random byte used in Florete with Saber.}}
\label{table:pseudo-rand-florete-saber}
\begin{tabular}{cccc|cc}
\hline
{\color[HTML]{3531FF} } & {\color[HTML]{3531FF} } & \multicolumn{2}{c|}{{\color[HTML]{3531FF} }} & \multicolumn{2}{c}{{\color[HTML]{3531FF} Performance on Cortex-M4}} \\
{\color[HTML]{3531FF} } & {\color[HTML]{3531FF} } & \multicolumn{2}{c|}{\multirow{-2}{*}{{\color[HTML]{3531FF} pseudo-random bytes}}} & \multicolumn{2}{c}{{\color[HTML]{3531FF} (x1000 clock cycles)}} \\ \cline{3-6} 
\multirow{-3}{*}{{\color[HTML]{3531FF} \begin{tabular}[c]{@{}c@{}}Scheme\\ Name\end{tabular}}} & \multirow{-3}{*}{{\color[HTML]{3531FF} \begin{tabular}[c]{@{}c@{}}Security\\ level\end{tabular}}} & \multicolumn{1}{l}{{\color[HTML]{3531FF} matrix $\mathbf{A}$}} & \multicolumn{1}{l|}{{\color[HTML]{3531FF} secret $\mathbf{s}$ ($\mathbf{s'}$)}} & \multicolumn{1}{l}{{\color[HTML]{3531FF} matrix $\mathbf{A}$}} & \multicolumn{1}{l}{{\color[HTML]{3531FF} secret $\mathbf{s}$ ($\mathbf{s'}$)}} \\ \hline
\multicolumn{1}{l}{} & {\color[HTML]{FD6864} Low} & {\color[HTML]{000000} \textbf{704}} & {\color[HTML]{000000} \textbf{128}} & {\color[HTML]{000000} \textbf{68}} & {\color[HTML]{000000} \textbf{18}} \\
{\color[HTML]{036400} Florete} & {\color[HTML]{CB0000} Medium} & {\color[HTML]{000000} \textbf{960}} & {\color[HTML]{000000} \textbf{192}} & {\color[HTML]{000000} \textbf{83}} & {\color[HTML]{000000} \textbf{32}} \\
\multicolumn{1}{l}{} & {\color[HTML]{680100} High} & {\color[HTML]{000000} \textbf{1280}} & {\color[HTML]{000000} \textbf{256}} & {\color[HTML]{000000} \textbf{110}} & {\color[HTML]{000000} \textbf{34}} \\ \hline
\multicolumn{1}{l}{} & {\color[HTML]{FD6864} Low} & {\color[HTML]{000000} 1664} & {\color[HTML]{000000} 640} & {\color[HTML]{000000} 137} & {\color[HTML]{000000} 76} \\
{\color[HTML]{036400} Saber} & {\color[HTML]{CB0000} Medium} & {\color[HTML]{000000} 3744} & {\color[HTML]{000000} 768} & {\color[HTML]{000000} 313} & {\color[HTML]{000000} 72} \\
\multicolumn{1}{l}{} & {\color[HTML]{680100} High} & {\color[HTML]{000000} 6656} & {\color[HTML]{000000} 768} & {\color[HTML]{000000} 545} & {\color[HTML]{000000} 75} \\ \hline
\end{tabular}
\end{table}

The centered binomial distribution is proposed by Alkim \textit{et al.}~\cite{newhope} to replace the costly Gaussian distribution, which is hard to implement in constant-time. This distribution is used in NIST standardized Kyber and third-round finalist Saber. Therefore, we have decided to sample the secret using CBD. Here, we refrained from taking any aggressive decision for secret distribution, eg. fixing the hamming weight of the secret key like LWR-based scheme Round5~\cite{round5} or fixing the weight of the secret vector like NTRU Prime~\cite{ntru_prime}. This decision has been taken to avoid any new adversarial attack due to the choice of secret distribution. The Saber team proposed a lightweight version of Saber, named uSaber~\cite{saber_round_3} $2$ bits secret key coefficient. More specifically, they use a uniform distribution over $2$ bits numbers. There is another lattice-based scheme proposed in the ongoing Korean PQC competition~\cite{kpqc}, called Smaug~\cite{smaug_kem}. The secret key of this specific scheme has each coefficient sampled from the set $\{-1,\ 0,\ 1\}$.

\subsection{Espada: MLWR based KEM}\label{sec:espada}

\begin{figure}[ht!]
\centering
  \includegraphics[scale=.6]{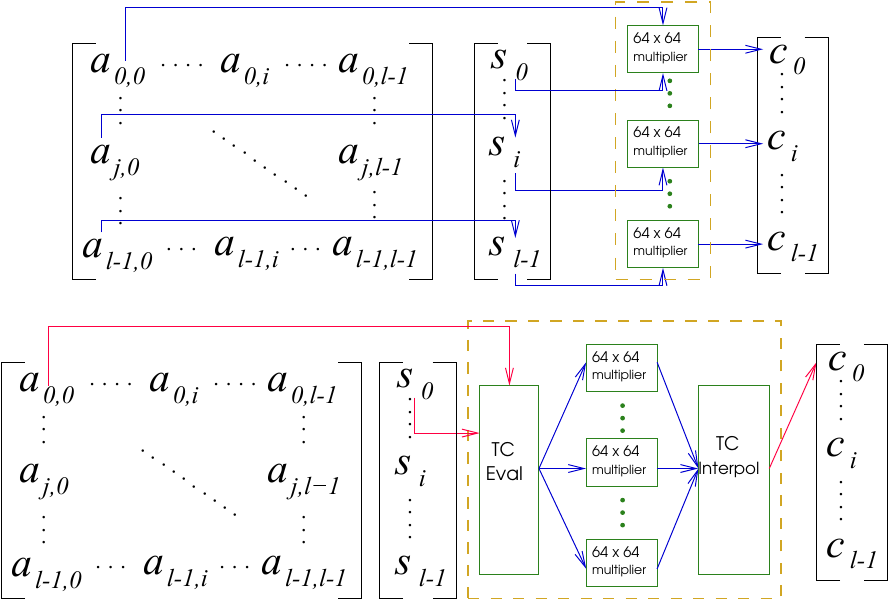}
  \caption{Comparison between the application of parallel $64\times 64$ polynomial multiplication in Espada (top) and Saber (bottom). The blue line represents parallel execution, and the red line denotes serial execution.}
  \label{fig:paraller_multiplication}
\end{figure}

The next LWR-based KEM in the Scabbard is Espada. It uses MLWR as a hard problem, and this KEM also takes advantage of module lattices like Saber and Kyber. 
Therefore the matrix $\mathbf{A}$ is an element of the ring $(\mathcal{R}_q^n)^{l\times l}$, and the secret vector $\mathbf{s}$ is an element of the ring $(\mathcal{R}_q^n)^{l}$. However, the underlying quotient ring is $\mathcal{R}_{q}^{64}$ ($n=64$) constructed by the help of cyclotomic polynomial $(x^{64}+1)$. Therefore the polynomial size of Espada is just $64$, which is very small compared to the size of the polynomial in Saber or Kyber (their polynomials are of size $256$). This design choice allows Espada to use less stack memory while executing in a reasonable amount of clock cycles when implemented in embedded devices, such as Cortex M4 microcontrollers. It can also be implemented in hardware with less area due to its small polynomial size. This scheme can also be implemented in hardware very fast by using multiple polynomial multiplication instances. It provides the flexibility to implement Espada using one or many polynomial multipliers depending on the application's requirements (as shown in Fig.~\ref{fig:paraller_multiplication}), which is not possible for Saber or Kyber.  
\newline

\noindent\textbf{Polynomial Multiplication}
\noindent The ring modulus $q=2^{15}$ for each version of Espada. Since $\epsilon_q (= \log_2(q))=15$ and we want to restrict each coefficient of the polynomial in $16$ bits of word length (described in Sec.~\ref{sec:florete}), we cannot use Toom-Cook 4-way multiplication for the $64\times 64$ polynomial multiplication of Espada. So, we use a combination of Karatsuba and schoolbook multiplication for the $64\times 64$ polynomial multiplications. During the implementation of the matrix-vector or vector-vector multiplication, we take advantage of the lazy interpolation technique~\cite{Saber_Time-memory}. However, the interpolation step in Karatsuba multiplication is smaller than the interpolation step in Toom-Cook $k$-way multiplication for $k>2$. The lazy interpolation technique helps to significantly improve the performance of matrix-vector and vector-vector multiplication because the vector dimension $l$ of Espada is large.

Saber or Kyber use $256\times 256$ polynomial multiplication, and utilizing multiple instances of this multiplication is expensive in hardware. In fact, Mera \textit{et al.}~\cite{MeraTKRV20} developed Saber's $256\times 256$ multiplier by using $7$ parallel $64\times 64$ polynomial multiplication instances together with an evaluation (TC Eval) and interpolation (TC Inter) steps of Toom-Cook $4$-way algorithm (as shown in Fig.~\ref{fig:paraller_multiplication}). \cite{MeraTKRV20} shows that the $64\times 64$ schoolbook multiplication is already very fast in hardware. However, this implementation uses $28$ DSP just for one $256\times 256$ multiplication. Therefore, using multiple such multiplications will make the whole design area expensive, and then there will not be much space left for other components. In Espada, the length of the polynomial is as small as $64$, so the vector dimension $l$ needs to be quite large in order to achieve the desired security. The values of $l$ are $10$, $12$, and $15$, corresponding to low, medium, and high-security versions of Espada (shown in Table~\ref{table:parameters}). Therefore, this scheme can also be implemented in hardware very fast by employing $l$ (the length of the vector) parallel $64\times 64$ polynomial multiplication instances while computing the matrix-vector (e.g.: $\mathbf{A\cdot s}$) and vector-vector (e.g.: $\mathbf{b^T\cdot s'}$) multiplication. In this work, we also utilize $64\times 64$ schoolbook multiplication as one polynomial multiplication in hardware in Espada. By design Espada has extremely parallelizable matrix-vector and vector-vector multiplication in hardware, and it aids Espada to achieve high throughput in hardware.   
\newline

\noindent\textbf{Message Encoding and Decoding}

\noindent The degree of the polynomial for all the versions of Espada is $64$, and the secret payload ($m'$) size is $256$ bits. So, we use one coefficient to hide multiple ($4$) bits of secret payloads and $B=4$. Here, the function $\mathtt{arrange\_msg}: \{0,\ 1\}^{256} \longrightarrow \{0,\ 1,\ 2,\ \ldots,\ 15\}^{64} $ is $\mathtt{arrange\_msg}(m')=m$ and $b\in \{ 0,\ 1,\ \ldots,\ 64 \}$, then $m[b] = m'[4*b]||m'[4*b+1]||m'[4*b+2]||m'[4*b+3]$. The function $\mathtt{original\_msg}: \{0,1,2,\ldots,15\}^{64}\longrightarrow \{0,\ 1\}^{256} $ is $\mathtt{original\_msg}(m'') = m'$ and $b\in \{ 0, 1, \ldots, 255 \}$ , then $m'[b] = (m''[b_1] \gg b_2)\&1$, where $b = 4*b_1+b_2$.
\newline

\vspace{2cm}
\noindent
\textbf{Secret Distribution}

\noindent In Espada, the coefficient of the secret $\mathbf{s}$ ($\mathbf{s'}$) is sampled according to the centered binomial distribution, $\beta_3$, for all the security versions. So, each secret coefficient is from the set $\{-3,\ -2,\ \cdots, 3\}$. These secret coefficients can be represented by $3$ bits, but the unpacking becomes fairly costly in that case. Therefore, the cost-effective way to store the secret $\mathbf{s}$ is to reserve $4$ bits for each coefficient. 

\subsection{Sable: an Alternate Saber}\label{sec:improved_saber}

Sable can be viewed as an improved version of Saber. Like Saber, Sable uses the MLWR structure, and the underlying quotient ring of Sable is the same as Saber $(x^{256}+1)$. In this scheme, we have readjusted the parameters of Saber. The modulus size of the quotient ring $q$ for Sable is $2^{11}$, and it is smaller than Saber's modulus $q=2^{13}$ (shown in Table~\ref{table:parameters}). The public-key modulus $p$ of Sable ($2^9$) is also smaller than Saber ($2^{10}$) for the low and medium security version, which assists in Sable having shorter public keys (given in Table~\ref{table:parameters}). 
\newline

\noindent\textbf{Polynomial Multiplication}

\noindent Sable can utilize the same $256\times 256$ polynomial multiplication, as the modulus is less than $13$ bits. Also, the polynomial multiplication of Sable is less costly than Saber in hardware, thanks to its smaller modulus. 
\newline

\noindent\textbf{Message Encoding and Decoding}

\noindent The degree of the polynomial for all the versions of Sable is $256$, which is the same as the secret payload ($m'$). Therefore, we use one coefficient for a single bit of secret payloads and $B=1$. The function $\mathtt{arrange\_msg}: \{0,\ 1\}^{256} \longrightarrow \{0,\ 1\}^{256} $ is $\mathtt{arrange\_msg}(m')=m'=m$ and the function $\mathtt{original\_msg}: \{0,\ 1\}^{256} \longrightarrow \{0,\ 1\}^{256} $ is $\mathtt{arrange\_msg}(m'')=m''=m'$. 
\newline

\noindent\textbf{Secret Distribution}

\noindent Like Florete, a coefficient of the secret vector is sampled from the CBD $\beta_2$ in every version of Sable. Therefore, one secret coefficient can be stored as a $2$ bits number, allowing Sable to have smaller secret keys. All these choices help Sable to reduce stack memory requirements when implemented in a microcontroller, and area requirements when implemented in hardware. More details regarding implementation are provided in Sec.~\ref{sec:SW_impl} \&~\ref{sec:HW_impl}.

%% file: sections/parameters.tex
\section{Parameter Set}\label{sec:parameters}

The lattice-based schemes whose hardness depends on the LWE problems or its variant, such as the (M/R)LWR problem, are solved by utilizing lattice reduction algorithms that construct a "sufficient orthogonal" basis from the given lattice. Currently the best-known algorithm for lattice reduction is the BKZ algorithm. Here, given one lattice or a basis of lattice, the attacker needs to find the block size or sub-lattice size required for recovering the shortest vector of the lattice while performing the BKZ algorithm. The security of a lattice-based scheme depends on the cost of the execution time of the BKZ algorithm on the underlying lattice. The BKZ algorithm also calls the shortest vector problem (SVP) solving oracle on sub-lattices. The cost of solving the LWE problem with block size $\beta$ depends on the number of SVP oracle calls made by the BKZ algorithm and the cost of solving each SVP for dimension $\beta$. This cost is approximately $2^{c\beta+o(\beta)}$~\cite{LWE_estimator}, where the value of $c$ is approximately $0.292$ in classical settings and $0.265$ with Grover's speed-up algorithm~\cite{DBLP:conf/stoc/Grover96} in quantum settings.

Dachman-Soled et al.~\cite{leaky_estimator} has introduced leaky-LWE-Estimator, the state-of-the-art toolkit to estimate the hardness of the underlying LWE problem for lattice-based schemes. This tool takes the $n=$ dimension of the lattice, $q=$ modulus, $D_e=$ error distribution, $D_s=$ secret distribution, and outputs the block size $\beta$. We have utilized this toolkit for the security estimation of our schemes. 
\textcolor{black}{
The post-quantum bit security is estimated as $0.265*\beta$~\cite{LWE_estimator_correct}, and classical bit security is estimated as $0.292*\beta$~\cite{DBLP:conf/soda/BeckerDGL16}. Since the post-quantum security is lower than classical security ($0.292*\beta$ bit secure, we have mentioned only the post-quantum (PQ) security of our schemes in Table~\ref{table:parameters}. 
}

\begin{table}[!ht]
\centering
\caption{Compare parameters and key sizes of Scabbard suite with Saber}
\label{table:parameters}
\begin{tabular}{lccccccccccc}
\hline
\multicolumn{1}{c}{{\color[HTML]{3531FF} \begin{tabular}[c]{@{}c@{}}{Scheme}\\ {Name}\end{tabular}}} & {\color[HTML]{3531FF} \begin{tabular}[c]{@{}c@{}}{Security}\\ {level}\end{tabular}} & \multicolumn{2}{c}{{\color[HTML]{3531FF} \begin{tabular}[c]{@{}c@{}}{Ring/Module}\\ {Parameters}\end{tabular}}} & {\color[HTML]{3531FF} \begin{tabular}[c]{@{}c@{}}{PQ}\\ {Security}\end{tabular}} & {\color[HTML]{3531FF} \begin{tabular}[c]{@{}c@{}}{Failure}\\ {probability}\end{tabular}} & \multicolumn{2}{c}{{\color[HTML]{3531FF} {Moduli}}} & {\color[HTML]{3531FF} \begin{tabular}[c]{@{}c@{}}{CBD}\\ {($\beta_\eta$)}\end{tabular}} & {\color[HTML]{3531FF} {Encoding}} & \multicolumn{2}{c}{{\color[HTML]{3531FF} \begin{tabular}[c]{@{}c@{}}{Key sizes for} \\ {KEM (Bytes)}\end{tabular}}} \\ \hline
                                                                                                                 & {\color[HTML]{036400} }                                                                         & {n:}                                                  & {512}                                                   & {\color[HTML]{CB0000} }                                                                      & {\color[HTML]{CB0000} }                                                                              & {$\epsilon_q$:}            & {11}             &                                                                                                     &                                         & {\color[HTML]{036400} {Public key:}}                                    & {608}                                     \\
                                                                                                                 & {\color[HTML]{FD6864} {Low}}                                                              &                                                             &                                                               & {\color[HTML]{CB0000} {$2^{104}$}}                                                     & {\color[HTML]{CB0000} {$2^{-138}$}}                                                            & {$\epsilon_p$:}            & {9}              & {$\eta=1$}                                                                                    & {B=1}                             & {\color[HTML]{036400} {Secret key:}}                                    & {800}                                     \\
                                                                                                                 & {\color[HTML]{036400} }                                                                         & {l:}                                                  & {1}                                                     & {\color[HTML]{CB0000} }                                                                      & {\color[HTML]{CB0000} }                                                                              & {$\epsilon_t$:}            & {2}              &                                                                                                     &                                         & {\color[HTML]{036400} {Ciphertext:}}                                    & {768}                                     \\ \cline{2-12} 
                                                                                                                 & {\color[HTML]{036400} }                                                                         & {n:}                                                  & {768}                                                   & {\color[HTML]{CB0000} }                                                                      & {\color[HTML]{CB0000} }                                                                              & {$\epsilon_q$:}            & {10}             &                                                                                                     &                                         & {\color[HTML]{036400} {Public key:}}                                    & {896}                                     \\
\multicolumn{1}{c}{{\color[HTML]{036400} {Florete}}}                                                       & {\color[HTML]{CB0000} {Medium}}                                                           &                                                             &                                                               & {\color[HTML]{CB0000} {$2^{157}$}}                                                     & {\color[HTML]{CB0000} {$2^{-131}$}}                                                            & {$\epsilon_p$:}            & {9}              & {$\eta=1$}                                                                                    & {B=1}                             & {\color[HTML]{036400} {Secret key:}}                                    & {1152}                                    \\
                                                                                                                 & {\color[HTML]{036400} }                                                                         & {l:}                                                  & {1}                                                     & {\color[HTML]{CB0000} }                                                                      & {\color[HTML]{CB0000} }                                                                              & {$\epsilon_t$:}            & {3}              &                                                                                                     &                                         & {\color[HTML]{036400} {Ciphertext:}}                                    & {1248}                                    \\ \cline{2-12} 
                                                                                                                 & {\color[HTML]{036400} }                                                                         & {n:}                                                  & {1024}                                                  & {\color[HTML]{CB0000} }                                                                      & {\color[HTML]{CB0000} }                                                                              & {$\epsilon_q$:}            & {10}             &                                                                                                     &                                         & {\color[HTML]{036400} {Public key:}}                                    & {1184}                                    \\
                                                                                                                 & {\color[HTML]{680100} {High}}                                                             &                                                             &                                                               & {\color[HTML]{CB0000} {$2^{220}$}}                                                     & {\color[HTML]{CB0000} {$2^{-165}$}}                                                            & {$\epsilon_p$:}            & {9}              & {$\eta=1$}                                                                                    & {B=1}                             & {\color[HTML]{036400} {Secret key:}}                                    & {1504}                                    \\
                                                                                                                 & {\color[HTML]{036400} }                                                                         & {l:}                                                  & {1}                                                     & {\color[HTML]{CB0000} }                                                                      & {\color[HTML]{CB0000} }                                                                              & {$\epsilon_t$:}            & {4}              &                                                                                                     &                                         & {\color[HTML]{036400} {Ciphertext:}}                                    & {1792}                                    \\ \hline
                                                                                                                 & {\color[HTML]{036400} }                                                                         & {n:}                                                  & {64}                                                    & {\color[HTML]{CB0000} }                                                                      & {\color[HTML]{CB0000} }                                                                              & {$\epsilon_q$:}            & {15}             &                                                                                                     &                                         & {\color[HTML]{036400} {Public key:}}                                    & {1072}                                    \\
                                                                                                                 & {\color[HTML]{FD6864} {Low}}                                                              &                                                             &                                                               & {\color[HTML]{CB0000} {$2^{101}$}}                                                     & {\color[HTML]{CB0000} {$2^{-148}$}}                                                            & {$\epsilon_p$:}            & {13}             & {$\eta=3$}                                                                                    & {B=4}                             & {\color[HTML]{036400} {Secret key:}}                                    & {1456}                                    \\
                                                                                                                 & {\color[HTML]{036400} }                                                                         & {l:}                                                  & {10}                                                    & {\color[HTML]{CB0000} }                                                                      & {\color[HTML]{CB0000} }                                                                              & {$\epsilon_t$:}            & {2}              &                                                                                                     &                                         & {\color[HTML]{036400} {Ciphertext:}}                                    & {1088}                                    \\ \cline{2-12} 
                                                                                                                 & {\color[HTML]{036400} }                                                                         & {n:}                                                  & {64}                                                    & {\color[HTML]{CB0000} }                                                                      & {\color[HTML]{CB0000} }                                                                              & {$\epsilon_q$:}            & {15}             &                                                                                                     &                                         & {\color[HTML]{036400} {Public key:}}                                    & {1280}                                    \\
\multicolumn{1}{c}{{\color[HTML]{036400} {Espada}}}                                                        & {\color[HTML]{CB0000} {Medium}}                                                           &                                                             &                                                               & {\color[HTML]{CB0000} {$2^{128}$}}                                                     & {\color[HTML]{CB0000} {$2^{-167}$}}                                                            & {$\epsilon_p$:}            & {13}             & {$\eta=3$}                                                                                    & {B=4}                             & {\color[HTML]{036400} {Secret key:}}                                    & {1728}                                    \\
                                                                                                                 & {\color[HTML]{036400} }                                                                         & {l:}                                                  & {12}                                                    & {\color[HTML]{CB0000} }                                                                      & {\color[HTML]{CB0000} }                                                                              & {$\epsilon_t$:}            & {3}              &                                                                                                     &                                         & {\color[HTML]{036400} {Ciphertext:}}                                    & {1304}                                    \\ \cline{2-12} 
                                                                                                                 & {\color[HTML]{036400} }                                                                         & {n:}                                                  & {64}                                                    & {\color[HTML]{CB0000} }                                                                      & {\color[HTML]{CB0000} }                                                                              & {$\epsilon_q$:}            & {15}             &                                                                                                     &                                         & {\color[HTML]{036400} {Public key:}}                                    & {1592}                                    \\
                                                                                                                 & {\color[HTML]{680100} {High}}                                                             &                                                             &                                                               & {\color[HTML]{CB0000} {$2^{168}$}}                                                     & {\color[HTML]{CB0000} {$2^{-162}$}}                                                            & {$\epsilon_p$:}            & {13}             & {$\eta=3$}                                                                                    & {B=4}                             & {\color[HTML]{036400} {Secret key:}}                                    & {2136}                                    \\
                                                                                                                 & {\color[HTML]{036400} }                                                                         & {l:}                                                  & {15}                                                    & {\color[HTML]{CB0000} }                                                                      & {\color[HTML]{CB0000} }                                                                              & {$\epsilon_t$:}            & {5}              &                                                                                                     &                                         & {\color[HTML]{036400} {Ciphertext:}}                                    & {1632}                                    \\ \hline
                                                                                                                 & {\color[HTML]{036400} }                                                                         & {n:}                                                  & {256}                                                   & {\color[HTML]{CB0000} }                                                                      & {\color[HTML]{CB0000} }                                                                              & {$\epsilon_q$:}            & {11}             &                                                                                                     &                                         & {\color[HTML]{036400} {Public key:}}                                    & {608}                                     \\
                                                                                                                 & {\color[HTML]{FD6864} {Low}}                                                              &                                                             &                                                               & {\color[HTML]{CB0000} {$2^{104}$}}                                                     & {\color[HTML]{CB0000} {$2^{-139}$}}                                                            & {$\epsilon_p$:}            & {9}              & {$\eta=1$}                                                                                    & {B=1}                             & {\color[HTML]{036400} {Secret key:}}                                    & {800}                                     \\
                                                                                                                 & {\color[HTML]{036400} }                                                                         & {l:}                                                  & {2}                                                     & {\color[HTML]{CB0000} }                                                                      & {\color[HTML]{CB0000} }                                                                              & {$\epsilon_t$:}            & {2}              &                                                                                                     &                                         & {\color[HTML]{036400} {Ciphertext:}}                                    & {672}                                     \\ \cline{2-12} 
                                                                                                                 & {\color[HTML]{036400} }                                                                         & {n:}                                                  & {256}                                                   & {\color[HTML]{CB0000} }                                                                      & {\color[HTML]{CB0000} }                                                                              & {$\epsilon_q$:}            & {11}             &                                                                                                     &                                         & {\color[HTML]{036400} {Public key:}}                                    & {896}                                     \\
\multicolumn{1}{c}{{\color[HTML]{036400} {Sable}}}                                                         & {\color[HTML]{CB0000} {Medium}}                                                           &                                                             &                                                               & {\color[HTML]{CB0000} {$2^{169}$}}                                                     & {\color[HTML]{CB0000} {$2^{-143}$}}                                                            & {$\epsilon_p$:}            & {9}              & {$\eta=1$}                                                                                    & {B=1}                             & {\color[HTML]{036400} {Secret key:}}                                    & {1152}                                    \\
                                                                                                                 & {\color[HTML]{036400} }                                                                         & {l:}                                                  & {3}                                                     &                                                                                              & {\color[HTML]{CB0000} }                                                                              & {$\epsilon_t$:}            & {4}              &                                                                                                     &                                         & {\color[HTML]{036400} {Ciphertext:}}                                    & {1024}                                    \\ \cline{2-12} 
                                                                                                                 & {\color[HTML]{036400} }                                                                         & {n:}                                                  & {256}                                                   & {\color[HTML]{CB0000} }                                                                      & {\color[HTML]{CB0000} }                                                                              & {$\epsilon_q$:}            & {11}             &                                                                                                     &                                         & {\color[HTML]{036400} {Public key:}}                                    & {1312}                                    \\
                                                                                                                 & {\color[HTML]{680100} {High}}                                                             &                                                             &                                                               & {\color[HTML]{CB0000} {$2^{203}$}}                                                     & {\color[HTML]{CB0000} {$2^{-208}$}}                                                            & {$\epsilon_p$:}            & {10}             & {$\eta=1$}                                                                                    & {B=1}                             & {\color[HTML]{036400} {Secret key:}}                                    & {1632}                                    \\
                                                                                                                 & {\color[HTML]{036400} }                                                                         & {l:}                                                  & {4}                                                     & {\color[HTML]{CB0000} }                                                                      & {\color[HTML]{CB0000} }                                                                              & {$\epsilon_t$:}            & {2}              &                                                                                                     &                                         & {\color[HTML]{036400} {Ciphertext:}}                                    & {1376}                                    \\ \hline
                                                                                                                 & {\color[HTML]{036400} }                                                                         & {n:}                                                  & {256}                                                   & {\color[HTML]{CB0000} }                                                                      & {\color[HTML]{CB0000} }                                                                              & {$\epsilon_q$:}            & {13}             &                                                                                                     &                                         & {\color[HTML]{036400} {Public key:}}                                    & {672}                                     \\
                                                                                                                 & {\color[HTML]{FD6864} {Low}}                                                              &                                                             &                                                               & {\color[HTML]{CB0000} {$2^{107}$}}                                                     & {\color[HTML]{CB0000} {$2^{-120}$}}                                                            & {$\epsilon_p$:}            & {10}             & {$\eta=5$}                                                                                    & {B=1}                             & {\color[HTML]{036400} {Secret key:}}                                    & {992}                                     \\
                                                                                                                 & {\color[HTML]{036400} }                                                                         & {l:}                                                  & {2}                                                     & {\color[HTML]{CB0000} }                                                                      & {\color[HTML]{CB0000} }                                                                              & {$\epsilon_t$:}            & {2}              &                                                                                                     &                                         & {\color[HTML]{036400} {Ciphertext:}}                                    & {736}                                     \\ \cline{2-12} 
                                                                                                                 & {\color[HTML]{036400} }                                                                         & {n:}                                                  & {256}                                                   & {\color[HTML]{CB0000} }                                                                      & {\color[HTML]{CB0000} }                                                                              & {$\epsilon_q$:}            & {13}             &                                                                                                     &                                         & {\color[HTML]{036400} {Public key:}}                                    & {992}                                     \\
\multicolumn{1}{c}{{\color[HTML]{036400} {Saber}}}                                                         & {\color[HTML]{CB0000} {Medium}}                                                           &                                                             &                                                               & {\color[HTML]{CB0000} {$2^{172}$}}                                                     & {\color[HTML]{CB0000} {$2^{-136}$}}                                                            & {$\epsilon_p$:}            & {10}             & {$\eta=4$}                                                                                    & {B=1}                             & {\color[HTML]{036400} {Secret key:}}                                    & {1440}                                    \\
                                                                                                                 & {\color[HTML]{036400} }                                                                         & {l:}                                                  & {3}                                                     & {\color[HTML]{CB0000} }                                                                      & {\color[HTML]{CB0000} }                                                                              & {$\epsilon_t$:}            & {3}              &                                                                                                     &                                         & {\color[HTML]{036400} {Ciphertext:}}                                    & {1088}                                    \\ \cline{2-12} 
                                                                                                                 & {\color[HTML]{036400} }                                                                         & {n:}                                                  & {256}                                                   & {\color[HTML]{CB0000} }                                                                      & {\color[HTML]{CB0000} }                                                                              & {$\epsilon_q$:}            & {13}             &                                                                                                     &                                         & {\color[HTML]{036400} {Public key:}}                                    & {1312}                                    \\
                                                                                                                 & {\color[HTML]{680100} {High}}                                                             &                                                             &                                                               & {\color[HTML]{CB0000} {$2^{236}$}}                                                     & {\color[HTML]{CB0000} {$2^{-165}$}}                                                            & {$\epsilon_p$:}            & {10}             & {$\eta=3$}                                                                                    & {B=1}                             & {\color[HTML]{036400} {Secret key:}}                                    & {1760}                                    \\
                                                                                                                 & {\color[HTML]{036400} }                                                                         & {l:}                                                  & {4}                                                     &                                                                                              &                                                                                                      & {$\epsilon_t$:}            & {5}              &                                                                                                     &                                         & {\color[HTML]{036400} {Ciphertext:}}                                    & {1472}                                    \\ \hline
                                                                                                                 & {\color[HTML]{036400} }                                                                         & {n:}                                                  & {256}                                                   & {\color[HTML]{CB0000} }                                                                      & {\color[HTML]{CB0000} }                                                                              & {$q$:}                     & {3329}           & {$\eta_1=3$}                                                                                  &                                         & {\color[HTML]{036400} {Public key:}}                                    & {800}                                     \\
                                                                                                                 & {\color[HTML]{FD6864} {Low}}                                                              &                                                             &                                                               & {\color[HTML]{CB0000} {$2^{107}$}}                                                     & {\color[HTML]{CB0000} {$2^{-139}$}}                                                            & {$\epsilon_p$:}            & {10}             &                                                                                                     & {B=1}                             & {\color[HTML]{036400} {Secret key:}}                                    & {1632}                                    \\
                                                                                                                 & {\color[HTML]{036400} }                                                                         & {l:}                                                  & {2}                                                     & {\color[HTML]{CB0000} }                                                                      & {\color[HTML]{CB0000} }                                                                              & {$\epsilon_t$:}            & {3}              & {$\eta_2=2$}                                                                                  &                                         & {\color[HTML]{036400} {Ciphertext:}}                                    & {768}                                     \\ \cline{2-12} 
                                                                                                                 & {\color[HTML]{036400} }                                                                         & {n:}                                                  & {256}                                                   & {\color[HTML]{CB0000} }                                                                      & {\color[HTML]{CB0000} }                                                                              & {$q$:}                     & {3329}           & {$\eta_1=2$}                                                                                  &                                         & {\color[HTML]{036400} {Public key:}}                                    & {1184}                                    \\
\multicolumn{1}{c}{{\color[HTML]{036400} {Kyber}}}                                                         & {\color[HTML]{CB0000} {Medium}}                                                           &                                                             &                                                               & {\color[HTML]{CB0000} {$2^{166}$}}                                                     & {\color[HTML]{CB0000} {$2^{-164}$}}                                                            & {$\epsilon_p$:}            & {10}             &                                                                                                     & {B=1}                             & {\color[HTML]{036400} {Secret key:}}                                    & {2400}                                    \\
                                                                                                                 & {\color[HTML]{036400} }                                                                         & {l:}                                                  & {3}                                                     & {\color[HTML]{CB0000} }                                                                      & {\color[HTML]{CB0000} }                                                                              & {$\epsilon_t$:}            & {3}              & {$\eta_2=2$}                                                                                  &                                         & {\color[HTML]{036400} {Ciphertext:}}                                    & {1088}                                    \\ \cline{2-12} 
                                                                                                                 & {\color[HTML]{036400} }                                                                         & {n:}                                                  & {256}                                                   & {\color[HTML]{CB0000} }                                                                      & {\color[HTML]{CB0000} }                                                                              & {$q$:}                     & {3329}           & {$\eta_1=2$}                                                                                  &                                         & {\color[HTML]{036400} {Public key:}}                                    & {1568}                                    \\
                                                                                                                 & {\color[HTML]{680100} {High}}                                                             &                                                             &                                                               & {\color[HTML]{CB0000} {$2^{232}$}}                                                     & {\color[HTML]{CB0000} {$2^{-174}$}}                                                            & {$\epsilon_p$:}            & {10}             &                                                                                                     & {B=1}                             & {\color[HTML]{036400} {Secret key:}}                                    & {3168}                                    \\
                                                                                                                 & {\color[HTML]{036400} }                                                                         & {l:}                                                  & {4}                                                     &                                                                                              &                                                                                                      & {$\epsilon_t$:}            & {4}              & {$\eta_2=2$}                                                                                  &                                         & {\color[HTML]{036400} {Ciphertext:}}                                    & {1568}                                    \\ \hline
\end{tabular}
\end{table}

We present the parameter sets of our schemes for three security levels in Table~\ref{table:parameters}. For security level $1$, PQ security of each of the KEMs is $\geq 2^{100}$, for security level $3$, PQ security is $\geq 2^{128}$, and for security level $5$, PQ security is $\geq 2^{160}$. 
LWE-based cryptosystem has another security factor which is failure probability. 
However, another type of attack is possible on LWR or LWR-based cryptographic schemes that exploit failure probability during decryption.
As mentioned in Sec.~\ref{sec:genericLWR}, the failure probability should be $\leq 2^{-S}$, where $S$ is the security of the KEM to maintain the IND-CCA security of the KEM. Therefore, for security level $1, 3, 5$ in contrast with the PQ security, the failure probability we maintain $\leq 2^{-100}, \leq 2^{-128}, \leq 2^{-160}$, respectively for each of the KEMs. NIST security levels $1$, $3$, and $5$ are represented by low, medium, and high-security levels in Table~\ref{table:parameters}.  

For comparing the key sizes of our schemes with Saber and Kyber, we also include the parameter sets of Saber in Table~\ref{table:parameters}. The public key and secret key sizes of Florete are smaller than Saber, while the size of the ciphertext is slightly larger for Florete than Saber for all three security levels. The public key and secret key sizes of Florete are also smaller than Kyber for all three security versions. Even, the size of the ciphertext is the same for the low-security version of Florete and Kyber. Due to larger moduli and vector dimensions, the public key, secret key, and ciphertext sizes are bigger in Espada than in Saber for the same security level. However, the secret key size of Espada is smaller in Espada than in Kyber for all the security versions. In the case of any of the three security levels of Sable, the sizes of the public key, secret key, and ciphertext are smaller than in the case of the same security level of Saber and Kyber.

%% file: sections/software_implementation.tex
\section{Software Implementation}\label{sec:SW_impl}

In this section, we describe the implementation results of our schemes on the software platforms. We have implemented Scabbard's schemes on general-purpose intel processors using C and advanced vector instructions (AVX2). We also implemented Scabbard's schemes on the NIST-recommended ARM Cortex-M4 platform. As most of the PQC schemes are implemented in these two software platforms, we can compare the implementation results of our schemes with the state-of-the-art schemes and demonstrate the efficiency of our scheme.  

\subsection{Results in C and AVX2}

To implement our schemes on general-purpose intel processors using C and advanced vector instructions (AVX2), we use the GCC 6.5 compiler and optimization flags-O3. We also used -fomit-frame-pointer on an Intel (R) Core (TM) i7-6600 CPU running in 2.60GHz, and disabled hyperthreading, turbo-boost, and multicore support in our system following the standard practice. 
The performance results of Scabbard's schemes in portable C and AVX2 implementations are presented in Table~\ref{tab:C_AVX_performace}. For comparison, we also include the performances of C and AVX2 implementation of NIST's third-round finalist Saber and NIST's standard Kyber together with BSI recommended~\cite{bsi_PQC} Frodo~\cite{DBLP:conf/ccs/BosCDMNNRS16} (also in consideration of ISO for standard~\cite{gsma_PQC}) in the tables. This table also compares KPQC schemes Smaug~\cite{smaug_kem}, NTRU+~\cite{ntru_plus_kem}, and Tiger~\cite{Tiger_kem}, which have been advanced to the second-round~\cite{kpqc}. 

\begin{table}[!ht]
\centering
\caption{Comparing performance of Scabbard schemes with Saber and kyber in portable C and AVX2 implementations}
\label{tab:C_AVX_performace}
\begin{tabular}{ccrrr|rrr}
\hline
{\color[HTML]{3531FF} }                              & {\color[HTML]{3531FF} }                                                                           & \multicolumn{3}{c|}{{\color[HTML]{3531FF} C (X1000 clock cycles)}}                                                                                         & \multicolumn{3}{c}{{\color[HTML]{3531FF} AVX (X1000 clock cycles)}}                                                                                       \\ \cline{3-8} 
\multirow{-2}{*}{{\color[HTML]{3531FF} Scheme Name}} & \multirow{-2}{*}{{\color[HTML]{3531FF} \begin{tabular}[c]{@{}c@{}}Security\\ level\end{tabular}}} & \multicolumn{1}{c}{{\color[HTML]{3531FF} KeyGen}} & \multicolumn{1}{c}{{\color[HTML]{3531FF} Encaps}} & \multicolumn{1}{c|}{{\color[HTML]{3531FF} Decaps}} & \multicolumn{1}{c}{{\color[HTML]{3531FF} KeyGen}} & \multicolumn{1}{c}{{\color[HTML]{3531FF} Encaps}} & \multicolumn{1}{c}{{\color[HTML]{3531FF} Decaps}} \\ \hline
\multicolumn{1}{l}{}                                 & {\color[HTML]{FD6864} Low}                                                                        & 43                                                & 66                                                & 83                                                 & 31                                                & 43                                                & 47                                                \\
{\color[HTML]{036400} Florete}                       & {\color[HTML]{CB0000} Medium}                                                                     & 64                                                & 104                                               & 143                                                & 45                                                & 66                                                & 75                                                \\
\multicolumn{1}{l}{}                                 & {\color[HTML]{680100} High}                                                                       & 80                                                & 140                                               & 181                                                & 52                                                & 83                                                & 97                                                \\ \hline
\multicolumn{1}{l}{}                                 & {\color[HTML]{FD6864} Low}                                                                        & 159                                               & 173                                               & 193                                                & 148                                               & 158                                               & 153                                               \\
{\color[HTML]{036400} Espada}                        & {\color[HTML]{CB0000} Medium}                                                                     & 224                                               & 234                                               & 232                                                & 203                                               & 215                                               & 210                                               \\
\multicolumn{1}{l}{}                                 & {\color[HTML]{680100} High}                                                                       & 336                                               & 351                                               & 352                                                & 310                                               & 324                                               & 317                                               \\ \hline
{\color[HTML]{036400} }                              & {\color[HTML]{FD6864} Low}                                                                        & 55                                                & 69                                                & 76                                                 & 38                                                & 45                                                & 41                                                \\
{\color[HTML]{036400} Sable}                         & {\color[HTML]{CB0000} Medium}                                                                     & 101                                               & 126                                               & 137                                                & 63                                                & 74                                                & 71                                                \\
{\color[HTML]{036400} }                              & {\color[HTML]{680100} High}                                                                       & 173                                               & 230                                               & 226                                                & 97                                                & 113                                               & 110                                               \\ \hline
{\color[HTML]{036400} }                              & {\color[HTML]{FD6864} Low}                                                                        & 64                                                & 81                                                & 92                                                 & 44                                                & 51                                                & 49                                                \\
{\color[HTML]{036400} Saber}~\cite{saber_round_3}                         & {\color[HTML]{CB0000} Medium}                                                                     & 116                                               & 143                                               & 154                                                & 73                                                & 85                                                & 82                                                \\
{\color[HTML]{036400} }                              & {\color[HTML]{680100} High}                                                                       & 188                                               & 222                                               & 244                                                & 107                                               & 122                                               & 120                                               \\ \hline
{\color[HTML]{036400} }                              & {\color[HTML]{FD6864} Low}                                                                        & 113                                               & 150                                               & 176                                                & 26                                                & 38                                                & 29                                                \\
{\color[HTML]{036400} Kyber}~\cite{Kyber-Kem}                         & {\color[HTML]{CB0000} Medium}                                                                     & 185                                               & 238                                               & 268                                                & 41                                                & 53                                                & 40                                                \\
{\color[HTML]{036400} }                              & {\color[HTML]{680100} High}                                                                       & 301                                               & 341                                               & 382                                                & 49                                                & 66                                                & 51                                                \\ \hline
\multicolumn{1}{l}{}                                 & {\color[HTML]{FD6864} Low}                                                                        & 1237                                              & 1382                                              & 1383                                               & -                                                 & -                                                 & -                                                 \\
{\color[HTML]{036400} Frodo}~\cite{DBLP:conf/ccs/BosCDMNNRS16}                         & {\color[HTML]{CB0000} Medium}                                                                     & 2654                                              & 2819                                              & 2509                                               & -                                                 & -                                                 & -                                                 \\
\multicolumn{1}{l}{}                                 & {\color[HTML]{680100} High}                                                                       & 4225                                              & 4238                                              & 4465                                               & -                                                 & -                                                 & -                                                 \\ \hline
\multicolumn{1}{l}{}                                 & {\color[HTML]{FD6864} Low}                                                                        & 89                                                & 83                                                & 93                                                 & 48                                                & 37                                                & 47                                                \\
{\color[HTML]{036400} Smaug}~\cite{smaug_kem}                         & {\color[HTML]{CB0000} Medium}                                                                     & 157                                               & 148                                               & 158                                                & 73                                                & 59                                                & 74                                                \\
\multicolumn{1}{l}{}                                 & {\color[HTML]{680100} High}                                                                       & 251                                               & 251                                               & 267                                                & 127                                               & 115                                               & 128                                               \\ \hline
{\color[HTML]{036400} }                              & {\color[HTML]{FD6864} Low}                                                                        & 339                                               & 110                                               & 164                                                & 18                                                & 15                                                & 12                                                \\
{\color[HTML]{036400} NTRU+}~\cite{ntru_plus_kem}                         & {\color[HTML]{CB0000} Medium}                                                                     & 335                                               & 154                                               & 233                                                & 16                                                & 18                                                & 16                                                \\
{\color[HTML]{036400} }                              & {\color[HTML]{680100} High}                                                                       & 358                                               & 180                                               & 277                                                & 14                                                & 20                                                & 18                                                \\ \hline
{\color[HTML]{036400} }                              & {\color[HTML]{FD6864} Low}                                                                        & 89                                                & 80                                                & 78                                                 & -                                                 & -                                                 & -                                                 \\
{\color[HTML]{036400} Tiger}~\cite{Tiger_kem}                         & {\color[HTML]{CB0000} Medium}                                                                     & 104                                               & 122                                               & 127                                                & -                                                 & -                                                 & -                                                 \\
{\color[HTML]{036400} }                              & {\color[HTML]{680100} High}                                                                       & 123                                               & 162                                               & 176                                                & -                                                 & -                                                 & -                                                 \\ \hline
\end{tabular}
\end{table}

\textcolor{black}{As we can see from Table~\ref{tab:C_AVX_performace}}, the performance of all three algorithms (key generation, encapsulation, and decapsulation) of Florete and Sable is better than all the other schemes, including Kyber, Frodo, and Saber, for all the security versions on C. All three algorithms of Florete and Sable perform better than Saber and Smaug in the AVX2 implementation as well for all the security versions. Algorithms of Espada take approximately twice as many clock cycles as Saber for all the security levels due to the use of more pseudo-random numbers and more $64\times64$ multiplications. However, the slowdown factor for Espada's performance compared to Saber's decreases as the security order increases because the underlying lattice rank $l\times n$ decreases in Espada compared to Saber as the security increases. 

\subsection{Results in Cortex-M4}
\textcolor{black}{
We have implemented Scabbard's schemes on the NIST-recommended 32-bit ARM Cortex-M4 microcontroller (STM32F407-DISCOVERY development board) using the PQM4~\cite{PQM4} framework. For compilation, we have used \texttt{arm-none-eabi-gcc} compiler version $4.9.3$. The PQM4 library uses a 24 MHz system clock to calculate clock cycles. 
The results of the implementations of Scabbard's schemes in a Cortex-M4 platform are presented in Table~\ref{performance-breakdown-on-m4}. We have also included the clock cycles spent in hashing, polynomial multiplication, and the remaining operations in this table. 
We have compared our implementations of Scabbard with the state-of-the-art schemes in Table~\ref{tab:M4_performance_stack}. This table contains two implementations of Saber, one with NTT multiplication (SaberNTT~\cite{saber_round_3}) and another with Toom-cook multiplication (Saber~\cite{saber_round_3}). For a fair comparison, we have also included implementation results of two versions of Kyber~\cite{PQM4} and Frodo~\cite{Frodo_kem_m4} (i) Kyber-Speed $\&$ Frodo-Speed: optimized to achieve speed, and (ii) Kyber-Stack $\&$ Frodo-Stack: optimized to reduce stack memory usage. }

\begin{table}[!ht]
\centering
\caption{\textcolor{black}{Spent cycles of Scabbard schemes in hashing, polynomial multiplication, and other operations on the Cortex-M4 platform.}}
\label{performance-breakdown-on-m4}
\resizebox{1\columnwidth}{!}{
\begin{tabular}{ccrrr|rrr|rrr|rrr}
\hline
{\color[HTML]{3531FF} } & {\color[HTML]{3531FF} } & \multicolumn{3}{c|}{{\color[HTML]{3531FF} Total Performance on M4}} & \multicolumn{3}{c|}{{\color[HTML]{3531FF} Hashing}} & \multicolumn{3}{c|}{{\color[HTML]{3531FF} Polynomial multiplication}} & \multicolumn{3}{c}{{\color[HTML]{3531FF} Other components}} \\
{\color[HTML]{3531FF} } & {\color[HTML]{3531FF} } & \multicolumn{3}{c|}{{\color[HTML]{3531FF} (x1000 clock cycles)}} & \multicolumn{3}{c|}{{\color[HTML]{3531FF} (x1000 clock cycles)}} & \multicolumn{3}{c|}{{\color[HTML]{3531FF} (x1000 clock cycles)}} & \multicolumn{3}{c}{{\color[HTML]{3531FF} (x1000 clock cycles)}} \\ \cline{3-14} 
\multirow{-3}{*}{{\color[HTML]{3531FF} \begin{tabular}[c]{@{}c@{}}Scheme \\ Name\end{tabular}}} & \multirow{-3}{*}{{\color[HTML]{3531FF} \begin{tabular}[c]{@{}c@{}}Security\\ level\end{tabular}}} & \multicolumn{1}{c}{{\color[HTML]{3531FF} KeyGen}} & \multicolumn{1}{c}{{\color[HTML]{3531FF} Encaps}} & \multicolumn{1}{c|}{{\color[HTML]{3531FF} Decaps}} & \multicolumn{1}{c}{{\color[HTML]{3531FF} KeyGen}} & \multicolumn{1}{c}{{\color[HTML]{3531FF} Encaps}} & \multicolumn{1}{c|}{{\color[HTML]{3531FF} Decaps}} & \multicolumn{1}{c}{{\color[HTML]{3531FF} KeyGen}} & \multicolumn{1}{c}{{\color[HTML]{3531FF} Encaps}} & \multicolumn{1}{c|}{{\color[HTML]{3531FF} Decaps}} & \multicolumn{1}{c}{{\color[HTML]{3531FF} KeyGen}} & \multicolumn{1}{c}{{\color[HTML]{3531FF} Encaps}} & \multicolumn{1}{c}{{\color[HTML]{3531FF} Decaps}} \\ \hline
\multicolumn{1}{l}{} & {\color[HTML]{FD6864} Low} & 299 & 536 & 606 & {\color[HTML]{333333} 158} & {\color[HTML]{333333} 261} & {\color[HTML]{333333} 183} & {\color[HTML]{333333} 121} & {\color[HTML]{333333} 243} & {\color[HTML]{333333} 364} & {\color[HTML]{333333} 20} & {\color[HTML]{333333} 32} & {\color[HTML]{333333} 59} \\
{\color[HTML]{036400} Florete} & {\color[HTML]{CB0000} Medium} & 439 & 815 & 957 & {\color[HTML]{333333} 209} & {\color[HTML]{333333} 362} & {\color[HTML]{333333} 259} & {\color[HTML]{333333} 202} & {\color[HTML]{333333} 405} & {\color[HTML]{333333} 607} & {\color[HTML]{333333} 28} & {\color[HTML]{333333} 48} & {\color[HTML]{333333} 91} \\
\multicolumn{1}{l}{} & {\color[HTML]{680100} High} & 598 & 1,131 & 1,357 & {\color[HTML]{333333} 259} & {\color[HTML]{333333} 463} & {\color[HTML]{333333} 335} & {\color[HTML]{333333} 300} & {\color[HTML]{333333} 600} & {\color[HTML]{333333} 900} & {\color[HTML]{333333} 39} & {\color[HTML]{333333} 68} & {\color[HTML]{333333} 122} \\ \hline
\multicolumn{1}{l}{} & {\color[HTML]{FD6864} Low} & 1,659 & 1,859 & 1804 & {\color[HTML]{333333} 1,029} & {\color[HTML]{333333} 1,170} & {\color[HTML]{333333} 1,054} & {\color[HTML]{333333} 482} & {\color[HTML]{333333} 530} & {\color[HTML]{333333} 579} & {\color[HTML]{333333} 148} & {\color[HTML]{333333} 159} & {\color[HTML]{333333} 171} \\
{\color[HTML]{036400} Espada} & {\color[HTML]{CB0000} Medium} & 2,342 & 2,566 & 2,497 & {\color[HTML]{333333} 1,442} & {\color[HTML]{333333} 1,596} & {\color[HTML]{333333} 1,455} & {\color[HTML]{333333} 694} & {\color[HTML]{333333} 752} & {\color[HTML]{333333} 810} & {\color[HTML]{333333} 206} & {\color[HTML]{333333} 218} & {\color[HTML]{333333} 232} \\
\multicolumn{1}{l}{} & {\color[HTML]{680100} High} & 3,577 & 3,859 & 3,779 & {\color[HTML]{333333} 2,181} & {\color[HTML]{333333} 2,372} & {\color[HTML]{333333} 2,206} & {\color[HTML]{333333} 1,085} & {\color[HTML]{333333} 1,157} & {\color[HTML]{333333} 1,230} & {\color[HTML]{333333} 311} & {\color[HTML]{333333} 330} & {\color[HTML]{333333} 343} \\ \hline
{\color[HTML]{036400} } & {\color[HTML]{FD6864} Low} & 381 & 558 & 568 & {\color[HTML]{333333} 205} & {\color[HTML]{333333} 296} & {\color[HTML]{333333} 218} & {\color[HTML]{333333} 148} & {\color[HTML]{333333} 222} & {\color[HTML]{333333} 296} & {\color[HTML]{333333} 28} & {\color[HTML]{333333} 40} & {\color[HTML]{333333} 54} \\
{\color[HTML]{036400} Sable} & {\color[HTML]{CB0000} Medium} & 745 & 1,005 & 1,031 & {\color[HTML]{333333} 363} & {\color[HTML]{333333} 491} & {\color[HTML]{333333} 388} & {\color[HTML]{333333} 333} & {\color[HTML]{333333} 444} & {\color[HTML]{333333} 555} & {\color[HTML]{333333} 49} & {\color[HTML]{333333} 70} & {\color[HTML]{333333} 88} \\
{\color[HTML]{036400} } & {\color[HTML]{680100} High} & 1,251 & 1,593 & 1,622 & {\color[HTML]{333333} 583} & {\color[HTML]{333333} 749} & {\color[HTML]{333333} 608} & {\color[HTML]{333333} 592} & {\color[HTML]{333333} 741} & {\color[HTML]{333333} 889} & {\color[HTML]{333333} 76} & {\color[HTML]{333333} 103} & {\color[HTML]{333333} 125} \\ \hline
\multicolumn{1}{l}{} & {\color[HTML]{FD6864} Low} & 306 & 431 & 419 & {\color[HTML]{333333} 204} & {\color[HTML]{333333} 294} & {\color[HTML]{333333} 218} & {\color[HTML]{333333} 66} & {\color[HTML]{333333} 94} & {\color[HTML]{333333} 132} & {\color[HTML]{333333} 36} & {\color[HTML]{333333} 43} & {\color[HTML]{333333} 69} \\
{\color[HTML]{036400} SableNTT} & {\color[HTML]{CB0000} Medium} & 568 & 742 & 730 & {\color[HTML]{333333} 370} & {\color[HTML]{333333} 497} & {\color[HTML]{333333} 395} & {\color[HTML]{333333} 129} & {\color[HTML]{333333} 167} & {\color[HTML]{333333} 222} & {\color[HTML]{333333} 69} & {\color[HTML]{333333} 78} & {\color[HTML]{333333} 113} \\
\multicolumn{1}{l}{} & {\color[HTML]{680100} High} & 924 & 1,149 & 1,124 & {\color[HTML]{333333} 599} & {\color[HTML]{333333} 763} & {\color[HTML]{333333} 624} & {\color[HTML]{333333} 213} & {\color[HTML]{333333} 260} & {\color[HTML]{333333} 331} & {\color[HTML]{333333} 112} & {\color[HTML]{333333} 126} & {\color[HTML]{333333} 169} \\ \hline
\end{tabular}}
\end{table}

\begin{table}[!ht]
\centering
\caption{Comparing performance and stack memory requirement of Scabbard schemes with Saber and Kyber on Cortex-M4 platform}
\label{tab:M4_performance_stack}
\begin{tabular}{lcrrr|rrr}
\hline
\multicolumn{1}{c}{{\color[HTML]{3531FF} }} & {\color[HTML]{3531FF} } & \multicolumn{3}{c|}{{\color[HTML]{3531FF} Performance}} & \multicolumn{3}{c}{{\color[HTML]{3531FF} Stack memory}} \\
\multicolumn{1}{c}{{\color[HTML]{3531FF} }} & {\color[HTML]{3531FF} } & \multicolumn{3}{c|}{{\color[HTML]{3531FF} (x1000 clock cycles)}} & \multicolumn{3}{c}{{\color[HTML]{3531FF} (bytes)}} \\ \cline{3-8} 
\multicolumn{1}{c}{\multirow{-3}{*}{{\color[HTML]{3531FF} \begin{tabular}[c]{@{}c@{}}Scheme \\ Name\end{tabular}}}} & \multirow{-3}{*}{{\color[HTML]{3531FF} \begin{tabular}[c]{@{}c@{}}Security\\ level\end{tabular}}} & \multicolumn{1}{c}{{\color[HTML]{3531FF} KeyGen}} & \multicolumn{1}{c}{{\color[HTML]{3531FF} Encaps}} & \multicolumn{1}{c|}{{\color[HTML]{3531FF} Decaps}} & \multicolumn{1}{c}{{\color[HTML]{3531FF} KeyGen}} & \multicolumn{1}{c}{{\color[HTML]{3531FF} Encaps}} & \multicolumn{1}{c}{{\color[HTML]{3531FF} Decaps}} \\ \hline
 & {\color[HTML]{FD6864} Low} & 299 & 536 & 606 & 8,256 & 8,392 & 8,392 \\
\multicolumn{1}{c}{{\color[HTML]{036400} Florete}} & {\color[HTML]{CB0000} Medium} & 439 & 815 & 957 & 18,252 & 18,420 & 18,420 \\
 & {\color[HTML]{680100} High} & 598 & 1,131 & 1,357 & 25,408 & 25,608 & 25,608 \\ \hline
 & {\color[HTML]{FD6864} Low} & 1,659 & 1,859 & 1804 & 2,544 & 1,960 & 1,840 \\
\multicolumn{1}{c}{{\color[HTML]{036400} Espada}} & {\color[HTML]{CB0000} Medium} & 2,342 & 2,566 & 2,497 & 2,896 & 2,120 & 2,000 \\
 & {\color[HTML]{680100} High} & 3,577 & 3,859 & 3,779 & 3,424 & 2,360 & 2,240 \\ \hline
\multicolumn{1}{c}{{\color[HTML]{036400} }} & {\color[HTML]{FD6864} Low} & 381 & 558 & 568 & 5,672 & 5,928 & 5,432 \\
\multicolumn{1}{c}{{\color[HTML]{036400} Sable}} & {\color[HTML]{CB0000} Medium} & 745 & 1,005 & 1,031 & 6,184 & 5,992 & 5,496 \\
\multicolumn{1}{c}{{\color[HTML]{036400} }} & {\color[HTML]{680100} High} & 1,251 & 1,593 & 1,622 & 6,696 & 6,056 & 5,560 \\ \hline
 & {\color[HTML]{FD6864} Low} & 306 & 431 & 419 & 5,548 & 6,220 & 6,228 \\
\multicolumn{1}{c}{{\color[HTML]{036400} SableNTT}} & {\color[HTML]{CB0000} Medium} & 568 & 742 & 730 & 6,564 & 7,244 & 7,252 \\
 & {\color[HTML]{680100} High} & 924 & 1,149 & 1,124 & 7,596 & 8,276 & 8,284 \\ \hline
 & {\color[HTML]{FD6864} Low} & 454 & 631 & 643 & 6,060 & 6,020 & 6,028 \\
\multicolumn{1}{c}{{\color[HTML]{036400} Saber~\cite{saber_round_3}}} & {\color[HTML]{CB0000} Medium} & 856 & 1,106 & 1,121 & 6,572 & 6,540 & 6,548 \\
 & {\color[HTML]{680100} High} & 1,382 & 1,694 & 1,726 & 7,084 & 7,052 & 7,060 \\ \hline
 & {\color[HTML]{FD6864} Low} & 351 & 481 & 452 & 5,628 & 6,308 & 6,316 \\
\multicolumn{1}{c}{{\color[HTML]{036400} SaberNTT~\cite{Saber_ntt_opt_implementation}}} & {\color[HTML]{CB0000} Medium} & 644 & 820 & 773 & 6,652 & 7,332 & 7,340 \\
 & {\color[HTML]{680100} High} & 992 & 1,203 & 1,149 & 7,676 & 8,348 & 8,356 \\ \hline
 & {\color[HTML]{FD6864} Low} & 434 & 530 & 477 & 4,320 & 5,424 & 5,432 \\
\multicolumn{1}{c}{{\color[HTML]{036400} Kyber-Speed~\cite{PQM4}}} & {\color[HTML]{CB0000} Medium} & 707 & 863 & 783 & 5,344 & 6,456 & 6,472 \\
 & {\color[HTML]{680100} High} & 1,123 & 1,316 & 1,210 & 6,400 & 7,496 & 7,512 \\ \hline
 & {\color[HTML]{FD6864} Low} & 434 & 532 & 478 & 2,248 & 2,336 & 2,352 \\
\multicolumn{1}{c}{{\color[HTML]{036400} Kyber-Stack~\cite{PQM4}}} & {\color[HTML]{CB0000} Medium} & 707 & 867 & 788 & 2,784 & 2,856 & 2,872 \\
{\color[HTML]{036400} } & {\color[HTML]{680100} High} & 1,127 & 1,324 & 1,219 & 3,296 & 3,368 & 3,392 \\ \hline
 & {\color[HTML]{FD6864} Low} & 75,000 & 85,000 & 84,000 & 12,516 & 14,468 & 14,476 \\
\multicolumn{1}{c}{{\color[HTML]{036400} Frodo-speed~\cite{Frodo_kem_m4}}} & {\color[HTML]{CB0000} Medium} & 169,000 & 186,000 & 185,000 & 18,572 & 19,860 & 19,868 \\
 & {\color[HTML]{680100} High} & 309,000 & 345,000 & 344,000 & 25,196 & 25,764 & 25,772 \\ \hline
 & {\color[HTML]{FD6864} Low} & 223,000 & 293,000 & 294,000 & 7,948 & 6,668 & 6,460 \\
\multicolumn{1}{c}{{\color[HTML]{036400} Frodo-stack~\cite{Frodo_kem_m4}}} & {\color[HTML]{CB0000} Medium} & 1,103,000 & 1,296,000 & 1,296,000 & 5,444 & 4,796 & 4,596 \\
 & {\color[HTML]{680100} High} & 2,003,000 & 2,380,000 & 2,379,000 & 6,916 & 5,532 & 5,324 \\ \hline
\end{tabular}
\end{table}

We have used optimized Toom-Cook-based polynomial multiplication for Florete, Espada, and Sable. These optimized polynomial multiplications are generated using the software package provided by Kannwischer et al.~\cite{polymul-z2mx-m4}. It can generate optimized assembly code for different combinations of Toom-Cook-based polynomial multiplications. As mentioned earlier, LWR-based schemes directly cannot use NTT. Later, Chung et al.~\cite{chung_kannwischer_NTT_friendly} showed that Saber could use the NTT-based polynomial multiplication over a big prime field so that the absolute magnitude of the largest possible number occurs from the polynomial multiplication is smaller than the big prime. \cite{chung_kannwischer_NTT_friendly} also showed that Saber with NTT-based polynomial multiplication (SaberNTT) performs better than Saber with Toom-Cook-based polynomial multiplication. Afterward, Abdulrahman et al.~\cite{Saber_ntt_opt_implementation} further improved the NTT-based polynomial multiplication of Saber using multi-moduli NTT. To show that Scabbard's schemes can be optimized for speed using NTT-based polynomial multiplication, we have implemented a version of Sable that uses NTT-based polynomial multiplication. We have implemented a multi-moduli NTT-based Sable with the help of the multi-moduli NTT-based implementation of Saber. We call this NTT-based Sable as SableNTT in Table~\ref{tab:M4_performance_stack}.

We can observe from Table~\ref{tab:M4_performance_stack} that the \texttt{KeyGen} algorithm of Florete performs $34\%$, $49\%$, $57\%$ faster than Saber, $31\%$, $38\%$, $47\%$ faster than Kyber-Speed, and $99.6\%$, $99.7\%$, $99.8\%$ faster than Frodo-Speed for low, medium, and high-security versions, respectively. The \texttt{Encaps} algorithm of Florete performs $15\%$, $26\%$, $33\%$ faster than Saber, and $99.4\%$, $99.6\%$, $99.7\%$ faster than Frodo-Speed for low, medium, and high-security versions, respectively. The \texttt{Encaps} algorithm of Florete performs $6\%$, $14\%$ better compared to Kyber-Speed for medium, and high-security versions, respectively. The \texttt{Decaps} algorithm of Florete performs $6\%$, $15\%$, $21\%$ faster than Saber, and $99.3\%$, $99.5\%$, $99.6\%$ faster than Frodo-speed for low, medium, and high-security versions, respectively. However, one thing to note is that the improvement of the performance of the \texttt{KeyGen}, \texttt{Encaps}, and \texttt{Decaps} algorithms for Florete against Saber increases as security increases. Also, the performance improvement of the \texttt{KeyGen}, \texttt{Encaps} algorithms for Florete against Kyber-Speed increases as security increases, and the slowdown factor for the \texttt{Decaps} algorithm of Florete against Kyber decreases as security increases. 

The stack memory requirements for the implementations of low, medium, and high-security versions of the \texttt{KeyGen} algorithm in Espada are respectively $58\%$, $56\%$, $52\%$ lower than Saber and $68\%$, $47\%$, $50\%$ lower than Frodo-Stack. The \texttt{KeyGen} algorithm of Espada requires more stack memory than Kyber\footnote{The latest Kyber uses a different technique to generate matrix $A$ and $A^T$ during the matrix-vector multiplication than Espada. In Kyber, each polynomial of the matrix $A$ and $A^T$ can be generated independently from SHAKE-128 with a slightly different version of the seed. Therefore, each polynomial of the matrix $A$ can be generated run-time during matrix-vector multiplication using just-in-time strategy~\cite{saber_on_arm} and only one polynomial space is required for the matrix $A$ and $A^T$. However, in Espada, the whole matrix $A$ is generated from a single seed. So, between matrix $A$ and $A^T$, the matrix $A$ can utilize the maximum benefit from the just-in-time strategy. Here, the matrix $A$ needs one polynomial to store the whole matrix, but the matrix $A^T$  needs a vector of polynomial space for the entire matrix. Therefore, the \texttt{KeyGen} algorithm of Espada requires more stack memory than Kyber. However, Kyber's matrix generation technique can also be used in Espada. Then, like \texttt{Encaps} algorithm, \texttt{KeyGen} algorithm of Espada will require less stack memory than Kyber.}. For implementations of low, medium, and high-security versions of the \texttt{Encaps} algorithm in Espada, the stack memory requirements are respectively $67\%$, $68\%$, $67\%$ lower than Saber, $16\%$, $26\%$, $30\%$ lower than Kyber, and $71\%$, $56\%$, $57\%$ lower than Frodo-Stack. The stack memory requirements in implementations of low, medium, and high-security versions of the \texttt{Decaps} algorithm in Espada are respectively $69\%$, $69\%$, $68\%$ lower than Saber, $22\%$, $30\%$, $34\%$ lower than Kyber, and $72\%$, $56\%$, $58\%$ lower than Frodo-Stack.

The \texttt{KeyGen} algorithm of Sable performs at least $9\%$ faster than Saber and at least $99.5\%$ faster than Frodo-Speed. The \texttt{Encaps} and \texttt{Decaps} algorithms of Sable perform at least $6\%$ faster than Saber and at least $99.3\%$ faster than Frodo-Speed. All the algorithms of Sable also use less stack memory compared to Saber for all the security versions. The NTT-based version of Sable is named SableNTT. The \texttt{KeyGen} algorithm of SableNTT performs $13\%$, $12\%$, $7\%$ faster than SaberNTT, $29\%$, $20\%$, $18\%$ faster than Kyber-Speed, and $99.6\%$, $99.7\%$, $99.7\%$ faster than Frodo-Speed for low, medium, and high-security versions, respectively. The \texttt{Encaps} algorithm of SableNTT performs $10\%$, $10\%$, $4\%$ faster than SaberNTT, $19\%$, $14\%$, $13\%$ faster than Kyber-Speed, and $99.5\%$, $99.6\%$, $99.7\%$ faster than Frodo-Speed for low, medium, and high-security versions, respectively. The \texttt{Decaps} algorithm of SableNTT performs $7\%$, $6\%$, $2\%$ faster than SaberNTT, $12\%$, $7\%$, $7\%$ faster than Kyber-Speed, and $99.5\%$, $99.6\%$, $99.7\%$ faster than Frodo-speed for low, medium, and high-security versions, respectively. Also, all three algorithms of SableNTT also need less stack memory compared to SaberNTT for all the security versions.

%% file: sections/hardware_implementation.tex
\section{Hardware Implementation}\label{sec:HW_impl}
The following section describes our design decisions for the unified implementations of Scabbard (medium security version) in hardware, followed by a discussion and comparison of the results. We will demonstrate how taking hardware efficiency into account during the design cycle of cryptographic schemes leads to efficient implementations on hardware platforms.

All schemes in the Scabbard suite (Florete, Espada and Sable) are LWR-based KEMs and consist of three main routines: (i) key-generation (\texttt{LWR.KEM.KeyGen}), (ii) encapsulation (\texttt{LWR.KEM.Encaps}), and (iii) decapsulation (\texttt{LWR.KEM.Decaps}). As outlined in Sec.~\ref{sec:preliminaries} and \ref{section:scabbard}, these share several fundamental operations, including polynomial multiplication, hashing, pseudo-random number generation, and binomial sampling. We explore the trade-off between high speed, low area and high flexibility for the full-hardware implementation of the KEM operations by (re-)using common building blocks where possible. Where possible, we optimize our implementation to meet our design objectives, outlined in Sec.~\ref{section:scabbard}.

\subsection{High-level Architecture}
\begin{figure}[ht!]
\centering
  \includegraphics[scale=.4]{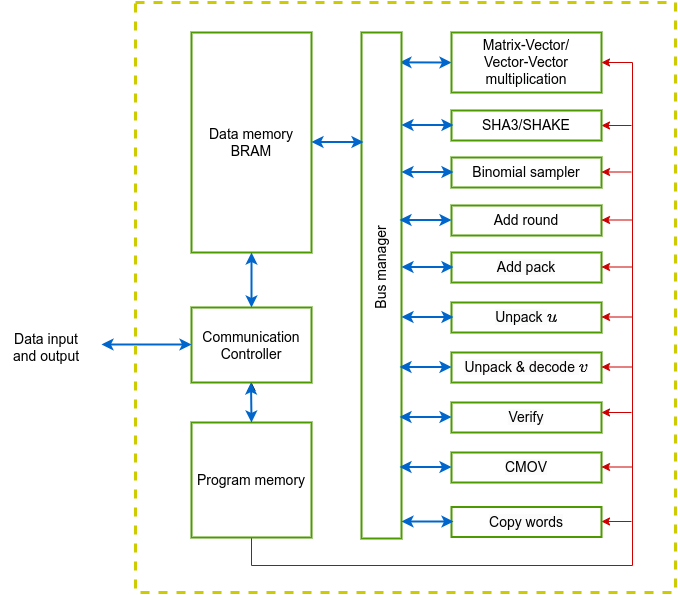}
  \caption{The high-level architecture diagram of the instruction set processor for schemes of Scabbard. The blue line symbolizes the data bus, and the red line indicates the control signal.}
  \label{fig:full_architecture}
\end{figure}

We follow a hardware (HW) only design methodology as opposed to a HW/SW co-design strategy. In a HW/SW co-design, only the most computationally expensive operations (i.e. polynomial multiplier) are implemented on hardware, providing high flexibility at the cost of reduced performance.
In our implementations, all the building blocks reside in hardware, as we prioritize speed. Yet, our implementation remains flexible (e.g. support for key generation, encapsulation, and decapsulation) by targeting an \textit{instruction-set coprocessor architecture} (ISA), as proposed in \cite{Sujoy_Saber_h/w}. Such a unified architecture offers instruction-level flexibility and modularity. A high-level diagram of the ISA for all schemes of Scabbard is shown in Fig.~\ref{fig:full_architecture}. As the CCA-secure KEM routines for Florete, Espada, and Sable follow a common framework (Sec.~\ref{sec:genericLWR}), the high-level architecture of their hardware implementations are similar. The differences between implementations of sub-blocks of different schemes are explicitly listed, and our design methodology is explained, if applicable. We also focus on the particularities of the different polynomial multipliers for Florete, Espada, and Sable.

The coprocessor is controlled by loading the program memory with the microcode of the protocol (e.g. key generation). The instruction words are 35-bit, consisting of a 5-bit wide instruction code, and 3$\times$10-bit data addresses, of which two are for input operands and one for the result. The algorithms and instructions are designed to not include conditional branching, to prevent timing-based side-channel attacks. The main communication controller interacts with the individual building blocks, which are designed to be constant-time. As a result, each of the implementations of all KEM operations takes a fixed amount of time.

\subsection{Data Memory}
Both input data and results for each of the operations are read from/written back to the data memory, which is implemented using BRAM tiles. The medium security versions of all Scabbard schemes require at minimum an 8KB memory size such that all KEM routines can be computed. The word size is 64-bit, as this allows for easy integration of the ISA co-processor with a 32-bit or 64-bit host computer. We ensure data is optimally packed inside the 64-bit words, and all individual blocks maximally exploit this format.

\subsubsection{Espada}
In order to be able to store the public matrix $\mathbf{A}$, generated from the public seed$_A$ using an $XOF$, we instantiate an additional $\pm$ 17KB of \textit{parallel} data memory. We design and implement it to consist of $l=12$ parallel memory banks, which each store $n=64$ coefficients. It allows for our polynomial multiplier to maximally exploit its parallel nature by simultaneously reading from and writing to all $l$ memory banks in a Single-Instruction Multiple-Data (SIMD) fashion during matrix-vector multiplication (Fig.~\ref{fig:espada_mult_hw}).

\subsection{SHA3/SHAKE}
The Scabbard suite relies on the HW-friendly Keccak sponge function (FIPS 202) \cite{fips_keccak} through the hash functions SHA3-256 and SHA3-512 and the extendable output function SHAKE-128 for generating pseudorandom numbers. The SHA3/SHAKE block is implemented using the open-source high-speed implementation of the Kecak core, designed by the Keccak Team \cite{keccak_hw}. Around this sits the SHA3/SHAKE wrapper from the open-source implementation of Saber on hardware \cite{Sujoy_Saber_h/w}.

All data padding and extraction operations are performed in the wrapper in hardware, controlled by a second instruction from the program memory. The input/output data length is flexible and first specified through 2$\times$16-bit fields, followed by the data and result operand addresses. The SHA3/SHAKE block consumes around 5,900 LUTs and 3,127 FFs, or up to 35$\%$ of total area utilization of the full HW implementation of Sable. Also, during the decapsulation operation, up to 21$\%$ of total execution time (1,521 clock cycles for Sable) is required for Keccak-related operations. For Florete, where the polynomial multiplier is a more performance-critical component, the SHA3/SHAKE block accounts for around 20$\%$ of total area utilization. For a full decapsulation operation, 4$\%$ of total execution time is required for Keccak-related operations. We argue that instantiating a single Keccak core in hardware is a good compromise, as we achieve high speed, and this building block is already an area-expensive component. 

\subsubsection{Florete} As this scheme is based on the RLWR hard problem, the ring/modulus parameter $l$ is equal to 1. The public matrix $\mathbf{A}$ is a polynomial and an element of the ring $\mathcal{R}_q^n$. Compared to Sable and Espada, where $\mathbf{A}$ is a matrix, the generation of this public value is much cheaper for Florete in hardware. Only 426 clock cycles are required in total for all SHAKE-128 operations during encapsulation and decapsulation.

\subsection{Binomial Sampler}
In Scabbard, the secret coefficients are drawn from a centered binomial distribution with parameter $\mu$. A $\mu$-bit pseudo-random string $r[\mu-1:0]$ is split in two parts, and the Hamming weight \texttt{HW()} of each is subtracted. More specifically, \texttt{HW($r[\frac{\mu}{2}-1:0]$) - HW($r[\mu - 1 :\frac{\mu}{2}]$)} is computed. As proposed in \cite{Sujoy_Saber_h/w}, the sampler is implemented as a combinatorial block with an input and output buffer. The output samples are in sign-magnitude representation. 

\subsubsection{Florete \& Sable}
For both schemes $\mu=2$, meaning the secret coefficients are in $[-1:1]$ (two bits), and 32 samples can be directly stored in a 64-bit data word. First, a 64-bit pseudo-random word is loaded from the data memory, stored in a buffer and then 32 samples are generated in parallel. The 64-bit result is transferred from the output buffer to the global data memory and repeated until the full secret polynomial is generated.

\subsubsection{Espada}
As $\mu=6$, which is not a divisor of 64, the input buffer is 192-bit since \texttt{lcm}$(6,64)=192$. Three 64-bit, pseudo-random strings are loaded to the input registers, after which 16 4-bit samples are computed twice in a row. Generating 16 output samples requires $96=2\cdot3\cdot16$ bits, meaning the process is repeated twice until the input buffer is filled again.

\subsection{Polynomial Multiplication}
The following section discusses the particularities of our polynomial multiplier design for Florete, Espada, and Sable. As Scabbard was designed to be polynomial arithmetic-friendly, we use off-the-shelf and state-of-the-art polynomial multipliers to demonstrate their efficiency in hardware.
We integrate the multipliers in our high-speed hardware design and modify them so they support both inner-product and matrix-vector polynomial multiplications on the same hardware. Depending on the instruction loaded from the program memory, the appropriate control signals are set, and the operation is performed.

During operation, typically, the secret polynomial $\mathbf{s}$ is first loaded from the data memory, unpacked, and stored in a LUT-based buffer. Secondly, the polynomial multiplicand $\mathbf{a}$ is loaded into an input buffer. For the matrix-vector multiplication $\mathbf{A\cdot s}$ or $\mathbf{A}^T\mathbf{\cdot s}$, the coefficients are $\epsilon_q$ bits and generated by \texttt{SHAKE-128}, 64 bits at a time and continuously stored in the data memory. The input buffer unpacks the coefficients which may be split across different words in data memory, to 16-bit operands for the multipliers. The processing starts as soon as the first few coefficients are available in this buffer, parallelizing the computation and data transfers, as proposed in \cite{Sujoy_Saber_h/w}. For the inner-product calculations $\mathbf{u}^T\mathbf{\cdot s}$ or $\mathbf{b}^T\mathbf{\cdot s'}$, the coefficients of the polynomial multiplicand are $\epsilon_p$-bit wide and zero-padded up to (and stored as) 16-bit coefficients. Four are loaded from the data memory at once (in one 64-bit word) and directly stored in the polynomial multiplicand registers. After the computation has finished, the coefficients of the resulting polynomial are zero-padded to 16-bit, packed into 64-bit data words, and stored in the data memory.

\subsubsection{Florete}
In Florete, a 768$\times$768 polynomial multiplication is required, which we decompose into smaller (256$\times$256) polynomial multiplications by implementing Toom-Cook 3-way evaluation and interpolation in hardware. These acts as a wrapper around five 256$\times$256 polynomial multipliers, which we all instantiate in parallel. A benefit of our approach, using Toom-Cook 3, is that our implementation can reuse any state-of-the-art 256$\times$256 polynomial multiplier available in literature \cite{multiplier256,MeraTKRV20,Sujoy_Saber_h/w}.

\begin{figure}[ht!]
\centering
  \includegraphics[scale=.55]{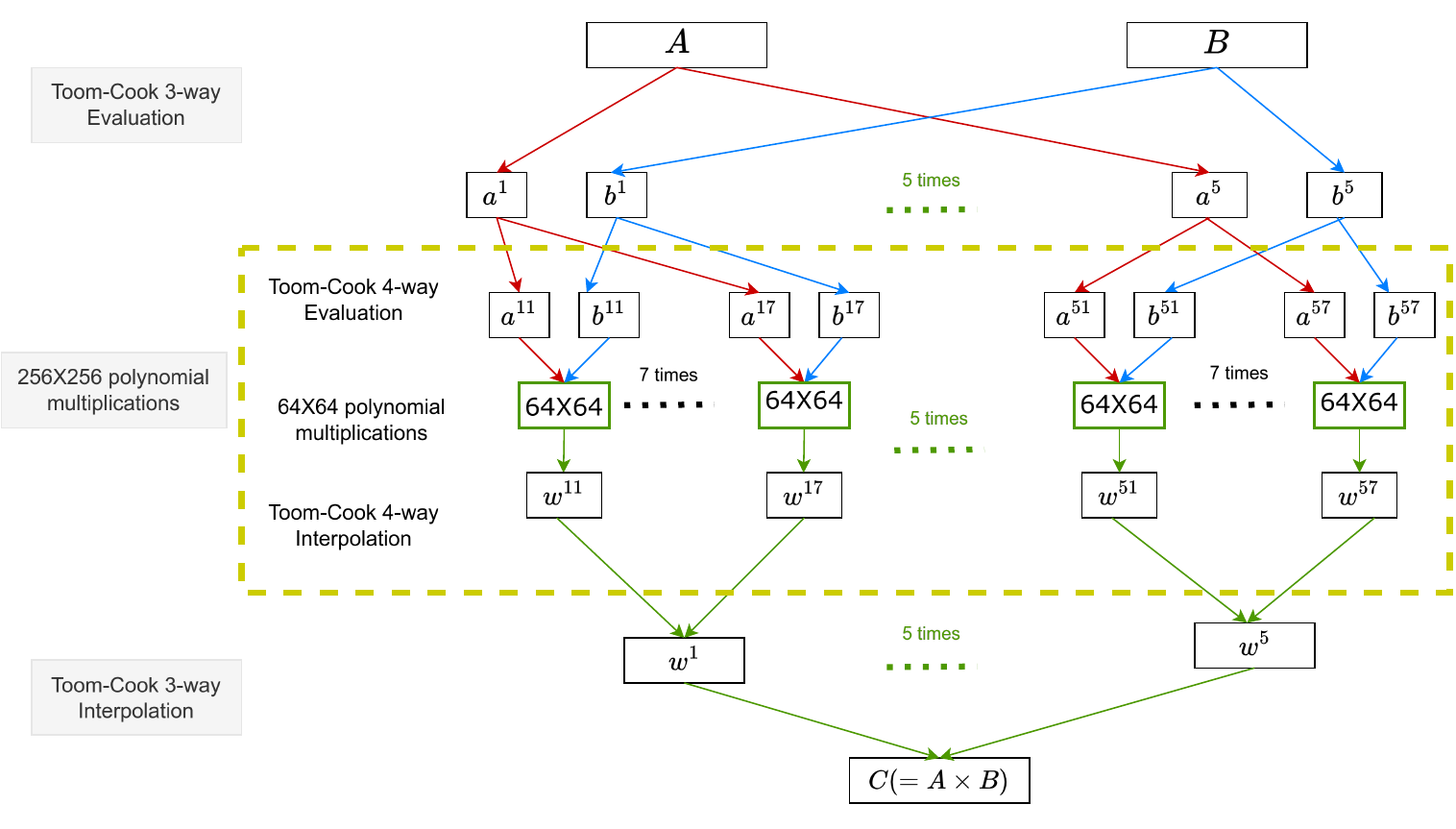}
  \caption{Polynomial multiplication of Florete. 768$\times$768 multiplication is decomposed into 35 64$\times$64 polynomial multiplications, using Toom-Cook 3-way and Toom-Cook 4-way.}
  \label{fig:florete_mult_hw}
\end{figure}

In order to further exploit parallelism in our hardware implementation, we break down the polynomial multiplication further using Toom-Cook 4-way evaluation and interpolation as proposed in \cite{MeraTKRV20}. Hence, each 256$\times$256 polynomial multiplier consists of 7 parallel 64$\times$64 polynomial multipliers, which all execute in parallel. In total, $7*5=35$ 64$\times$64 polynomial multiplications are instantiated and performed in parallel (Fig.~\ref{fig:florete_mult_hw}).

As a result, a 768$\times$768 multiplication takes as long as a single 256$\times$256 multiplication at the cost of a five times larger area. In between the different stages, intermediate and accumulated results are stored in LUT-based buffers. The evaluation and interpolation datapath are pipelined.

\subsubsection{Espada}
\begin{figure}[ht!]
\centering
  \includegraphics[scale=.4]{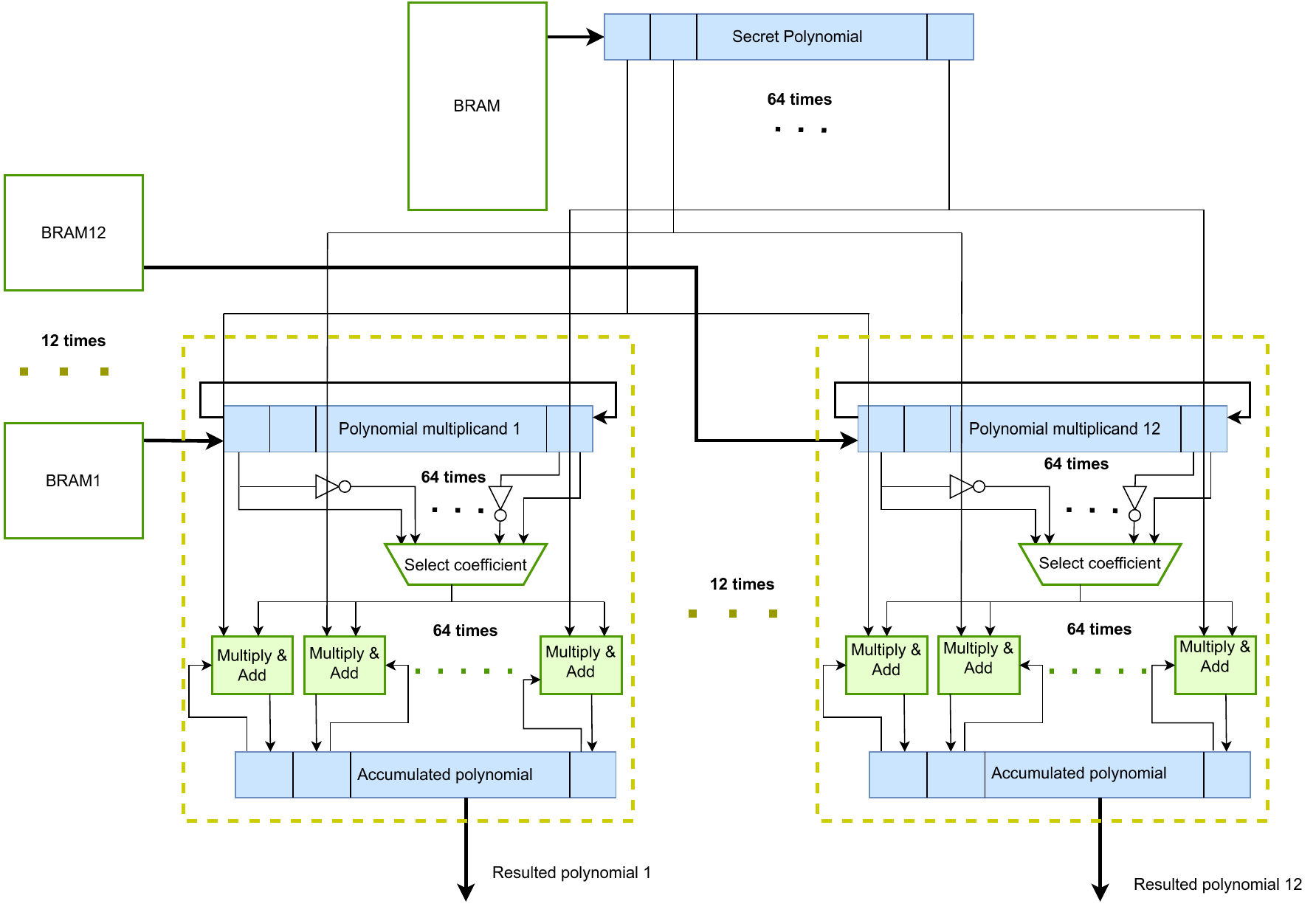}
  \caption{Polynomial multiplication for Espada. Blue colored blocks represent register and the multiplication and add block is colored green. The dotted yellow colored block represents a $64\times 64$ polynomial multiplication. Here, we perform 12 such multiplication in parallel.}
  \label{fig:espada_mult_hw}
\end{figure}

Our Espada multiplier is designed to exploit the inherent parallelism of the scheme and its matrix-vector multiplication. We do not require any evaluation/interpolation steps to break down a large polynomial multiplication due to the choice of parameters and choice of module lattices.
More specifically, as can be observed in Fig.~\ref{fig:espada_mult_hw}, $l$ 64$\times$64 polynomial multipliers are instantiated in parallel ($l=12$ for medium security level). During the matrix-vector multiplication, each multiplier is fed with one row of the public matrix of dimension $l \times l$ in parallel. During the inner-product operation, only one multiplier is active. In both cases, the corresponding secret polynomial is the same for all multipliers and is loaded first in a small LUT-based buffer. 

Before the computation starts, the polynomial multiplicands are loaded to small LUT-based buffers, instantiated for each of the multipliers. 
These allow for each of the multipliers to perform read and write operations during computation. 
In order to minimize the overhead of loading the large public matrix $\mathbf{A}$ from data memory and writing back the results of all $l$ multipliers to global memory, we instantiate an additional data memory consisting of $l=12$ banks. During the generation of the public matrix using SHAKE-128, these are filled in a sequential manner, containing one row of the matrix each. The multipliers read the polynomial multiplicands from these banks in parallel and write their results to $l$ banks in parallel. As such, not only the computation but also the read and write operations are parallelized (on the matrix-vector level), bringing the performance close to state-of-the-art.

For both Florete and Espada, our architecture leaves space to optimize the 64$\times$64 multipliers for area or performance. Our parametrizable design allows us to select the number of arithmetic units (implemented using DSP units), achieving higher performance at the cost of higher area utilization. We choose 4 DSP units per polynomial multiplier, in order to keep area cost at a reasonable level. As a result, our latency is high compared to our Sable implementation, as our multiplier requires $16*64$ clock cycles to perform a full 64$\times$64 multiplication. 

\subsubsection{Sable}
\begin{figure}[ht!]
\centering
  \includegraphics[scale=.5]{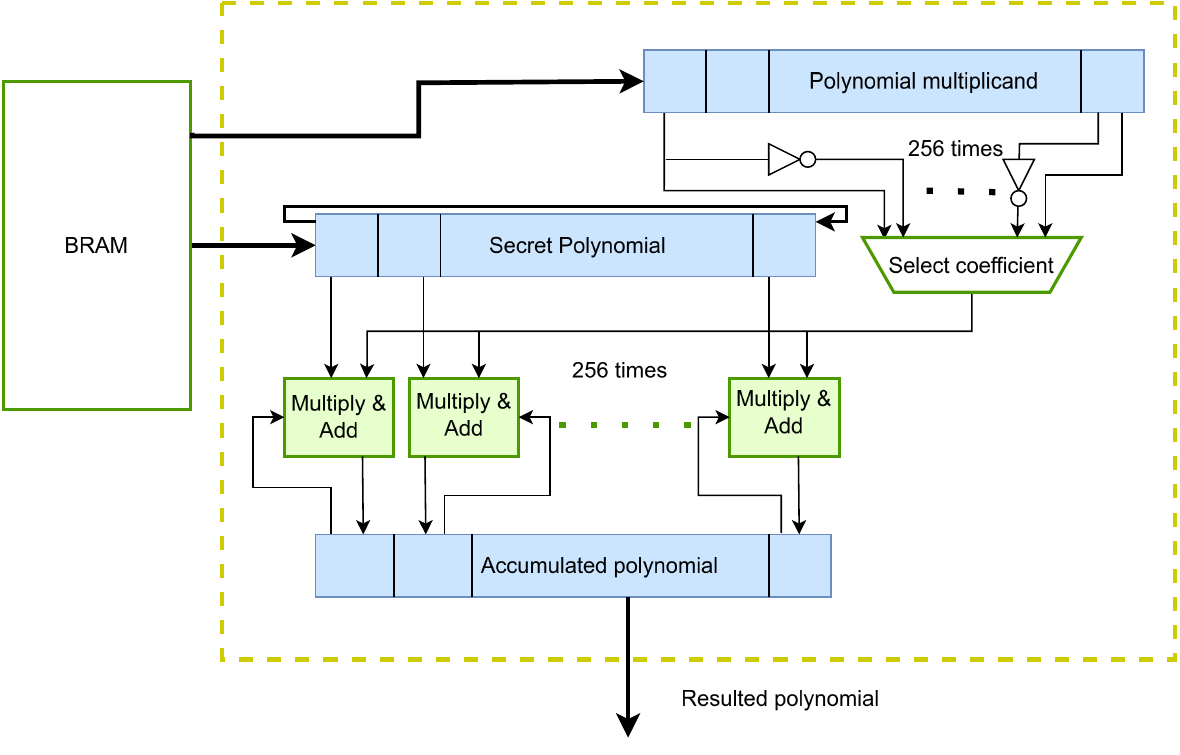}
  \caption{Polynomial multiplication for Sable. Blue colored blocks represent register and the multiplication and add block is colored green.}
  \label{fig:sable_mult_hw}
\end{figure}

The Sable multiplier is based on the high-speed Saber implementation in \cite{Sujoy_Saber_h/w} but optimized for the Sable parameters. As a result, the area requirements are reduced without a performance loss as the property of 2-bit secrets is exploited. Our proposed architecture is drawn in Fig.~\ref{fig:sable_mult_hw}. The custom `Multiply \& Add' arithmetic unit (in green) is instantiated 256 times, resulting in a full parallel polynomial multiplication. The arithmetic unit is a combinatorial block, thus, a full $N=256$ polynomial multiplication requires only 256 cycles. 

Before computation, the entire 512-bit secret polynomial $\mathbf{s}$ is loaded in a LUT-based shift register, which allows for the negacyclic convolution to be performed in-place and access to all secret coefficients at once. The nega-cyclic left-shift operation moves each secret coefficient from position $i$ to $i+1$ and the last secret to the first position after modular subtraction from zero. As the secret coefficients use sign-magnitude representation, only a simple sign flip is required.

The full polynomial multiplicand $\mathbf{a}$ is also loaded into an input buffer, from which four coefficients at a time are processed by the arithmetic units. For matrix-vector multiplication, as $\epsilon_q=11$, 11-bit coefficients are unpacked to four 16-bit coefficients by the input buffer, and 64-bit data words are fed to the MAC units. For the inner-product calculations, four zero-padded coefficients are stored in one data-memory word ($\epsilon_p=11$), which are directly wired to the MAC units.

The result of the arithmetic unit is stored in the output accumulator buffer. This value is reset or preserved, depending on whether an inner product or full row-column multiplication (matrix-vector multiplication) is computed. Upon completion, the resulting polynomial is stored in the global data memory. 

The custom arithmetic unit is optimized for Sable's 2-bit secrets: only if \texttt{LSB(}$s_i$) is 1 the accumulated result will be updated. The most significant bit of the secret determines if $a_i$ is added or subtracted from the result.

\subsection{AddRound, AddPack, and Unpack/Decode}
All three schemes in the Scabbard suite use power-of-two moduli $p=2^{\epsilon_p}$ and $q=2^{\epsilon_q}$. In hardware, this translates to modular reduction and rounding being essentially free as they consist of shifting, re-wiring, adding, and bit-selecting. However, as the exact parameters are rarely selected as multiples of 8, low-level bit manipulations and small memory buffers are required. We fine-tune our implementation to minimize additional area utilization.

\subsection{Remaining sub-blocks}
The \textit{Verify} module compares the received ciphertext and re-encrypted ciphertext during the decapsulation, word-by-word, and stores the result in a flag register. The \textit{CMOV} module copies either the shared session key $K$ or a pseudo-random string to a specified location based on this flag. The data move is constant-time.
\subsubsection{Florete} As $l=1$ for Florete, being based on the RLWR hard problem, $\mathbf{A}$ consists of a single polynomial. As a result, no matrix transpose is required during key generation, resulting reducing the cycle count. 
\subsubsection{Espada} An additional module, \textit{CopyTranspose}, is added in order to efficiently transpose the $l \times l$ matrix $\mathbf{A}$. Still, it is relatively more expensive compared to other Scabbard schemes and Saber.

\subsection{Performance Evaluation}
Our full-hardware ISA is described in mixed Verilog and VHDL and compiled using Xilinx Vivado 2021.1 (default strategies) for the target platform Xilinx ZCU102 board, containing an Ultrascale+ XCZU9EG-2FFVB1156 FPGA and Arm Cortex-A53 host processor.
Before a KEM operation, all operand data is transferred from the host processor to the coprocessor at once, then all computations are performed on the FPGA, and the result is read back by the host processor. 


\subsubsection{Timing Results}
We first give a detailed breakdown of the cycle counts for the individual low-level operations and total cycle count in Table \ref{tab:FPGA_cycles_timing_results}. Numbers for our implementation of all Scabbard schemes and the Saber implementation (using 256 MAC units \& multipliers) from \cite{Sujoy_Saber_h/w} are provided and compared. Table \ref{tab:FPGA_time_timing_results} shows the total execution times for our hardware implementations, calculated at 150MHz using Vivado simulation.
We compare our designs of the Scabbard suite, which are based on variants of the LWR problem, with Saber, as it is the most well-known LWR-based scheme. For \texttt{Keygen}/\texttt{Encaps}/\texttt{Decaps} operations, our Sable implementation requires 13/11/10$\%$ fewer clock cycles. This is mainly due to our optimized multiplier design's relaxed requirements for sampling pseudo-random numbers compared to Saber. Our multiplier, similar to the Saber design, uses 256 MAC units and multipliers, which allows for the best direct comparison. It is clear that our design decisions lead to improved performance in hardware.

Additionally, all Scabbard schemes benefit from their choice of secret distribution, which results in more efficient vector sampling. For all KEM operations, Florete/Espada/Sable require 84/16/84\% fewer clock cycles compared to Saber, respectively.

In all Scabbard KEM operations, the time spent performing polynomial multiplications is significant: 85/85/86\%, 69/82/87\% and 55/59/60\% of total \texttt{Keygen}/\texttt{Encaps}/\texttt{Decaps} cycle counts. Our Sable multiplier is optimized for the 2-bit secrets and consists of 256 MAC units and multipliers, resulting in a low total latency. 

Both Espada and Florete rely on 64$\times$64 polynomial multipliers, which are implemented using only 4 DSP units. Increasing the DSP units of each multiplier will bring their performance closer to the state-of-the-art at the cost of increased area utilization. Our implementation prioritizes area cost while still achieving reasonable latency overhead. Notice that for Espada, due to our parallelized design, the latency for a complete matrix-vector multiplication and inner product are identical, as $l$ polynomial multipliers are instantiated in parallel. 

The second significant factor in the execution time of Scabbard are all Keccak-based functions: SHA3-256, SHA3-512, and SHAKE-128. For key generation, encapsulation, and decapsulation, this is 10/8/4\%, 24/16/10\%, and 28/28/21\% of total cycle counts, respectively. 

Compared to Saber, our Florete implementation requires 67/70/70\% fewer clock cycles during  \texttt{Keygen}/\texttt{Encaps}/\texttt{Decaps} operations. Our Sable HW implementation requires 22/24/24\% fewer clock cycles compared to Saber.
A large contributing factor to the Espada cycle counts is its randomness requirement (SHAKE-128) for the generation of $\mathbf{A}$ (around 5K clock cycles). 

\begin{table}[!ht]
\centering
\caption{Total cycles spent in low-level operations for Scabbard schemes and Saber \cite{Sujoy_Saber_h/w} using Vivado simulation (Medium security parameters).}
\label{tab:FPGA_cycles_timing_results}
\begin{tabular}{lc|rrr}
\hline
\multicolumn{1}{c}{{\color[HTML]{3531FF} }}                              & {\color[HTML]{3531FF} }                                                                           & \multicolumn{3}{c}{{\color[HTML]{3531FF} Cycle Count}}                                                                                                                                                 \\ 
\multicolumn{1}{c}{\multirow{-2}{*}{{\color[HTML]{3531FF} Instruction}}} & \multirow{-2}{*}{{\color[HTML]{3531FF} \begin{tabular}[c]{@{}c@{}}Scheme\\ Name\end{tabular}}} & \multicolumn{1}{c}{{\color[HTML]{3531FF} KeyGen}} & \multicolumn{1}{c}{{\color[HTML]{3531FF} Encaps}} & \multicolumn{1}{c}{{\color[HTML]{3531FF} Decaps}} \\ \hline
                                    & {\color[HTML]{036400} Florete} & 200 & 589 &  333 \\
                                    & {\color[HTML]{036400} Espada} & 272 & 661 & 333  \\
\multicolumn{1}{c}{{SHA3-256}}      & {\color[HTML]{036400} Sable} & 200 & 541 & 387\\
                                    & & & &\\
                                    & {\color[HTML]{036400} Saber} &  339 & 585 & 303 \\ \hline
                                    
                                    & {\color[HTML]{036400} Florete} & 0 & 65 & 68  \\
                                    & {\color[HTML]{036400} Espada} & 0  & 65 & 68  \\
\multicolumn{1}{c}{{SHA3-512}}      & {\color[HTML]{036400} Sable} & 0 & 65 & 68 \\
                                    & & & &\\
                                    & {\color[HTML]{036400} Saber} &  0 & 62 & 62 \\ \hline
                                    
                                    & {\color[HTML]{036400} Florete} & 489 & 426 & 426  \\
                                    & {\color[HTML]{036400} Espada} & 5,352 & 5,286 & 5,286  \\
\multicolumn{1}{c}{{SHAKE-128}}      & {\color[HTML]{036400} Sable} & 1,135 & 1,066  & 1,066 \\
                                    & & & &\\
                                    & {\color[HTML]{036400} Saber} &  1,461 & 1,403 & 1,403 \\ \hline
                                    
                                    & {\color[HTML]{036400} Florete} & 28 & 28 & 28  \\
                                    & {\color[HTML]{036400} Espada} & 147 & 147 & 147  \\
\multicolumn{1}{c}{{Vector sampling}}      & {\color[HTML]{036400} Sable} & 25 & 28 & 28 \\
                                    & & & &\\
                                    & {\color[HTML]{036400} Saber} &  176 & 176 & 176 \\ \hline
                                    
                                    & {\color[HTML]{036400} Florete} & 6,051 & 12,102 &  18,153 \\
                                    & {\color[HTML]{036400} Espada} & 15,826 & 31,652 &  47,478 \\
\multicolumn{1}{c}{{Polynomial multiplications}}      & {\color[HTML]{036400} Sable} & 2,598 & 3,464 & 4,330 \\
                                    & & & &\\
                                    & {\color[HTML]{036400} Saber} &  2,685 & 3,592 & 4,484 \\ \hline
                                    
                                    & {\color[HTML]{036400} Florete} & 318 & 954 & 2,087  \\
                                    & {\color[HTML]{036400} Espada} & 1,441 & 767 & 1,511  \\
\multicolumn{1}{c}{{Remaining operations}}      & {\color[HTML]{036400} Sable} &  783 & 738 & 1,385 \\
                                    & & & &\\
                                    & {\color[HTML]{036400} Saber} &  792 & 800 & 1,606 \\ \hline \hline
                                    
                                        & {\color[HTML]{036400} Florete} & 7,086 & 14,164 & 21,095  \\
                                        & {\color[HTML]{036400} Espada} & 23,038 & 38,578 & 54,823  \\
\multicolumn{1}{c}{{Total cycles}}      & {\color[HTML]{036400} Sable} & 4,741 & 5,902  & 7,264 \\ 
                                        & & & &\\
                                    & {\color[HTML]{036400} Saber} & 5,453  & 6,618 & 8,034 \\     \hline                            
\end{tabular}
\end{table}

\begin{table}[!ht]
\centering
\caption{Execution times for Scabbard schemes and Saber \cite{Sujoy_Saber_h/w} using Vivado simulation calculated at 150 MHz 
(Medium security parameters).}
\label{tab:FPGA_time_timing_results}
\begin{tabular}{lc|rrr}
\hline
\multicolumn{1}{c}{{\color[HTML]{3531FF} }}                              & {\color[HTML]{3531FF} }                                                                           & \multicolumn{3}{c}{{\color[HTML]{3531FF} Time ($\mu$s)}}                                                                                                                                                 \\ 
\multicolumn{1}{c}{\multirow{-2}{*}{{\color[HTML]{3531FF} Instruction}}} & \multirow{-2}{*}{{\color[HTML]{3531FF} \begin{tabular}[c]{@{}c@{}}Scheme\\ Name\end{tabular}}} & \multicolumn{1}{c}{{\color[HTML]{3531FF} KeyGen}} & \multicolumn{1}{c}{{\color[HTML]{3531FF} Encaps}} & \multicolumn{1}{c}{{\color[HTML]{3531FF} Decaps}} \\ \hline
                                                    & {\color[HTML]{036400} Florete} & 47.24 & 94.43 & 140.63  \\
                                        & {\color[HTML]{036400} Espada} & 153.59 & 257.19 &  365.49 \\
\multicolumn{1}{c}{{Total time at 150 MHz}}      & {\color[HTML]{036400} Sable} & 31.61  & 39.35 & 48.43 \\ 
                                                & & & &\\
                                    & {\color[HTML]{036400} Saber} & 36.35  & 44.12 & 53.56 \\ 

         \hline                                                                
\end{tabular}
\end{table}


\subsubsection{Area Results}
In Table \ref{tab:FPGA_area_results} a detailed breakdown of the area utilization of the full instruction-set coprocessor architecture of all Scabbard schemes and Saber is provided. We include numbers of the internal building blocks. 

Our full HW Sable implementation requires 27\% less LUTs compared to the state-of-the-art Saber implementation and 15\% more FFs. The increase in registers is related to the choice of $\epsilon_p=9$ and $\epsilon_q=11$, which result in non-multiples of 8-bit operands. As a result, larger intermediate buffers are required to temporarily store data operands during conversion to 16-bit operands. Our flexible Espada implementation utilizes 20\% fewer LUTs and requires 14 BRAM tiles (to store $l$ rows of $\mathbf{A}$) instead of 2 in the Saber implementation. On the matrix-vector level, our Espada implementation is fully parallelized, consisting of $l=12$ multipliers and intermediate buffers.

For all implementations, the poly-vector multiplier is the largest contributor at 75/66\%, 39/62\%, and 58/49\% of total LUT/FF counts (and all DSP units) for Florete, Espada, and Sable, respectively.
Still, due to making hardware-aware design choices, our multipliers perform well compared to prior art. Our Sable multiplier utilizes 43\% fewer LUTs and only 9\% more FFs due to its small secret coefficients and custom shift register architecture. Our Espada design requires 58\% fewer LUTs and around two times as many FFs. This is due to the fact that each of the $l$ multipliers requires intermediate buffers for the unpacking of public matrix $\mathbf{A}$ to 16-bit coefficients in parallel.
Our Florete multiplier design consists of a full Toom-Cook 3 and 4-way evaluation and interpolation (pipelined) datapath, with 35 64$\times$64 multipliers. Because of the massive parallelization, our implementation utilizes 21\% more LUTs and 2 times more FFs compared to Saber.
By choosing only 4 DSP units per multiplier, we keep the area utilization at a reasonable level. However, our flexible design allows to use of any 64$\times$64 polynomial multiplier, including increasing the DSP units per multiplier.

\begin{table}[!ht]
\centering
\caption{Area results for full HW implementation of Scabbard schemes and Saber \cite{Sujoy_Saber_h/w}, with clock frequency constraint set to 250MHz in Vivado (Medium security parameters).}
\label{tab:FPGA_area_results}
\begin{tabular}{lc|rrrr}
\hline
\multicolumn{1}{c}{{\color[HTML]{3531FF} }}                              & {\color[HTML]{3531FF} }                                                                           & \multicolumn{3}{c}{{\color[HTML]{3531FF}}}                                                                                                                                                 \\
\multicolumn{1}{c}{\multirow{-2}{*}{{\color[HTML]{3531FF} Block}}} & \multirow{-2}{*}{{\color[HTML]{3531FF} \begin{tabular}[c]{@{}c@{}}Scheme\\ Name\end{tabular}}} & \multicolumn{1}{c}{{\color[HTML]{3531FF} LUTs}} & \multicolumn{1}{c}{{\color[HTML]{3531FF} FFs}} & \multicolumn{1}{c}{{\color[HTML]{3531FF} DSPs}} & \multicolumn{1}{c}{{\color[HTML]{3531FF} BRAMs}} \\ \hline
                                    & {\color[HTML]{036400} Florete} & 5,834 & 3,126 & 0 & 0  \\
                                    & {\color[HTML]{036400} Espada} & 5,964 & 3,127 & 0 & 0   \\
\multicolumn{1}{c}{{SHA3/SHAKE}}      & {\color[HTML]{036400} Sable} & 5,831 & 3,127 & 0 & 0 \\
                                    & & & &\\
                                    & {\color[HTML]{036400} Saber} &  5,113 & 3,068 & 0 & 0 \\ \hline
                                    
                                    & {\color[HTML]{036400} Florete} & 79 & 86 & 0 & 0  \\
                                    & {\color[HTML]{036400} Espada} & 253 & 282 & 0 & 0  \\
\multicolumn{1}{c}{{Binomial Sampler}}      & {\color[HTML]{036400} Sable} & 67 & 86 & 0 & 0  \\
                                    & & & &\\
                                    & {\color[HTML]{036400} Saber} &  92 & 88 & 0 & 0 \\ \hline
                                    
                                    & {\color[HTML]{036400} Florete} & 21,143 & 10,613 & 140 & 0  \\
                                    & {\color[HTML]{036400} Espada} & 7,286  & 11,662 & 48 & 0  \\
\multicolumn{1}{c}{{Poly-vector multiplier}}      & {\color[HTML]{036400} Sable} & 9,841 & 5,523 & 0 & 0 \\
                                    & & & &\\
                                    & {\color[HTML]{036400} Saber} &  17,429 & 5,083 & 0 & 0 \\ \hline
                                    
                                    & {\color[HTML]{036400} Florete} & 1,225 & 2,204 & 0 & 0  \\
                                    & {\color[HTML]{036400} Espada} & 5,238 & 3,752 & 0 & 0  \\
\multicolumn{1}{c}{{Other blocks}}      & {\color[HTML]{036400} Sable} & 1,353 & 2,544 & 0 & 0 \\
                                    & & & &\\
                                    & {\color[HTML]{036400} Saber} &  1,052 & 1,566 & 0 & 0 \\ \hline \hline
                                    
                                        & {\color[HTML]{036400} Florete} & 28,281 & 16,029 & 140 & 2  \\
                                        & {\color[HTML]{036400} Espada} & 18,741 & 18,823 &  48 &  14 \\
\multicolumn{1}{c}{{Full co-processor}}      & {\color[HTML]{036400} Sable} & 17,092  & 11,280 & 0 & 2 \\ 
                                        & & & &\\
                                    & {\color[HTML]{036400} Saber} & 23,686  & 9,805 & 0 & 2 \\    \hline                             
\end{tabular}
\end{table}

\subsubsection{Comparisons with existing implementations}
Our high-speed and highly flexible architectures for Florete, Espada, and Sable are compared with recent hardware implementations of other post-quantum KEM schemes in Table \ref{tab:comparisonHW}. It is important to note that different hardware implementations target different schemes, security levels, platforms, or design methodologies. As a result, a fair and direct comparison is not always possible. The timing results of our implementation are derived from the Vivado simulation, calculated at 250MHz.

The fairest comparison is of our high-speed Sable implementation with the Saber implementation by \cite{Sujoy_Saber_h/w}. Both are high-speed designs and implement the polynomial multiplier with 256 MAC units and multipliers, costing 256 cycles in total. Due to the hardware-aware design decisions, Sable requires 13/11/10$\%$ fewer clock cycles and has a lower area utilization.

Our Espada and Florete implementations are targeting a trade-off between high-speed and low-area. More specifically, the multiplier architecture around the 64$\times$64 polynomial multipliers is highly parallelized and pipelined, allowing for efficient data transfers. We implement the 64$\times$64 multipliers with only 4 DSP units per multiplier in order to reduce area utilization, requiring 64*16 clock cycles for complete multiplication. Our flexible design allows for the DSP count to be increased and directly reduces the total latency up to a factor of 16 or to be replaced with any available off-the-shelf designs.

Frodo KEM is implemented in hardware by How et al.~\cite{FrodoHW} and uses dedicated data paths for \texttt{KeyGen}, \texttt{Encaps} and \texttt{Decaps}. The security of Frodo is based on the standard LWE problem, meaning the computationally expensive matrix-vector multiplications need to be computed several times. As a result, the latency of KEM operations is significantly higher compared to ring or module lattice-based schemes, like Scabbard or Saber. The high-speed Kyber implementation targets an extremely high operating frequency (450MHz), which translates into a faster execution time.

Compared to the high-speed NTRU prime hardware implementation by Peng et al.~\cite{NTRUHW}, Sable outperforms both in performance and area utilization. Florete achieves faster (total) execution time at lower area utilization, mainly due to the expensive \texttt{KeyGen} of NTRU Prime. Espada is slower yet has significantly lower area utilization, which is our design goal nonetheless.
Compared to the low area NTRU prime hardware implementation, all Scabbard implementations significantly outperform NTRU performance-wise. However, this NTRU implementation has a lower area utilization than any other work listed in Table \ref{tab:comparisonHW}.

\textcolor{black}{More recently, several designs tailored for ASIC have been published. Ghosh et al.~\cite{saberASIC} implemented NIST Round 3 Saber~\cite{saber_round_3} in TSMC 65nm technology, targeting a low power consumption. Their crypto accelerator runs at 160 MHz and occupies 0.158 mm$^2$ and is 2.07/0.76/4.92 times slower compared to our implementations. We also highlight a unified Dilithium/Kyber ASIC implementation by Aikata et al.~\cite{kaliASIC}, occupying 0.263 mm$^2$ (TSMC 28 nm) and which utilizes multiple clock domains. As they utilize an advanced technology node, their design can run at 2 GHz (compared to 250 MHz) and still our Sable implementation is only $9\%$ slower in total execution time.}

\begin{table}[!ht]
\centering
\caption{\textcolor{black}{Overview and comparison of Scabbard schemes and existing hardware implementations of CCA-secure KEM schemes. (Medium security level.)}}
\label{tab:comparisonHW}
\begin{tabular}{l|cccc}
\hline
\multicolumn{1}{c|}{{\color[HTML]{3531FF} }}   & \multicolumn{4}{c}{{\color[HTML]{3531FF}}}  \\
\multicolumn{1}{c|}{{\color[HTML]{3531FF} }}   & \multicolumn{4}{c}{{\color[HTML]{3531FF}}}  \\
\multicolumn{1}{c|}{\multirow{-3}{*}{{\color[HTML]{3531FF} Implementation}}} & 
\multicolumn{1}{c}{\multirow{-3}{*}{{\color[HTML]{3531FF} Platform}}} & 
\multicolumn{1}{c}{\multirow{-3}{*}{{\color[HTML]{3531FF} \begin{tabular}[c]{@{}c@{}}Time in $\mu$s\\ (\texttt{KeyGen}/\texttt{Encaps}/\texttt{Decaps})\end{tabular}}}} & 
\multicolumn{1}{c}{\multirow{-3}{*}{{\color[HTML]{3531FF} \begin{tabular}[c]{@{}c@{}}Frequency\\ (MHz)\end{tabular}}}} & 
\multicolumn{1}{c}{\multirow{-3}{*}{{\color[HTML]{3531FF} \begin{tabular}[c]{@{}c@{}}Area\\ (LUT/FF/DSP/BRAM)\\(or $mm^2$ for ASIC)\end{tabular}}}} \\ \hline

\multicolumn{1}{c|}{{\color[HTML]{036400} Florete}}  & UltraScale+ & 28.3/56.7/84.4 & 250 & 28.2K/16.0K/140/2  \\
\multicolumn{1}{c|}{{\color[HTML]{036400} Espada}}  & UltraScale+ & 92.2/154.3/219.3 & 250 & 18.7K/18.8K/48/14  \\
\multicolumn{1}{c|}{{\color[HTML]{036400} Sable}}  & UltraScale+ & 18.9/23.6/29.0 & 250 & 17.0K/11.2K/0/2  \\
\multicolumn{1}{c|}{{\color[HTML]{036400} Saber}~\cite{Sujoy_Saber_h/w}}  & UltraScale+ & 21.8/26.5/32.1 & 250 & 23.6K/9.8K/0/2 \\
\multicolumn{1}{c|}{{\color[HTML]{036400} Kyber}~\cite{NTRUHW}}  & UltraScale+ & 5.9/8.3/10.9 & 450 & 10.6K/10.5K/6/6.5\\
\multicolumn{1}{c|}{{\color[HTML]{036400} Frodo}~\cite{FrodoHW}}  & Artix-7 & 45K/45K/47K & 167 & $\approx$7.7K/3.5K/1/24\\
\multicolumn{1}{c|}{{\color[HTML]{036400} NTRU Prime (High Speed)}~\cite{Peng2022StreamlinedNP}}  & UltraScale+ & 224.7/17.3/38.6 & 285 & 40.1K/26.4K/36.5/31\\
\multicolumn{1}{c|}{{\color[HTML]{036400} NTRU Prime (Low Area)}~\cite{Peng2022StreamlinedNP}}  & UltraScale+ & 2.2K/100.8/302.6 & 285 & 9.2K/4.4K/8.5/18\\
\multicolumn{1}{c|}{{\color[HTML]{036400} Saber}~\cite{saberASIC}}  & ASIC (65 nm) & 89/117/146 & 160 & 0.158\\  \multicolumn{1}{c|}{{\color[HTML]{036400} Kyber}~\cite{kaliASIC}}  & ASIC (28 nm) & 6.18/11.09/47.89 & 2K& 0.263\\  \hline                                                             
\end{tabular}
\end{table}

From our experimental performance evaluation we can conclude that our Scabbard coprocessors have fast computation time compared to other lattice-based KEM hardware implementations, with moderate area utilization. In the case of Sable, which utilizes the same multiplier architecture as the Saber implementation in \cite{Sujoy_Saber_h/w}, our implementation requires 13\%/11\%/10$\%$ fewer clock cycles for KEM operations and lower area utilization.
By considering hardware design choices during the design of the schemes themselves, efficient implementations have been achieved. Our full-hardware instruction-set coprocessor architectures result in high-speed and highly flexible implementations. We leave support for multiple parameter sets and security versions in a single hardware implementation as future work.


%% file: sections/sca.tex
\section{Physical attack analysis}
\textcolor{black}{
Physical attacks have been demonstrated to be very potent against even for mathematically secure cryptographic algorithms~\cite{Kocher96,KocherJJ99,KoeuneS04,BonehDL97}, and lattice-based cryptography is no exception~\cite{HCY2019,RBS+2020,ACL+2020,GJN2020,catinca_sca}. Therefore, physical attack analysis is one of the essential measures that ought to be carried out before deploying cryptographic algorithms in the real-world. Physical attacks can be divided into two categories depending on their properties: (i) passive attacks, which include timing-based side-channel attacks~\cite{Kocher96,GJN2020}, power-based side-channel attacks~\cite{KocherJJ99,HCY2019,ACL+2020,catinca_sca,DBLP:journals/tecs/AydinATGO21}, side-channel attacks based on electromagnetic radiation (EM)~\cite{KoeuneS04,RBS+2020}, etc., and (ii) active attacks, which include fault-injection attacks~\cite{BonehDL97,RBS+2020,puja-rowhammer-fault,carry-your-fault}.}

\textcolor{black}{
Timing-based SCA is mitigated with constant-time implementations, where the execution time of the cryptographic algorithm does not depend on the secret data. All of our software and hardware implementations of Scabbard are in constant-time. It has been accomplished by avoiding secret-data-dependent operations (such as division), secret-data-dependent conditional operations (if-else), or secret-data-dependent memory access. Recently, Bernstein et al.~\cite{DBLP:journals/iacr/BernsteinBBCCKKPRT24} have proposed that the modular divisions used in Kyber's implementation are vulnerable to timing SCA due to division by prime modulus. These timing SCA do not apply to our schemes as our divisions are merely shift operations.} 

\textcolor{black}{
The implementations of Scabbard are susceptible to power- or EM-based SCA like other lattice-based schemes, e.g., Kyber~\cite{RBS+2020}, Saber~\cite{catinca_sca}, NewHope~\cite{RBS+2020}, Frodo~\cite{DBLP:journals/tecs/AydinATGO21}, NTRU Prime~\cite{HCY2019}, etc. More specifically, as our schemes use similar constructions as Saber, most of the power- or EM-based SCA shown on Saber are also applicable to Scabbard's schemes. For example,~\cite{catinca_sca} has shown correlation power analysis based SCA on the Toom-Cook-based polynomial multiplication of Saber; similar attacks are possible on the schemes of Scabbard. 
}

\textcolor{black}{
Masking~\cite{ChaJutRaoRoh1999} or shuffling~\cite{DBLP:conf/acns/HerbstOM06} are utilized to thwart power-based or EM-based SCA. Between these two countermeasures, masking is a provably secure countermeasure and is usually integrated with lattice-based cryptographic algorithms to prevent SCA~\cite{HO_mask_Saber,BronchainC22}. Although shuffling is a low-cost countermeasure compared to masking, it has been shown to be insufficient~\cite{RaviPJDB24} to prevent SCA when shuffling is used alone. Therefore, the overhead reduction in incorporating masking countermeasures into lattice-based KEMs is one of the crucial steps. One way to achieve this is by exploring masking-friendly design elements of the existing lattice-based KEMs.}    

\textcolor{black}{
In masking, secret dependent variables (e.g., $x$) are split into multiple separate shares (for the first-order masking, $x$ is divided into two shares $x_1$ and $x_2$). Then, all the operations of the cryptographic algorithms are performed independently on all the shares. To achieve efficiency, masking LWE/LWR-based KEMs requires two kinds of masking techniques, i$)$ arithmetic masking ($x=x_1+x_2 \bmod{q}$) and ii$)$ Boolean masking ($x=x_1\oplus x_2$). Masked LWE/LWR-based KEMs mainly use the following components: (i) masked polynomial arithmetic (modular addition, subtraction, and multiplication), (ii) masked compression, (iii) masked message decoding and encoding function, (iv) masked CBD, (v) masked Keccak (used in SHA3-512 and SHAKE-128), and (vi) masked ciphertext comparison. From all the components, (ii) Masked compression, (iii) masked message decoding and encoding function, (iv) masked CBD, and (vi) masked ciphertext comparison require either arithmetic to Boolean (A2B) conversion or Boolean to arithmetic (B2A) conversion. A2B or B2A conversions are one of the performance hefty operations introduced solely due to masking, making the masked components that use them expensive in terms of performance. However, this performance cost heavily depends on the parameters of the LWE/LWR-based KEMs. For example, the modulus $q$ is usually chosen to be prime for LWE-based schemes for being able to use NTT, whereas it is mostly a power-of-2 for LWR-based schemes. A2B and B2A conversions are much cheaper for a power-of-2 modulus compared to a prime modulus with the same bit length.}
\textcolor{black}{Also, the smaller parameters in our schemes reduce the performance overhead of masking components. More elaborated observations regarding the design choices of the schemes of Scabbard and their effect on masking have been shown in~\cite{masked_scabbard}. We have presented some of these results in Appendix~\ref{appendix}. Overall, all three schemes of Scabbard outperform Kyber on the Cortex-M4 platform when masking countermeasures are integrated.}

\textcolor{black}{
Masking LWE/LWR-based KEMs can prevent some of the fault-injection attacks (FIA), such as safe-error attacks~\cite{BettaleMR21}. However, several FIA have been proposed on masked implementation of LWE/LWR-based KEMs~\cite{PesslP21,HermelinkPP21,Delvaux22,XagawaIUTH21,puja-rowhammer-fault,carry-your-fault}. In fact, \cite{carry-your-fault} demonstrates that masking introduces new attack surfaces for the FIA in the LWE/LWR-based KEMs. The FIA in the context of LWE/LWR-based schemes can be primarily divided into two categories: (i) ineffective fault attacks and (ii) FIA at the ciphertext comparison. }
\textcolor{black}{In ineffective fault attacks, the secret data-dependent behavioral changes of the decapsulation procedure of the KEMs upon fault injection on a specific variable leak information regarding the secret key. Basically, in these attacks, injected fault changes the value of the targeted variable, which causes decapsulation failure for some values of the targeted variable. For the other values of the targeted variable, the injected fault doesn't affect the final outcome \textit{i.e.} decapsulation success. This phenomenon leaks information regarding the targeted variable's value, which depends on the secret key. }
\textcolor{black}{Some examples of such fault attacks are~\cite{PesslP21,HermelinkPP21,Delvaux22,carry-your-fault}. In the FIA at the ciphertext comparison~\cite{XagawaIUTH21,puja-rowhammer-fault}, the last equality checking in the decapsulation procedure between re-encrypted ciphertext and received public ciphertext is bypassed. Fundamentally, this removes the Fujisaki-Okamoto transform \textit{i.e} changes a CCA-secure KEM scheme to a CPA-secure KEX scheme. It forces the decapsulation to succeed even in cases where decapsulation would have failed in the normal scenario. Therefore, the adversary can retrieve the long-term secret key from the decapsulation process with the help of specially crafted input ciphertexts.}

\textcolor{black}{Scabbard's schemes are also vulnerable to the fault-injection attacks discussed here. Recently, some works~\cite{RaviCDB24,BerthetTDS23} have proposed detection-based countermeasures against FIA on LWE/LWR-based KEMs. These countermeasures can be integrated into Scabbard with small adjustments. However, further research is needed to verify the effectiveness and cost of these countermeasures on Scabbard.}

%% file: sections/conclusion.tex
\section{Conclusion}

We provide a suite of three LWR-based KEMs by exploring possible design choices and the parameter set. Our study improves the state-of-the-art lattice-based post-quantum cryptography in aspects of software and hardware implementations. We show that the choice of design primitives heavily affects the scheme's efficiency on the software and hardware platforms. In fact, the design choices of a lattice-based KEM also affect the performance overhead of the scheme's secure implementations. \textcolor{black}{The work in~\cite{masked_scabbard} experimentally demonstrated that the schemes of Scabbard outperform Kyber on the Cortex-m4 platform when side-channel countermeasure masking is integrated.} In this work, we consider implementation aspects during the design of a scheme, which results in a more efficient scheme while providing a similar level of security. 
\textcolor{black}{Our result opens a new research direction for LWR-based lightweight secure PQC KEMs, which can be extended to LWE-based KEMs. In fact, a new lightweight MLWE-based scheme, Rudraksh~\cite{cryptoeprint:2024/1170}, has been proposed by performing a similar module-space exploration strategy on LWE-based primitives. We believe this research will benefit other LWE-/LWR-based primitives such as lattice-based digital signature schemes, lightweight schemes, group-key exchange schemes, etc. We have left these as potential future work.}

%% file: sections/appendix.tex
\section{Performance of masked Scabbard}~\label{appendix}
\textcolor{black}{
Kundu et al.~\cite{masked_scabbard} integrated masking countermeasures on the medium security version (NIST-3) of all the schemes of Scabbard. It also proposes proof-of-concept implementations on the ARM Cortex-M4 platform using the PQM4 framework~\cite{PQM4}. The test vector leakage assessment hasn't been performed and is left as future work. Table~\ref{tab:Florete_performance},~\ref{tab:Espada_performance}, and~\ref{tab:Sable_performance} present performance results of masked Florete, Espada, and Sable, respectively. In Table~\ref{tab:masked_state-of-the-art_performance}, masked implementations of Scabbard have been compared with the state-of-the-art implementations of LWE/LWR-based KEMs, including Kyber.}
\begin{table}[!ht]
\centering
\caption{\textcolor{black}{Performance of components of Florete on Cortex-M4}~\cite{masked_scabbard}}
\label{tab:Florete_performance}
\resizebox{.9\textwidth}{!}{%
\begin{tabular}{llllllllrrrrrrrrrr}
\hline
                      &  & \multicolumn{5}{l}{}                                                  &  & \multicolumn{10}{c}{\color[HTML]{3531FF}{x1000 clock cycles}}                                                                                                                                                                                                                                                                                            \\ \cline{9-18} 
                      &  & \multicolumn{5}{c}{\color[HTML]{3531FF}{Order}}                                             &  & \multicolumn{1}{c|}{\color[HTML]{3531FF}{Unmask}}                & \multicolumn{3}{c|}{\color[HTML]{3531FF}{1st}}                                                                       & \multicolumn{3}{c|}{\color[HTML]{3531FF}{2nd}}                                                                         & \multicolumn{3}{c}{\color[HTML]{3531FF}{3rd}}                                                      \\ \hline
\multicolumn{7}{l}{{ \color[HTML]{036400}{\textbf{Florete CCA-KEM-Decapsulation}}}}                &  & \multicolumn{1}{r|}{954}                   &                      & 2,621                 & \multicolumn{1}{r|}{(2.74x)}                    &                      & 4,844                   & \multicolumn{1}{r|}{(5.07x)}                    &                      & 7,395                   & (7.75x)                     \\
\multicolumn{1}{l|}{} &  & \multicolumn{5}{l}{\textbf{CPA-PKE-Decryption}}                       &  & \multicolumn{1}{r|}{248}                   &                      & 615                   & \multicolumn{1}{r|}{(2.47x)}                    &                      & 1,107                   & \multicolumn{1}{r|}{(4.46x)}                    &                      & 1,651                   & (6.65x)                     \\
\multicolumn{1}{l|}{} &  & \multicolumn{1}{l|}{} &  & \multicolumn{3}{l}{Polynomial arithmetic}  &  & \multicolumn{1}{r|}{241}                   &                      & 461                   & \multicolumn{1}{r|}{(1.91x)}                    &                      & 690                     & \multicolumn{1}{r|}{(2.86x)}                    &                      & 917                     & (3.80x)                     \\
\multicolumn{1}{l|}{} &  & \multicolumn{1}{l|}{} &  & \multicolumn{3}{l}{Compression}            &  & \multicolumn{1}{r|}{}                      &                      &                       & \multicolumn{1}{r|}{}                           &                      &                         & \multicolumn{1}{r|}{}                           &                      &                         &                             \\
\multicolumn{1}{l|}{} &  & \multicolumn{1}{l|}{} &  & \multicolumn{3}{l}{$original\_msg$}        &  & \multicolumn{1}{r|}{\multirow{-2}{*}{6}}   & \multicolumn{1}{l}{} & \multirow{-2}{*}{153} & \multicolumn{1}{r|}{\multirow{-2}{*}{(25.50x)}} & \multicolumn{1}{l}{} & \multirow{-2}{*}{416}   & \multicolumn{1}{r|}{\multirow{-2}{*}{(69.33x)}} & \multicolumn{1}{l}{} & \multirow{-2}{*}{734}   & \multirow{-2}{*}{(122.33x)} \\
\multicolumn{1}{l|}{} &  & \multicolumn{5}{l}{\textbf{Hash $\mathcal{G}$ (SHA3-512)}}            &  & \multicolumn{1}{r|}{13}                    &                      & 123                   & \multicolumn{1}{r|}{(9.46x)}                    &                      & 242                     & \multicolumn{1}{r|}{(18.61x)}                   &                      & 379                     & (29.15x)                    \\
\multicolumn{1}{l|}{} &  & \multicolumn{5}{l}{\textbf{CPA-PKE-Encryption}}                       &  & \multicolumn{1}{r|}{554}                   &                      & 1,744                 & \multicolumn{1}{r|}{(3.14x)}                    &                      & 3,354                   & \multicolumn{1}{r|}{(6.05x)}                    &                      & 5,225                   & (9.43x)                     \\
\multicolumn{1}{l|}{} &  & \multicolumn{1}{l|}{} &  & \multicolumn{3}{l}{Secret generation}      &  & \multicolumn{1}{r|}{29}                    &                      & 427                   & \multicolumn{1}{r|}{(14.72x)}                   &                      & 982                     & \multicolumn{1}{r|}{(33.86x)}                   &                      & 1,663                   & (57.34x)                    \\
\multicolumn{1}{l|}{} &  & \multicolumn{1}{l|}{} &  & \multicolumn{1}{l|}{} &  & XOF (SHAKE-128) &  & \multicolumn{1}{r|}{25}                    &                      & 245                   & \multicolumn{1}{r|}{(9.80x)}                    &                      & 484                     & \multicolumn{1}{r|}{(19.36x)}                   &                      & 756                     & (30.24x)                    \\
\multicolumn{1}{l|}{} &  & \multicolumn{1}{l|}{} &  & \multicolumn{1}{l|}{} &  & CBD ($\beta_1$) &  & \multicolumn{1}{r|}{4}                     &                      & 182                   & \multicolumn{1}{r|}{(45.50x)}                   &                      & 497                     & \multicolumn{1}{r|}{(124.25x)}                  &                      & 907                     & (226.75x)                   \\
\multicolumn{1}{l|}{} &  & \multicolumn{1}{l|}{} &  & \multicolumn{3}{l}{Polynomial arithmetic}  &  & \multicolumn{1}{r|}{}                      &                      &                       & \multicolumn{1}{r|}{}                           &                      &                         & \multicolumn{1}{r|}{}                           &                      &                         &                             \\
\multicolumn{1}{l|}{} &  & \multicolumn{1}{l|}{} &  & \multicolumn{3}{l}{$arrange\_msg$}         &  & \multicolumn{1}{r|}{}                      & \multicolumn{1}{l}{} & \multirow{-2}{*}{943} & \multicolumn{1}{r|}{}                           & \multicolumn{1}{l}{} & \multirow{-2}{*}{1,357} & \multicolumn{1}{r|}{}                           & \multicolumn{1}{l}{} & \multirow{-2}{*}{1,783} &                             \\
\multicolumn{1}{l|}{} &  & \multicolumn{1}{l|}{} &  & \multicolumn{3}{l}{Polynomial Comparison}  &  & \multicolumn{1}{r|}{\multirow{-3}{*}{524}} & \multicolumn{1}{c}{} & 373                   & \multicolumn{1}{r|}{\multirow{-3}{*}{(2.51x)}}  & \multicolumn{1}{l}{} & 1,014                   & \multicolumn{1}{r|}{\multirow{-3}{*}{(4.52x)}}  & \multicolumn{1}{l}{} & 1,778                   & \multirow{-3}{*}{(6.79x)}   \\
\multicolumn{1}{l|}{} &  & \multicolumn{5}{l}{Other operations}                                  &  & \multicolumn{1}{r|}{138}                   &                      & 139                   & \multicolumn{1}{r|}{(1.00x)}                    &                      & 140                     & \multicolumn{1}{r|}{(1.01x)}                    &                      & 140                     & (1.01x)                     \\ \hline
\end{tabular}%
}
\end{table}

\begin{table}[!ht]
\centering
\caption{\textcolor{black}{Performance of components of Espada on Cortex-M4}~\cite{masked_scabbard}}
\label{tab:Espada_performance}
\resizebox{.9\textwidth}{!}{%
\begin{tabular}{llllllllrrrrrrrrrr}
\hline
                      &  & \multicolumn{5}{l}{}                                                  &  & \multicolumn{10}{c}{\color[HTML]{3531FF}{x1000 clock cycles}}                                                                                                                                                                                                                                                                                            \\ \cline{9-18} 
                      &  & \multicolumn{5}{c}{\color[HTML]{3531FF}{Order}}                                             &  & \multicolumn{1}{c|}{\color[HTML]{3531FF}{Unmask}}                & \multicolumn{3}{c|}{\color[HTML]{3531FF}{1st}}                                                                       & \multicolumn{3}{c|}{\color[HTML]{3531FF}{2nd}}                                                                         & \multicolumn{3}{c}{\color[HTML]{3531FF}{3rd}}                                                      \\ \hline
\multicolumn{7}{l}{{ \color[HTML]{036400}{\textbf{Espada CCA-KEM-Decapsulation}}}}                 &  & \multicolumn{1}{r|}{2,422}                   &                      & 4,335                   & \multicolumn{1}{r|}{(1.78x)}                    &                      & 6,838                   & \multicolumn{1}{r|}{(2.82x)}                     &                      & 9,861                   & (4.07x)                     \\
\multicolumn{1}{l|}{} &  & \multicolumn{5}{l}{\textbf{CPA-PKE-Decryption}}                       &  & \multicolumn{1}{r|}{70}                      &                      & 137                     & \multicolumn{1}{r|}{(1.95x)}                    &                      & 230                     & \multicolumn{1}{r|}{(3.28x)}                     &                      & 324                     & (4.62x)                     \\
\multicolumn{1}{l|}{} &  & \multicolumn{1}{l|}{} &  & \multicolumn{3}{l}{Polynomial arithmetic}  &  & \multicolumn{1}{r|}{69}                      &                      & 116                     & \multicolumn{1}{r|}{(1.68x)}                    &                      & 170                     & \multicolumn{1}{r|}{(2.46x)}                     &                      & 225                     & (3.26x)                     \\
\multicolumn{1}{l|}{} &  & \multicolumn{1}{l|}{} &  & \multicolumn{3}{l}{Compression}            &  & \multicolumn{1}{r|}{}                        &                      &                         & \multicolumn{1}{r|}{}                           &                      &                         & \multicolumn{1}{r|}{}                            &                      &                         &                             \\
\multicolumn{1}{l|}{} &  & \multicolumn{1}{l|}{} &  & \multicolumn{3}{l}{$original\_msg$}        &  & \multicolumn{1}{r|}{\multirow{-2}{*}{0.4}}   & \multicolumn{1}{l}{} & \multirow{-2}{*}{20}    & \multicolumn{1}{r|}{\multirow{-2}{*}{(50.00x)}} & \multicolumn{1}{l}{} & \multirow{-2}{*}{60}    & \multicolumn{1}{r|}{\multirow{-2}{*}{(150.00x)}} & \multicolumn{1}{l}{} & \multirow{-2}{*}{99}    & \multirow{-2}{*}{(247.50x)} \\
\multicolumn{1}{l|}{} &  & \multicolumn{5}{l}{\textbf{Hash $\mathcal{G}$ (SHA3-512)}}            &  & \multicolumn{1}{r|}{13}                      &                      & 123                     & \multicolumn{1}{r|}{(9.46x)}                    &                      & 243                     & \multicolumn{1}{r|}{(18.69x)}                    &                      & 379                     & (29.15x)                    \\
\multicolumn{1}{l|}{} &  & \multicolumn{5}{l}{\textbf{CPA-PKE-Encryption}}                       &  & \multicolumn{1}{r|}{2,215}                   &                      & 3,950                   & \multicolumn{1}{r|}{(1.78x)}                    &                      & 6,240                   & \multicolumn{1}{r|}{(2.81x)}                     &                      & 9,031                   & (4.07x)                     \\
\multicolumn{1}{l|}{} &  & \multicolumn{1}{l|}{} &  & \multicolumn{3}{l}{Secret generation}      &  & \multicolumn{1}{r|}{57}                      &                      & 748                     & \multicolumn{1}{r|}{(13.12x)}                   &                      & 1,650                   & \multicolumn{1}{r|}{(28.94x)}                    &                      & 3,009                   & (52.78x)                    \\
\multicolumn{1}{l|}{} &  & \multicolumn{1}{l|}{} &  & \multicolumn{1}{l|}{} &  & XOF (SHAKE-128) &  & \multicolumn{1}{r|}{51}                      &                      & 489                     & \multicolumn{1}{r|}{(9.58x)}                    &                      & 968                     & \multicolumn{1}{r|}{(18.98x)}                    &                      & 1,510                   & (29.60x)                    \\
\multicolumn{1}{l|}{} &  & \multicolumn{1}{l|}{} &  & \multicolumn{1}{l|}{} &  & CBD ($\beta_3$) &  & \multicolumn{1}{r|}{6}                       &                      & 259                     & \multicolumn{1}{r|}{(43.16x)}                   &                      & 681                     & \multicolumn{1}{r|}{(113.50x)}                   &                      & 1,498                   & (249.66x)                   \\
\multicolumn{1}{l|}{} &  & \multicolumn{1}{l|}{} &  & \multicolumn{3}{l}{Polynomial arithmetic}  &  & \multicolumn{1}{r|}{}                        &                      &                         & \multicolumn{1}{r|}{}                           &                      &                         & \multicolumn{1}{r|}{}                            &                      &                         &                             \\
\multicolumn{1}{l|}{} &  & \multicolumn{1}{l|}{} &  & \multicolumn{3}{l}{$arrange\_msg$}         &  & \multicolumn{1}{r|}{}                        & \multicolumn{1}{l}{} & \multirow{-2}{*}{2,865} & \multicolumn{1}{r|}{}                           & \multicolumn{1}{l}{} & \multirow{-2}{*}{3,593} & \multicolumn{1}{r|}{}                            & \multicolumn{1}{l}{} & \multirow{-2}{*}{4,354} &                             \\
\multicolumn{1}{l|}{} &  & \multicolumn{1}{l|}{} &  & \multicolumn{3}{l}{Polynomial Comparison}  &  & \multicolumn{1}{r|}{\multirow{-3}{*}{2,157}} & \multicolumn{1}{c}{} & 259                     & \multicolumn{1}{r|}{\multirow{-3}{*}{(1.44x)}}  & \multicolumn{1}{l}{} & 996                     & \multicolumn{1}{r|}{\multirow{-3}{*}{(2.12x)}}   & \multicolumn{1}{l}{} & 1,667                   & \multirow{-3}{*}{(2.79x)}   \\
\multicolumn{1}{l|}{} &  & \multicolumn{5}{l}{Other operations}                                  &  & \multicolumn{1}{r|}{124}                     &                      & 124                     & \multicolumn{1}{r|}{(1.00x)}                    &                      & 124                     & \multicolumn{1}{r|}{(1.00x)}                     &                      & 126                     & (1.01x)                     \\ \hline
\end{tabular}%
}
\end{table}

\begin{table}[!ht]
\centering
\caption{\textcolor{black}{Performance of components of Sable on Cortex-M4}~\cite{masked_scabbard}}
\label{tab:Sable_performance}
\resizebox{.9\textwidth}{!}{%
\begin{tabular}{llllllllrrrrrrrrrr}
\hline
                      &  & \multicolumn{5}{l}{}                                                  &  & \multicolumn{10}{c}{\color[HTML]{3531FF}{x1000 clock cycles}}                                                                                                                                                                                                                                                                                            \\ \cline{9-18} 
                      &  & \multicolumn{5}{c}{\color[HTML]{3531FF}{Order}}                                             &  & \multicolumn{1}{c|}{\color[HTML]{3531FF}{Unmask}}                & \multicolumn{3}{c|}{\color[HTML]{3531FF}{1st}}                                                                       & \multicolumn{3}{c|}{\color[HTML]{3531FF}{2nd}}                                                                         & \multicolumn{3}{c}{\color[HTML]{3531FF}{3rd}}                                                      \\ \hline
\multicolumn{7}{l}{{ \color[HTML]{036400}{\textbf{Sable CCA-KEM-Decapsulation}}}}                  &  & \multicolumn{1}{r|}{1,020}                 &                      & 2,431                   & \multicolumn{1}{r|}{(2.38x)}                    &                      & 4,348                   & \multicolumn{1}{r|}{(4.26x)}                    &                      & 6,480                   & (6.35x)                     \\
\multicolumn{1}{l|}{} &  & \multicolumn{5}{l}{\textbf{CPA-PKE-Decryption}}                       &  & \multicolumn{1}{r|}{130}                   &                      & 291                     & \multicolumn{1}{r|}{(2.23x)}                    &                      & 510                     & \multicolumn{1}{r|}{(3.92x)}                    &                      & 745                     & (5.73x)                     \\
\multicolumn{1}{l|}{} &  & \multicolumn{1}{l|}{} &  & \multicolumn{3}{l}{Polynomial arithmetic}  &  & \multicolumn{1}{r|}{128}                   &                      & 238                     & \multicolumn{1}{r|}{(1.85x)}                    &                      & 350                     & \multicolumn{1}{r|}{(2.73x)}                    &                      & 465                     & (3.63x)                     \\
\multicolumn{1}{l|}{} &  & \multicolumn{1}{l|}{} &  & \multicolumn{3}{l}{Compression}            &  & \multicolumn{1}{r|}{}                      &                      &                         & \multicolumn{1}{r|}{}                           &                      &                         & \multicolumn{1}{r|}{}                           &                      &                         &                             \\
\multicolumn{1}{l|}{} &  & \multicolumn{1}{l|}{} &  & \multicolumn{3}{l}{$original\_msg$}        &  & \multicolumn{1}{r|}{\multirow{-2}{*}{2}}   & \multicolumn{1}{l}{} & \multirow{-2}{*}{52}    & \multicolumn{1}{r|}{\multirow{-2}{*}{(26.00x)}} & \multicolumn{1}{l}{} & \multirow{-2}{*}{160}   & \multicolumn{1}{r|}{\multirow{-2}{*}{(80.00x)}} & \multicolumn{1}{l}{} & \multirow{-2}{*}{280}   & \multirow{-2}{*}{(140.00x)} \\
\multicolumn{1}{l|}{} &  & \multicolumn{5}{l}{\textbf{Hash $\mathcal{G}$ (SHA3-512)}}            &  & \multicolumn{1}{r|}{13}                    &                      & 123                     & \multicolumn{1}{r|}{(9.46x)}                    &                      & 242                     & \multicolumn{1}{r|}{(18.61x)}                   &                      & 379                     & (29.15x)                    \\
\multicolumn{1}{l|}{} &  & \multicolumn{5}{l}{\textbf{CPA-PKE-Encryption}}                       &  & \multicolumn{1}{r|}{764}                   &                      & 1,903                   & \multicolumn{1}{r|}{(2.49x)}                    &                      & 3,482                   & \multicolumn{1}{r|}{(4.55x)}                    &                      & 5,241                   & (6.85x)                     \\
\multicolumn{1}{l|}{} &  & \multicolumn{1}{l|}{} &  & \multicolumn{3}{l}{Secret generation}      &  & \multicolumn{1}{r|}{29}                    &                      & 427                     & \multicolumn{1}{r|}{(14.72x)}                   &                      & 984                     & \multicolumn{1}{r|}{(33.93x)}                   &                      & 1,666                   & (57.44x)                    \\
\multicolumn{1}{l|}{} &  & \multicolumn{1}{l|}{} &  & \multicolumn{1}{l|}{} &  & XOF (SHAKE-128) &  & \multicolumn{1}{r|}{25}                    &                      & 245                     & \multicolumn{1}{r|}{(9.80x)}                    &                      & 484                     & \multicolumn{1}{r|}{(19.36x)}                   &                      & 756                     & (30.24x)                    \\
\multicolumn{1}{l|}{} &  & \multicolumn{1}{l|}{} &  & \multicolumn{1}{l|}{} &  & CBD ($\beta_1$) &  & \multicolumn{1}{r|}{4}                     &                      & 182                     & \multicolumn{1}{r|}{(45.50x)}                   &                      & 499                     & \multicolumn{1}{r|}{(124.75x)}                  &                      & 909                     & (227.25x)                   \\
\multicolumn{1}{l|}{} &  & \multicolumn{1}{l|}{} &  & \multicolumn{3}{l}{Polynomial arithmetic}  &  & \multicolumn{1}{r|}{}                      &                      &                         & \multicolumn{1}{r|}{}                           &                      &                         & \multicolumn{1}{r|}{}                           &                      &                         &                             \\
\multicolumn{1}{l|}{} &  & \multicolumn{1}{l|}{} &  & \multicolumn{3}{l}{$arrange\_msg$}         &  & \multicolumn{1}{r|}{}                      & \multicolumn{1}{l}{} & \multirow{-2}{*}{1,187} & \multicolumn{1}{r|}{}                           & \multicolumn{1}{l}{} & \multirow{-2}{*}{1,640} & \multicolumn{1}{r|}{}                           & \multicolumn{1}{l}{} & \multirow{-2}{*}{2,086} &                             \\
\multicolumn{1}{l|}{} &  & \multicolumn{1}{l|}{} &  & \multicolumn{3}{l}{Polynomial Comparison}  &  & \multicolumn{1}{r|}{\multirow{-3}{*}{734}} & \multicolumn{1}{c}{} & 287                     & \multicolumn{1}{r|}{\multirow{-3}{*}{(2.00x)}}  & \multicolumn{1}{l}{} & 856                     & \multicolumn{1}{r|}{\multirow{-3}{*}{(3.40x)}}  & \multicolumn{1}{l}{} & 1,488                   & \multirow{-3}{*}{(4.86x)}   \\
\multicolumn{1}{l|}{} &  & \multicolumn{5}{l}{Other operations}                                  &  & \multicolumn{1}{r|}{112}                   &                      & 113                     & \multicolumn{1}{r|}{(1.00x)}                    &                      & 113                     & \multicolumn{1}{r|}{(1.00x)}                    &                      & 113                     & (1.00x)                     \\ \hline
\end{tabular}%
}

\end{table}

\begin{table}[!hbt]
\centering
\caption{\textcolor{black}{Comparing performance of masked Scabbard with the state-of-the-art}~\cite{masked_scabbard}}
\label{tab:masked_state-of-the-art_performance}
\resizebox{0.8\columnwidth}{!}{
\begin{tabular}{clrrrr|crr}
\hline
\multicolumn{2}{c}{{ }}                               & \multicolumn{4}{c|}{{ \color[HTML]{3531FF}{Performance}}}                                                                                                                                         & \multicolumn{3}{c}{{ \color[HTML]{3531FF}{\# Random numbers}}}                                                                  \\ \cline{3-9} 
\multicolumn{2}{c}{{ }}                               & \multicolumn{4}{c|}{\color[HTML]{3531FF}{(x1000 clock cycles)}}                                                                                                                                                       & \multicolumn{3}{c}{{ \color[HTML]{3531FF}{(bytes)}}}                                                                           \\ \cline{3-9} 
\multicolumn{2}{c}{\multirow{-3}{*}{{ \color[HTML]{3531FF}{Scheme}}}}       & \multicolumn{1}{c}{{ }} & \multicolumn{1}{c}{{ \color[HTML]{3531FF}{1st}}} & \multicolumn{1}{c}{{ \color[HTML]{3531FF}{2nd}}} & \multicolumn{1}{c|}{{ \color[HTML]{3531FF}{3rd}}} & { \color[HTML]{3531FF}{1st}} & \multicolumn{1}{c}{{ \color[HTML]{3531FF}{2nd}}} & \multicolumn{1}{c}{{ \color[HTML]{3531FF}{3rd}}} \\ \hline
\color[HTML]{036400}{ Florete} & \cite{masked_scabbard}                              & { }                     & 2,621                                          & 4,844                                          & 7,395                                           & \multicolumn{1}{r}{15,824} & 52,176                                         & 101,280                                        \\
\color[HTML]{036400}{ Espada}  & \cite{masked_scabbard}                              & { }                     & 4,335                                          & 6,838                                          & 9,861                                           & \multicolumn{1}{r}{11,496} & 39,320                                         & 85,296                                         \\
\color[HTML]{036400}{ Sable}   & \cite{masked_scabbard}                              & { }                     & \textbf{2,431}                                 & \textbf{4,348}                                 & \textbf{6,480}                                  & \multicolumn{1}{r}{12,496} & 39,152                                         & 75,232                                         \\
\color[HTML]{036400}{ Saber}   & \cite{HO_mask_Saber}                     & { }                     & 3,022                                          & 5,567                                          & 8,649                                           & \multicolumn{1}{r}{12,752} & 43,760                                         & 93,664                                         \\
\color[HTML]{036400}{ uSaber}  & \cite{HO_mask_Saber}                     & { }                     & 2,473                                          & 4,452                                          & 6,947                                           & \multicolumn{1}{r}{10,544} & 36,848                                         & 79,840                                         \\
\color[HTML]{036400}{ Kyber}   & \cite{BronchainC22}                      & \multicolumn{1}{l}{}                        & 10,018                                         & 16,747                                         & 24,709                                          & -                          & \multicolumn{1}{c}{-}                          & \multicolumn{1}{c}{-}                          \\ \hline
\end{tabular}
}
\end{table}
